\documentclass[12pt]{iopart}

\expandafter\let\csname equation*\endcsname\relax
\expandafter\let\csname endequation*\endcsname\relax

\usepackage{amsmath}
\usepackage{amsfonts}
\usepackage{amssymb}

\usepackage[utf8x]{inputenc}
\usepackage[T1]{fontenc}

\usepackage{hyperref}
\usepackage{graphicx}
\usepackage{lmodern} % fonts
\usepackage{esvect} % vector symbol
\usepackage{wrapfig} % wrapping figures
\usepackage{bm}
\usepackage{algorithm,algpseudocode}
\usepackage{listings}
\usepackage{accents} % underaccent

\usepackage{color}

\definecolor{mygreen}{rgb}{0,0.6,0}
\definecolor{mylightgray}{rgb}{0.95,0.95,0.95}
\definecolor{mygray}{rgb}{0.5,0.5,0.5}
\definecolor{mymauve}{rgb}{0.58,0,0.82}

\usepackage{enumitem}
\newlist{abbrv}{itemize}{1}
\setlist[abbrv,1]{label=,labelwidth=1in,align=parleft,itemsep=0.1\baselineskip,leftmargin=!}

\newcommand{\Minline}[1]{ \lstinline[language=Matlab]{#1} }

\usepackage{url}

\usepackage{tikz} % images
\usetikzlibrary{patterns,decorations}

\usepackage{tcolorbox}

\newcommand{\ket}[1]{ | \, #1 \rangle}
\newcommand{\bra}[1]{ \langle #1 \, |} 
\newcommand{\proj}[1]{\ket{#1}\bra{#1}} 
\newcommand{\kb}[2]{\ket{#1}\bra{#2}}
\newcommand{\bk}[2]{ \langle #1 | #2 \rangle}
\newcommand{\bkc}[2]{ \langle #1 , #2 \rangle}
\newcommand{\Frob}[2]{ #1 \bullet #2 }

\newcommand{\mean}[1]{ \langle #1 \rangle}
\newcommand{\standardDeviation}[1]{ \bm{\sigma} \left[ #1 \right]}

\newcommand{\Norm}[1]{ \left| \left| #1 \, \right| \right|} % norm
\newcommand{\SN}[1]{ \left| \left| #1 \, \right| \right|_2} % square norm
\newcommand{\FN}[1]{ \left| \left| #1 \, \right| \right|_F} % Frobenus norm
\newcommand{\Ab}[1]{ \left| #1 \, \right|} % absolute value

\DeclareMathOperator{\interior}{int}
\DeclareMathOperator{\core}{core}
\DeclareMathOperator{\cont}{cont}
\DeclareMathOperator{\Mat}{Mat}
\DeclareMathOperator{\Vector}{vec}

\DeclareMathOperator{\Diag}{Diag}
\DeclareMathOperator{\Herm}{Herm}
\DeclareMathOperator{\epi}{epi} % epigraph
\DeclareMathOperator{\hyp}{hyp} % hypograph
\DeclareMathOperator{\epiM}{\mathbf{epi}} % matrix epigraph
\DeclareMathOperator{\hypM}{\mathbf{hyp}} % matrix hypograph
\DeclareMathOperator{\Persp}{P} % perspective of a function
\DeclareMathOperator{\PerspM}{\mathbf{P}} % non-commutative perspective of a function
\DeclareMathOperator{\dom}{dom}
\DeclareMathOperator{\lsc}{lsc}
\DeclareMathOperator{\ball}{B}
\DeclareMathOperator{\argmin}{argmin}

\DeclareMathOperator{\var}{var}
\DeclareMathOperator{\cov}{cov}
\DeclareMathOperator{\Cov}{Cov}
\DeclareMathOperator{\corr}{corr}
\DeclareMathOperator{\Corr}{Corr}

\newcommand{\bigforall}{\mbox{\Large $\mathsurround=0pt\forall$}} 
\newcommand{\bigexists}{\mbox{\Large $\mathsurround=0pt\exists$}} 

\DeclareRobustCommand\openone{\leavevmode\hbox{\small1\normalsize\kern-.33em1}}%

\newcommand{\boundellipse}[3]% center, xdim, ydim
{(#1) ellipse (#2 and #3)
} % http://tex.stackexchange.com/questions/65018/how-to-draw-an-ellipse

\usepackage{etoolbox}
\makeatletter
\def\@mkboth#1#2{}
\newlength\appendixwidth
\preto\appendix{\addtocontents{toc}{\protect\patchl@section}}
\newcommand{\patchl@section}{%
	\settowidth{\appendixwidth}{\textbf{Appendix }}%
	\addtolength{\appendixwidth}{1.5em}%
	\patchcmd{\l@section}{1.5em}{\appendixwidth}{}{\ddt}%
}
\makeatother

\begin{document}
	
	\title[SDP and QI]{Semi-definite programming and quantum information}
	
	\author{Piotr Mironowicz}
	
	\address{Department of Physics, Stockholm University, S-10691 Stockholm, Sweden}
	\address{Faculty of Electronics, Telecommunications and Informatics, Gdańsk University of Technology, Narutowicza 11/12, 80-233 Gdańsk, Poland}
	\address{International Centre for Theory of Quantum Technologies, University of Gdańsk, Wita Stwosza 57, 80-308 Gdańsk, Poland}
	\ead{piotr.mironowicz@gmail.com}
	\vspace{10pt}
	\begin{indented}
		\item[]\date{\today}
	\end{indented}
	
	\begin{abstract}
		This paper presents a comprehensive exploration of semi-definite programming (SDP) techniques within the context of quantum information. It examines the mathematical foundations of convex optimization, duality, and SDP formulations, providing a solid theoretical framework for addressing optimization challenges in quantum systems. By leveraging these tools, researchers and practitioners can characterize classical and quantum correlations, optimize quantum states, and design efficient quantum algorithms and protocols. The paper also discusses implementational aspects, such as solvers for SDP and modeling tools, enabling the effective employment of optimization techniques in quantum information processing. The insights and methodologies presented in this paper have proven instrumental in advancing the field of quantum information, facilitating the development of novel communication protocols, self-testing methods, and a deeper understanding of quantum entanglement. Overall, this study offers a resource for researchers interested in the intersection of optimization and quantum information, opening up new avenues for exploration and breakthroughs in this rapidly evolving field.
	\end{abstract}
	
	% TODO: odkomentować później; teraz za dużo czasu trwa kompilacja
	\setcounter{tocdepth}{2}
	\tableofcontents

	\section{Introduction}
	
	% warto przeczytać wstęp do https://core.ac.uk/download/pdf/144146753.pdf
	% barrrdzo dużo materiałów pełnotekstowych jest tu https://www-user.tu-chemnitz.de/~helmberg/semidef.html
	% a to fajny artykuł przeglądowy: M. J. Todd Semidefinite Optimization 2001 (jest na dysku)
	% o tym jak się temat ma ogólnie do optymalizacji: https://neos-guide.org/content/optimization-taxonomy
	
	Optimization, in its various forms, has been a cornerstone of scientific and technological advancements across numerous disciplines. From engineering and economics to machine learning and operations research, optimization techniques have played a crucial role in solving complex problems and driving innovation. Over the years, different variants of optimization methods have emerged, each tailored to address specific problem structures and objectives~\cite{FiaccoMcCormick90,boyd1994linear,BL06,anjos2011handbook,beck2014introduction,nesterov2018lectures}. In recent decades, semi-definite programming (SDP) has emerged as a powerful variant of convex optimization, offering a versatile framework for solving optimization problems involving positive semi-definite matrices. SDP has found applications in diverse fields, including control theory, signal processing, combinatorial optimization, and quantum information theory~\cite{NN94,Terlaky96,Boyd04,gartner2012approximation}. Particularly, in the field of quantum information, SDP has proven to be an indispensable tool for characterizing and manipulating quantum correlations and probabilities~\cite{bengtsson2017geometry,watrous2018theory}. Quantum information theory deals with the fundamental principles governing the representation, transmission, and processing of information in quantum systems. It explores the unique properties of quantum mechanics to develop new paradigms for computation, communication, and cryptography. Quantum correlations, such as entanglement, and the manipulation of probabilities in quantum systems are essential components in designing quantum algorithms and protocols.
	
	This paper aims to provide a comprehensive study of SDP in the context of quantum information. The outline of the paper is as follows.
	We present first a mathematical framework for convex optimization, covering the necessary preliminaries and notation, and provide a software overview useful for implementing techniques discussed in this work. To aid researchers in the practical implementation of SDP, the paper provides an overview of software tools, solvers, and modeling techniques in sec.~\ref{sec:softwareAndIPM}. It discusses the different solvers available for solving SDP problems, as well as the modeling tools used to formulate and represent optimization problems. Sec.~\ref{sec:probDistr} contains a brief overview of the topic of probability distributions occurring in quantum mechanics and other important problems of the theory which can be effectively treated with SDP. The terms introduced there will be used in sec.~\ref{sec:basicTools}.
	
	The further discussion in sec.~\ref{sec:frameworkConvex} encompasses sets, spaces, cones, and functions, including important concepts like Fenchel conjugate and subgradient. Duality, a fundamental aspect of optimization, is explored extensively, shedding light on its role in problem formulations and solution methods.
	The theory of SDP is a focal point of this paper, as it enables the optimization of positive semi-definite matrices, which are fundamental objects in quantum information theory. Next, in sec.~\ref{sec:theorySDP}, the work delves into the definition and characterization of positive semi-definiteness, presenting various formulations of SDP problems, such as the canonical form, the Vandenberghe and Boyd form, the so-called SDPA form, and the Watrous symmetric form. It also discusses the duality of SDP, the treatment of complex variables in semidefinite problems imposing equality and inequality constraints, and the topic of the Schur complement and submatrices. Next, we concentrate on implementing SDPs, to provide a general understanding of the involved numerical methods. The paper also explores how solvers employ Interior Point Methods, highlighting their internal mechanisms, such as predictor-corrector methods, warm start strategies, and exploitation of the problem structure.
	
	In the following sec.~\ref{sec:basicTools}, the paper introduces basic tools and techniques in SDP that are specifically relevant to quantum information. These tools include semidefinite representations, separability criteria, Choi-Jamiołkowski isomorphism (state-channel duality), the sum of squares decomposition, and Lovász theta. Understanding and utilizing these tools are crucial for solving optimization problems involving quantum states, correlations, and probabilities.
	A significant portion of the paper focuses on the application of moment matrices in quantum information. Moment matrices play a vital role in capturing the correlations present in quantum systems. The paper explores correlation matrices and moment matrices, their mathematical properties, and their significance in optimization problems involving non-commuting variables. Hierarchical methods, such as the Navascués-Pironio-Acin (NPA) hierarchy, the so-called MLP hierarchy, and the Navascués-Vertesi hierarchy, are discussed in detail for optimizing probability distributions without or with dimension constraints. Additionally, the SWAP method for self-testing in quantum information is presented.
	
	In conclusion, this paper aims to bridge the gap between the theory and applications of SDP in the realm of quantum information. Providing a comprehensive study of the mathematical framework, theoretical foundations, and practical tools, it equips researchers and practitioners with the necessary knowledge and understanding to tackle optimization problems in quantum information. Through its exploration of SDP's role in characterizing and manipulating quantum correlations and probabilities, this paper intends to open up new avenues for advancements in quantum information theory and its applications.

	\subsection{Historical notes on optimization}
	
	% Sam początek trzeba uzupełnić np. o Kantorovicha
	% https://www.nobelprize.org/prizes/economic-sciences/1975/press-release/
	% A także wzmiankować dawnych ekonomistów:
	% http://econospeak.blogspot.com/2019/06/learning-origin-of-duality.html
	% https://en.wikipedia.org/wiki/Linear_programming#History
	% może coś uzupełnić z tego https://empowerops.com/en/blogs/2018/12/6/brief-history-of-optimization
	% a tu jest bardzo dobrze wszystko opisane: "Bibliography comments 1.9781611970791.bm.pdf" (na dysku)
	
	The historical development of linear optimization, linear programming (LP), can be traced back to the times of a critical need for optimal resource management during World War II. Soon after, in 1947, George Dantzig introduced the simplex method~\cite{Dantzig90}\index{simplex method}, marking a significant milestone in this field. The simplex method operates by starting at a vertex of a convex polytope representing feasible solutions and is gradually moving towards its extreme point. It is important to note that although the simplex method algorithm exhibits exponential worst-case complexity, it has demonstrated remarkable efficiency in practical problem-solving scenarios.
	To understand the underlying reason for the high complexity of the simplex method, it is necessary to examine the specific instances where the algorithm visits every vertex of the feasible region, leading to exponential worst-case complexity. The Klee-Minty problem, formulated in 1970~\cite{KleeMinty70}, serves as an illustrative example of such instances. During the 1960s and 1970s, there was an increasing recognition of the significance of computational complexity, fueling the pursuit of efficient algorithms with polynomial time complexity.
	
	In the 1960s, in the realm of nonlinear programming, it became a common practice to transform constrained problems into unconstrained ones through the utilization of the so-called \textit{barrier methods}~\cite{FiaccoMcCormick90}\index{barrier methods}. By introducing a specialized barrier function, it became possible to delineate a trajectory within the space of optimization variables known as the \textit{central path}\index{central path}, which could be traversed using the well-established Newton's method. However, the prevalence of barrier methods experienced a temporary decline in the 1970s. Meanwhile, in 1979, Khachian introduced the \textit{ellipsoid method}\index{ellipsoid method}, the first algorithm for LP with polynomial worst-case complexity~\cite{Khachian79}. Surprisingly, despite its favorable theoretical complexity, the ellipsoid method proved to be exceedingly slow when applied to most practical problems. Consequently, before 1984, two primary methods for LP existed:
	\begin{itemize}
		\item The simplex method, which possessed exponential worst-case complexity but demonstrated practical efficiency.
		\item The ellipsoid method, exhibited polynomial complexity but was notably inefficient in practice.
	\end{itemize}
	
	% Karmarkar's revolution
	The field of optimization has undergone a significant transformation with the introduction of interior point methods (IPM)\index{interior point method}. Before 1984, IPMs did not hold a prominent position until Karmarkar's groundbreaking paper, \textit{A new polynomial-time algorithm for linear programming}~\cite{Karmarkar84}, was published. Notably, it was demonstrated that IPMs were no less efficient than the simplex method for solving practical linear programming problems. This revelation of IPM's potential sparked what is often referred to as a \textit{revolution in optimization}~\cite{Wright05}, leading IPMs to be recognized as one of the most significant algorithms of all time. Before 1984, there existed only minimal connections between LP and nonlinear programming. However, it was soon discovered~\cite{GMSTW86} that the IPM method was equivalent to a logarithmic barrier method applied to LP. This equivalence enabled the development of a unified framework, based on barrier function methods, for analyzing both linear and nonlinear problems~\cite{NN92}.
	
	% Interior Point Method for convex problems and SDPs
	The next significant advancement in the field of IPM came with the independent work of Alizadeh~\cite{A91}, Nesterov, and Nemirovskii~\cite{NN92,NN94} in the late 1980s. They expanded the applicability of IPM to various convex optimization problems. Nesterov and Nemirovskii discovered that the key to utilizing IPM for convex problems lies in knowing a specific barrier function known as a self-concordant barrier~\cite{N04}. For practical implementation, it is essential that the first and second derivatives of the barrier function can be computed easily. Vandenberghe and Boyd~\cite{Boyd95} utilized the theory developed by Nesterov and Nemirovskii to apply the LP method given by Gonzaga and Todd~\cite{GT92} to SDPs.
	A self-concordant barrier\index{self-concordant barrier} refers to a smooth convex function defined within the interior of a given set. It diverges towards infinity as it approaches the boundary and, along with its derivatives, satisfy certain Lipschitz continuity conditions. Nesterov and Nemirovskii demonstrated that IPM can be applied to any set where such a barrier function can be formulated. Fortunately, a relatively computationally tractable self-concordant barrier is known for SDPs, \textit{viz.} $F(X) \equiv -\ln \det X$. For a comprehensive historical overview of the development and significance of SDP, detailed information can be found in several notable references such as~\cite{Terlaky96,FM00,Wright05,Gondzio12}. These works provide an in-depth exploration of SDPs, shedding light on its emergence as a powerful tool in modern optimization.
	The key property of SDP problems is the fact that they may be efficiently solved numerically using IPM, as sketched further in sec.~\ref{sec:IPM}, and at the same time they can express or approximate a tremendous range of scientific and engineering problems.

	\subsection{Preliminaries and notation}
	% https://en.wikipedia.org/wiki/List_of_numerical_analysis_topics#Basic_concepts_2
	
	We now briefly specify the notation used in this work. In some places, the notation used is overloaded, with the same symbols having different meanings. The reason is that the paper covers a variety of different fairy-specialized topics. We decided to keep the established notation characteristic for each of the specializations. We made an effort to ensure that this does not lead to any ambiguity.
	
	\subsubsection{Spaces and sets notation}~\\
	
	In this work, we denote by $\mathbb{N}$ the set of natural numbers (including $0$), $\mathbb{N}_{+}$ is the set of natural numbers (excluding $0$), $\mathbb{R}$ is the set of real number, and $\mathbb{C}$ is the set of complex numbers.
	% TODO: odkomentować, gdy np. MAX-CUT będzie opisywany: For bits, we define the negation $\neg 0 \equiv 1$ and $\neg 1 \equiv 0$.
	The sets of vectors composed of real numbers, non-negative real numbers, and complex numbers, with $k$ elements, are denoted by $\mathbb{R}^k$, $\mathbb{R}_{+}^k$, and $\mathbb{C}^k$, respectively. The set of real $k \times l$ matrices is denoted by $\mathbb{R}^{k \times l}$ and the set of real $n \times n$ symmetric matrices by $\mathbb{S}^n$. The set of real $n \times n$ symmetric matrices that are positive semidefinite (PSD) or positive definite (PD), (see sec.~\ref{sec:PSDdefinition} for the definitions) we denote with $\mathbb{S}^n_+$ or $\mathbb{S}^n_{++}$, $\mathbb{S}^n_{++} \subset \mathbb{S}^n_+ \subset \mathbb{S}^n$.
	Similarly, the set of complex $k \times l$ matrices is denoted by $\mathbb{C}^{k \times l}$, the set of Hermitian $n \times n$ matrices by $\mathbb{H}^n$, and its subset of PSD or PD matrices by $\mathbb{H}^n_+$ or $\mathbb{H}^n_{++}$, $\mathbb{H}^{n}_{++} \subset \mathbb{H}^{n}_+ \subset \mathbb{H}^{n}$. We refer to the pair of values $k$ and $l$ (for matrices of arbitrary size) or $n$ (for square matrices) as the size of the matrix. For $n \in \mathbb{N}_{+}$ we denote $[n] \equiv \{1, \cdots, n\}$.
	The relation $\succeq$ denotes the so-called L\"{o}wner's partial order\index{L\"{o}wner's partial order} of PSD matrices~\cite{lowner1934monotone,kraus1936konvexe}. For two symmetric or Hermitian matrices $A$ and $B$ we have $A \succeq B$ when $A - B$ is PSD.
	
	Banach spaces are denoted with letters $X, Y, \cdots$, and their continuous dual spaces with $X^{*}, Y^{*}, \cdots$. The real and complex Hilbert spaces are denoted with letters $\mathcal{X}, \mathcal{Y}, \mathcal{Z}, \cdots$, and by $\mathcal{X}^{*}, \mathcal{Y}^{*}, \mathcal{Z}^{*}, \cdots$ their dual spaces. Usually, we consider vector spaces of real or complex matrices with the Frobenius product, defined below in~\eqref{eq:Frob}, as the inner product; thus the spaces $\mathbb{R}^{k \times 1}$ and $\mathbb{C}^{k \times 1}$ are the ordinary $k$ dimensional real or complex Euclidean spaces, i.e. finite-dimensional Hilbert spaces.
	In some cases we use finite sets of symbols as indices of vectors or matrices; this will be particularly useful in the context of moment matrices, see sec.~\ref{sec:NPA}. For a set of symbols $\Sigma$ we use the standard convention of set theory to denote $\mathbb{C}^\Sigma$ ($\mathbb{R}^\Sigma$) the set of all functions from $\Sigma$ to $\mathbb{C}$ ($\mathbb{R}$). Since there exists a natural isomorphism between $\mathbb{C}^\Sigma$ and $\mathbb{C}^{\Ab{\Sigma}}$ ($\mathbb{R}^\Sigma$ and $\mathbb{R}^{\Ab{\Sigma}}$), all operations defined for the latter can be easily mapped to relevant operations on the former, with an arbitrary ordering of the symbols in $\Sigma$, that will be treated here as implicit.
	For a metric space $(X, d)$ we denote by $\ball_X(x,r)$ the \textit{closed ball} centered at $x \in X$ with radius $r$; $\ball_X \equiv \ball_X(0,1)$ is the unit closed ball. %BZ05p11
	
	% dual spaces
	Consider a Banach space $X$ over the field $F$.
	For a linear functional $x^{*}: X \rightarrow F$ we define its norm as $\Norm{x^{*}} \equiv \sup_{x \in X: \Norm{X} \leq 1} \Ab{x^{*}(x)}$. % Janich, Topologia, s.169
	We define $X^{*}$ as the continuous dual space\index{dual space}, i.e. the space of all linear continuous functionals on $X$ with this norm. % L06p294
	The weak topology of $X$ is the weakest topology in $X$ in that all elements of $X^{*}$ are continuous. % L06p66
	%Note that thus, for $X$, we consider two natural topologies, \textit{viz.} the one topology induced by the norm, and the weak topology. % Janich, Topologia, s.169
	For $X^{*}$ we consider also the weak\textsuperscript{*} topology, defined as the weakest topology on $X^{*}$ for that every element $x \in X$ corresponds to a continuous functional on $X^{*}$. % Wolfram Weal-* Topology
	We denote the \textit{bidual} spaces $X^{**} \equiv (X^{*})^{*}$. Spaces for that $X = X^{**}$ are called \textit{reflexive}. % Janich, Topologia, s.169
	It can be seen that every finite-dimensional normed space is reflexive.
	%We restrict our further considerations to $F = \mathbb{R}$.
	The action of a conjugate element $x^{*}$ on an element $x$, i.e. $x^{*}(x)$, is further denoted as $\bkc{x^{*}}{x}$ to make the linearity explicit. % L06p294
	This is to be contrasted with the inner (scalar) product $\bk{\cdot}{\cdot}$ for Hilbert spaces.
	
	To provide the most explicit formulations, we usually denote the elements of a conjugate space with the star symbol ${}^*$, e.g. $x^{*} \in \mathcal{X}^{*}$. The symbol is barely a notation suggesting an element of a conjugate space and has no algebraic meaning. Similarly, we optionally (with no special mathematical meaning) denote with $\cdot$ (dot) a matrix multiplication in places where it allows us to avoid ambiguity, especially to stress the presence of a scalar product of two vectors.
	Due to the specificity of our topic closely mixing the \textit{explicit} numerical representation of operators as arrays of (numerical) real values with the abstract complex Hilbert formalism of quantum mechanics, we decided to use both this ``dot'' notation of the scalar product, and the bra-ket notation, with the latter used in cases not directly related to the computer implementations.
	
	We denote by $\mathsf{L}\left[ \mathcal{X}, \mathcal{Y} \right]$ the set of all linear operators from the Hilbert space $\mathcal{X}$ to the Hilbert space $\mathcal{Y}$. For $\mathcal{X} = \mathbb{C}^n$ and $\mathcal{Y} = \mathbb{C}^m$ this set is isomorphic to the set of complex $m \times n$ matrices, $\mathbb{C}^{m \times n}$. We use the latter whenever our considerations are directly related to computer implementations. For $A \in \mathsf{L}\left[ \mathcal{X}, \mathcal{Y} \right]$ we define the \textit{adjoint operator}, $A^{\dagger} \in \mathsf{L}\left[ \mathcal{Y}, \mathcal{X} \right]$ as the unique operator satisfying the scalar product relation
	\begin{equation}
		\label{eq:adjointOperator}
		y^{\dagger} \cdot A x = \left( A^{\dagger} y \right)^{\dagger} \cdot x, % TODO: potrzebujemy żeby to było ogólniej, nie tylko na Euklidesa, ale też na Banacha, np. https://math.stackexchange.com/questions/2110405/hilbert-space-adjoint-vs-banach-space-adjoint
	\end{equation}
	or, in the bra-ket notation, $\bk{y}{A x} = \bk{A^{\dagger} y}{x}$, for all $x \in \mathcal{X}$ and $y \in \mathcal{Y}$. We use the abbreviation $\mathsf{L}\left[ \mathcal{X} \right] \equiv \mathsf{L}\left[ \mathcal{X}, \mathcal{X} \right]$.
	For real matrix space the adjoint is the \textit{transposition} denoted by $^T$; and for complex matrices it is the \textit{Hermitian conjugate} denoted by $^\dagger$.
	Thus we denote by $x^{\dagger} \in \mathcal{X}^{*}$ the unique linear operator $x^{\dagger}: \mathcal{X} \rightarrow F : x' \mapsto \bk{x}{x'}$, where $F = \mathbb{C}$ for complex Hilbert spaces; or by $^T$ with $F = \mathbb{R}$ for real Hilbert spaces. The symbol of $^\dagger$ used for real spaces is equivalent to $^T$.
	We define the set of all Hermitian operators acting on a complex Euclidean vector space $\mathcal{X}$ as $\Herm\left[\mathcal{X}\right] \equiv \left\{ X \in \mathsf{L}\left[ \mathcal{X} \right] : X^{\dagger} = X \right\}$.
%	Thus, for $x, x' \in \mathcal{X}$, we have $x^{\dagger} \in \mathcal{X}^{*}$ and $x^{\dagger} \cdot x' = \bk{x}{x'}$, where $\bk{\cdot}{\cdot}$ denotes the relevant inner product, in particular the Frobenius product $\Frob{}{}$, see sec.~\ref{sec:Frob}.
%	The usual convention, when the basis of the space is fixed, is also to write $x_2^{*} x_1$ as $\bk{x_2}{x_1}$ or $x^2 \cdot x^1$ for Hilbert space, with the latter restricted to Euclidean spaces.

	\subsubsection{Matrix conventions and the Frobenius product}~\\
	\label{sec:Frob}
	
	In this work, we use the MATLAB notation for elements of matrices. We recall a few examples. In this notation the expression $M_{k,l:m}$ means a submatrix consisting of the elements of the matrix $M$ within the row $k$ and with column indices in $\{l, l+1, \cdots, m\}$. The expression $M_{:,k}$ means the vector consisting of the $k$-th column of the matrix $M$. $M_{k,l}$ refers to the element in $k$-th row and $l$-th column. For a vector $v$ its $k$-th element is $v_k$. Vectors are represented by one-column matrices. Matrix elements are numbered from $1$. The identity matrix of size $d$ by $d$ is denoted by $\openone_{d}$; in some cases, instead of the size we specify a space $\mathcal{X}$ and then $\openone_{\mathcal{X}}$ denotes the identity operator on $\mathcal{X}$. The zero operator and zero matrix for all spaces is denoted with $0$. The Kronecker delta is denoted by $\delta_{i,j}$ and is equal $1$ for $i = j$ and $0$ otherwise. Trace operation is denoted as $\Tr[\cdot]$, and partial trace by $\Tr_{\{s_i\}_i}[\cdot]$, where $\{s_i\}_i$ enumerates the subsystems which are traced out. Similarly, the partial transposition of subsystems $\{s_i\}_i$ is denoted as $^{T_{\{s_i\}_i}}$.
	
	The function $\Vector(\cdot)$ defines a vector containing the elements of the given matrix in column-wise order. $\Mat(\cdot)$ is the inverse of this function. For example, we have
	\begin{equation}
		\Vector \left( \begin{bmatrix} a & c \\ b & d \end{bmatrix} \right) = \begin{bmatrix} a \\ b \\ c \\ d \end{bmatrix}.
	\end{equation}
	We also use the following standard convention in which upper-case letters denote matrices, and lower-case letters denote vectors of elements of the matrices, e.g. $x = \Vector(X) \in \mathbb{R}^{n^2}$ and $X = \Mat(x) \in \mathbb{R}^{n \times n}$.
	For two matrices $A, B \in \mathbb{R}^{m \times n}$ we define the relation $A \leq B$ to hold if and only if $\forall_{i = 1, \dots, m} \forall_{j = 1, \dots, n} A_{i,j} \leq B_{i,j}$.
	We define relations $A < B$, $A \geq B$ and $A > B$ in an analogous way. $\Diag[(d_i)_{i \in [n]}]$ is an $n$ by $n$ diagonal matrix with diagonal entries $d_i$.
	
	The \textit{Frobenius product}\index{Frobenius product} of two complex (or real) matrices, $A, B \in \mathbb{C}^{k \times l}$ (or $A, B \in \mathbb{R}^{k \times l}$) is defined as $\Tr(A^\dagger B)$ (or $\Tr(A^T B)$). We follow the convention common in the literature close to implementation issues, and usually denote the Frobenius product as $\Frob{A}{B}$~\cite{Fujisawa97,M97,AHO98,TTT98,T99,Sturm99,wang2009new,kheirfam2015adaptive}. It can be easily shown that
	\begin{equation}
		\label{eq:Frob}
		\Frob{A}{B} \equiv \Tr(A^\dagger B) = \Tr(A B^T) = \sum_{i = 1, \dots, k} \sum_{j = 1, \dots, l} A_{i,j}^{*} B_{i,j} = \Vector(A)^\dagger \cdot \Vector{B},
	\end{equation}
	and similarly for real matrices. Thus, for real matrices, the Frobenius product is the sum of the elements of the element-wise product of entries of two matrices. One can also show that for real symmetric $A$ and real antisymmetric $B$ we have $\Tr(A^T B) = \Tr(A B) = 0$. For real symmetric $A$ we have $\Frob{A}{B} = \Tr(A^\dagger B) = \Tr(A^T B) = \Tr(AB)$. The Frobenius product induces a \textit{Frobenius norm}\index{Frobenius norm} of a matrix, $\FN{\cdot}$ defined as $\FN{A} = \sqrt{\Tr(A^\dagger A)}$. This norm is called also a Hilbert-Schmidt norm. The Frobenius product is a direct generalization of the vector Euclidean product, as it can be seen from~\eqref{eq:Frob}, and the $\mathbb{C}^{k \times l}$ ($\mathbb{R}^{k \times l}$) with Frobienius inner product is isomorphic with the Euclidean space $\mathbb{C}^{kl}$ ($\mathbb{R}^{kl}$). For $A = \begin{bmatrix} A_{11} & A_{12} \\ A_{21} & A_{22} \end{bmatrix}$ and $B = \begin{bmatrix} B_{11} & B_{12} \\ B_{21} & B_{22} \end{bmatrix}$, where $A_{11}, B_{11} \in \mathbb{C}^{n_1 \times n_1}$, $A_{12}, B_{12} \in \mathbb{C}^{n_1 \times n_2}$, $A_{21}, B_{21} \in \mathbb{C}^{n_2 \times n_1}$, and $A_{22}, B_{22} \in \mathbb{C}^{n_2 \times n_2}$, for some $n_1$ and $n_2$, we have
	\begin{equation}
		\Tr(A^\dagger B) = \Tr(A_{11}^\dagger B_{11}) + \Tr(A_{12}^\dagger B_{12}) + \Tr(A_{21}^\dagger B_{21}) + \Tr(A_{22}^\dagger B_{22}).
	\end{equation}
	This equality easily generalizes to matrices $A = (A_{r,c})_{r,c}$ and $B = (B_{r,c})_{r,c}$ divided into arbitrary number of blocks, \textit{viz.} $\Frob{A}{B} = \sum_{r,c} \left( \Frob{A_{r,c}}{B_{r,c}} \right)$.
%	n1 = 3; n2 = 5;
%	A11 = rand(n1,n1) + 1i * rand(n1,n1);
%	A12 = rand(n1,n2) + 1i * rand(n1,n2);
%	A21 = rand(n2,n1) + 1i * rand(n2,n1);
%	A22 = rand(n2,n2) + 1i * rand(n2,n2);
%	B11 = rand(n1,n1) + 1i * rand(n1,n1);
%	B12 = rand(n1,n2) + 1i * rand(n1,n2);
%	B21 = rand(n2,n1) + 1i * rand(n2,n1);
%	B22 = rand(n2,n2) + 1i * rand(n2,n2);
%	A = [A11 A12; A21 A22];
%	B = [B11 B12; B21 B22];
%	t1 = trace(A' * B)
%	t2 = trace(A11' * B11) + trace(A12' * B12) + trace(A21' * B21) + trace(A22' * B22)
%	assert(abs(t1 - t2) < 1e-12)

	\subsection{Software overview, usage, and implementation}
	\label{sec:softwareAndIPM}
	
	The basic tool used to find solutions to SDP is SDP solvers.  Modeling languages are useful supporting software aiding in formulating SDPs to be passed to a solver.  We refer readers to \cite{Mittelmann12} for a comprehensive overview. From the experience of the author, most of the analysis involving SDPs in quantum information is conducted using either Python or Matlab language. The latter language has two major implementations, \textit{viz.} the software MATLAB from MathWorks~\cite{MATLAB2022} and its open-source alternative OCTAVE~\cite{octave}.

	%	\subsection{Solvers for semi-definite programming}
	%	\label{sec:software}
	
	% przegląd solverów
	% i sztuczek np. z https://yalmip.github.io/tutorials/
	% środowisko i solvery (Matlab/Octave, YALMIP/CVX, SeDuMi/SDPT3/Mosek, ncpol2sdpa, QDimSum)
	
	A standard choice for an SDP solver among the NPA community (see sec.~\ref{sec:NPA}) using Matlab seems \cite{NPA07,NPA08,MiguelVertesi} to be the SeDuMi solver \cite{SeDuMi,Sturm02} created by J.~F.~Sturm, currently developed and maintained by Imre Pólik and Oleksandr Romanko under the direction of prof.~Tamás Terlaky~\cite{SeDuMiGitHub}. This solver implements self-dual embedding IPM~\cite{cheng2006implementation}.
	Another SDP solver of particular interest in Matlab is SDPT3 solver \cite{SDPT3b,TTT12} implemented by Toh, Todd and T{\"u}t{\"u}nc{\"u}. It uses infeasible primal-dual IPM with so-called NT and HKM search directions (see sec.~\ref{sec:IPM} for the definition of the search directions). Other examples of SDP solvers include CSDP~\cite{CSDP} by Borchers, DSDP~\cite{DSDP}, and SDPA~\cite{SDPA}. The mentioned solvers are freely available, in most cases in open-source form. A very efficient commercial SDP solver is Mosek~\cite{mosek21}, possible to be used in Python and MATLAB, with a free license for academia. The solver particularly relevant for large problems is SDPNAL~\cite{yang2015sdpnal,sun2020sdpnal+} implementing a Newton-CG augmented Lagrangian method for SDP~\cite{zhao2010newton}.
	
	%TODO: odkomentować, gdy będzie więcej przykładów implementacji
	%The choice of SeDuMi and SDPT3 as reference solvers further in this paper is motivated in the case of SeDuMi by its popularity and in the case of SDPT3 by its scientific importance reflected by the number of papers referring to its mechanisms, and the number of their citations. At the moment of writing this work, among the mentioned solver, these two were the most cited.
	A popular family of solvers are the mentioned SDPA solvers by Fujisawa \textit{et al.}~\cite{SDPA,yamashita2003implementation,fujisawa2002sdpa,yamashita2010high,nakata2010numerical} that are using the Vandenberghe and Boyd, or the SDPA, form of SDP, see sec.~\ref{sec:BoydSDP}. The variants cover Matlab interface (SDPA-M), parallel implementation for large SDPs (SDPARA), higher precision arithmetics (SDPA-GMP\/QD\/DD), structural sparsity (SDPA-C), see~\cite{yamashita2012latest} of an overview. SDPA solver implements primal-dual IPM with Mehrotra type predictor-corrector, see sec.~\ref{sec:PredictorCorrector}.
	When deciding solver to be used, a performance benchmark should be consulted~\cite{Miltenberger_mittelmann-plots_-_Interactive_2021}. % ładnie prezentowane: https://mattmilten.github.io/mittelmann-plots/

	%	\subsection{Modelling tools for semi-definite programming}
	%	\label{sec:modellingTools}
	
	A plenty of papers~\cite{baccari2017efficient,pozas2019bounding,brown2019framework,smania2020experimental,chaturvedi2020quantum,bernards2020generalizing,brown2021computing,agresti2021experimental,chaturvedi2021characterising,frerot2022unveiling,chaturvedi2022extending,lin2022naturally} uses the Python package NCPOL2SPDA~\cite{wittek2015algorithm} by Peter Wittek, currently under maintenance by Peter J.~Brown~\cite{Ncpol2sdpaGitHub}. NCPOL2SDPA implements a framework for global polynomial optimization problems with SDP relaxations. The functionality of particular interest covers
	the NPA~\cite{NPA07,NPA08,pironio2010convergent} hierarchy for noncommutative operators;
	Lasserre’s hierarchy for commutative polynomials\cite{lasserre2001global}; % TODO: odkomentować referencje tu i nieopodal, gdy już będzie sekcja (see sec.~\ref{sec:Lasserre}) i pozostałe;
	the \textit{more randomness from the same data} technique~\cite{nieto2014using,bancal2014more}; % (see sec.~\ref{sec:moreRandomness});
	a hierarchy for bilevel polynomial optimization problem~\cite{jeyakumar2016convergent}; % (see sec.~\ref{sec:bilevel});
	the Moroder's hierarchy~\cite{moroder2013device}; % (see sec.~\ref{sec:Moroder});
	and a hierarchy of sufficient conditions for the steerability of bipartite quantum states~\cite{kogias2015hierarchy}. % (see sec.~\ref{sec:steering}).
	Note that NCPOL2SDPA is not an SDP solver but a modeling toolbox, used to reduce the human effort when formulating SDPs, and it requires a solver to be included separately.
	
	Popular modeling toolboxes to be used with Matlab language are YALMIP~\cite{yalmip} and CVX~\cite{grant2011cvx,grant2014cvx,guimaraes2015tutorial}. They allow using of various solvers, including SeDuMi, SDPT3, and Mosek. YALMIP can be supported with a package QDimSum (Symmetric SDP relaxations for qudits systems)~\cite{QDimSumGitHub} that implements the hierarchy~\cite{MiguelVertesi} using the symmetrization methods~\cite{aguilar2018connections,tavakoli2019enabling} to enhance the performance.
	In~\ref{sec:matlab:1stYALMIP} a sample simple execution is given with YALMIP.
	% TODO: odkomentować gdy będą: To provide the reader with a better idea of the role of the modeling software in the coding of SDPs, and to make it easier to choose and start using this type of software, in the Appendix we present some examples of implementation of problems typical for quantum information research. Listing~\ref{python:1stNPA-code} and Listing~\ref{python:1stNPA-results} provide examples of code implementing simple calculations regarding the CHSH Bell expression using NCPOL2SDPA and results thereof.Next, Listing~\ref{python:NPA-constraints-code} and Listing~\ref{python:NPA-constraints-results} show how to employ additional constraints to SDPs with this toolbox.
	
	At this stage we mention that models in YALMIP and many other modeling languages are interpreted as so-called dual problems~\cite{dualizeIt}, discussed in sec.~\ref{sec:formulations}. The dual form of SDP is given in~\eqref{SDP-dual}, where the SDP variable $Z$ is in a \textit{disaggregated} form, i.e. it is a matrix composed of linear combinations of scalar variables. This is to be contrasted with the primal form of SDP~\eqref{SDP-primal}, where the SDP variable $X$ is treated as a single matrix variable. There are two reasons, why modeling languages prefer the dual form over the primal form. The major reason is that symbolic manipulations are much easier when the variables are disaggregated. The other reason is that it was observed that in many different fields, the dual form is more natural to formulate the problems occurring in them, see e.g. Tab.~\ref{tab:NPAsize} in sec.~\ref{sec:NPA}.

	\subsection{Basic problems of quantum information}
	\label{sec:probDistr}
	
	Many useful functions that occur in quantum information belong to the family of the so-called semi-algebraic functions. These functions can be represented using SDP constraints, and thus are particularly relevant for this review. On the other hand, many other functions are not semi-algebraic, like the logarithm function used e.g. in the definitions of such quantities as entropies, including Shannon, quantum, or their relative or conditional variants. It would be beneficial to be able to express them, or at least their approximations as SDPs. It revealed that it is possible when the non-semi-algebraic function is approximated with a polynomial function, for instance with the support of one of the Gauss quadratures, e.g. the Radau quadrature. Recent results allowed the use of SDP to approximate the matrix logarithm function~\cite{fawzi2019semidefinite}, and as a result, the use of semidefinite programming to efficiently optimize expressions on various entropies~\cite{fawzi2018efficient}. One recent article~\cite{brown2021device} used these methods to determine the lower bounds of the conditional von Neumann entropy certified in a device-independent approach using an extended NPA method~\cite{pironio2010convergent} using the NCPOL2SDPA tool~\cite{wittek2015algorithm}. We provide a brief overview of the theory of semidefinite representations of semi-algebraic functions in sec.~\ref{sec:SDPreps}.
	
	An important problem in the investigation of the properties of quantum states is in determining whether they are separable or not. An $N$-partite state $\rho$ is called separable when it is written as a convex combination of product states, \textit{viz.} $\rho = \sum_i p_i \rho_1^i \otimes \cdots \otimes \rho_N^i$~\cite{werner1989quantum}. Determining whether a given state is separable or entangled based solely on the definition is a challenging task in practice. Thus, the so-called separability problem emerges as one of the fundamental issues in the study of entanglement. The famous Peres-Horodecki Positive Partial Transpose (PPT) criterion~\cite{peres1996separability,horodecki1996separability} provided a necessary condition for separability of states and says that if a bipartite state $\rho_{\mathcal{A}\mathcal{B}}$ is separable, then $\rho_{\mathcal{A}\mathcal{B}}^{T_B} \succeq 0$. Another attempt at this issue was~\cite{lewenstein1998separability}, where a constructive algorithm that enables the identification of the optimal separable approximation for any density matrix associated with a finite-dimensional composite quantum system was presented. The method established a condition for separability and provided a measure of entanglement. An important connection of the separability problem with SDP was the so-called DPS method given in~\cite{DPS02,DPS04}, which may be considered as a direct development of the PPT criterion, providing a hierarchy of SDP approximations discussed in sec.~\ref{sec:PPT}.
	
	Another notion crucial to quantum information is quantum channels~\cite{wilde_2017}. A quantum channel is a communication channel that transmits quantum information. Their formalism is general enough to term a channel any form of state evolution either in time or space governed by quantum mechanics. There are multiple different issues to be studied regarding quantum channels. One of the main research problems related to them is the characterization and classification of different types of channels, like depolarizing or amplitude-damping channels. Another research problem related to quantum channels is the development of methods for channel estimation and tomography, or the study of noisy and imperfect channels. In practice, all communication channels are subject to noise and imperfections that can degrade the quality of transmitted quantum states. Therefore, it is essential to develop methods for mitigating the effects of noise and imperfections on quantum communication. A basic tool used in modeling quantum channels, or more general maps linear $\mathsf{L}[\mathcal{H}_1, \mathcal{H}_2]$, for Hilbert spaces $\mathcal{H}_1$ and $\mathcal{H}_2$, is the Choi-Jamiołkowski isomorphism discussed in sec.~\ref{sec:C-J}.
	
	The Tsirelson bound also referred to as the Cirel'son bound, is a concept in quantum mechanics that holds significance in the investigation of quantum non-locality. It was first derived by Boris Tsirelson in 1980~\cite{cirel1980quantum}. In essence, the Tsirelson bound establishes a maximum level of correlation achievable between two distant quantum systems. This bound carries profound implications for our comprehension of quantum mechanics and its practical applications. In particular, if the correlations violate the Tsirelson bound, it implies that quantum physics cannot reproduce them. This observation significantly contributes to our understanding of entanglement. The Tsirelson bound finds practical utility in various applications of quantum information theory, including quantum cryptography and quantum teleportation. For instance, it enables the quantification of the requisite and attainable level of entanglement for secure communication through quantum cryptography. In sec.~\ref{sec:SoS} we briefly describe the sum of squares (SoS) technique and then show an example of how can it be applied to the derivation of the Tsirelson bound.
	
	A fundamental concept, closely related to the Tsirelson bound is quantum contextuality. It refers to the property of quantum systems where the outcome of a measurement depends on the context in which it is measured. In other words, the value of a quantum property is not determined by the property itself, but by the other properties with which it is measured. This means that the same quantum system can exhibit different properties depending on how it is measured, and this property has been shown to be essential for many quantum information processing tasks. One of the approaches to the analysis of contextuality was given in~\cite{cabello2010non,cabello2014graph} where a relationship with the so-called Lov{\'a}sz theta has been established. The method revealed to be very profound~\cite{leifer2014quantum,howard2014contextuality,regula2021fundamental,cabello2021converting,tilly2022variational,gupta2023quantum}. Here, in sec.~\ref{sec:Lovasz} we describe the SDP methods for Lov{\'a}sz theta and show how to relate it to contextuality.
	
	We will now briefly review the topic of Bell inequalities and Bell operators~\cite{bell1964einstein}, as this will be needed in many places in this work, especially for sec.~\ref{sec:NPA}. Bell inequalities are mathematical expressions that set a limit on certain probabilities, which cannot be violated in a classical physics framework but can be exceeded in quantum mechanics. The violation of Bell inequalities can be observed through experimental measurements, providing conclusive evidence that the behavior of the world cannot be explained solely by classical physics. Such groundbreaking experiments were conducted in the 1980s by Aspect \textit{et al.}~\cite{aspect1981experimental,aspect1982experimental,aspect1982experimental2}. A Bell experiment involves two or more separate parties who share a quantum state and perform measurements with different settings, without any form of communication between them. By conducting a series of such measurements and analyzing the collected data, it becomes possible to estimate the joint probabilities $\{P(a,b|x,y)\}$ of the outcomes conditioned on the settings. A bipartite Bell operator\index{Bell operator} is a linear functional that operates on a probability distribution defined over two parties or subsystems of the form $\sum_{a,b,x,y} \alpha_{a,b,x,y} P(a,b|x,y)$, where $\alpha_{a,b,x,y} \in \mathbb{R}$.
	
	Classical devices can be described using the following elements. The class $\mathfrak{L}$ represents local distributions, which are in accordance with classical physics. The statistical description of the pair of devices is given by $P(a,b|x,y) = \sum_{\lambda} P(\lambda) \cdot P_{A|X,\Lambda}(a|x,\lambda) \cdot P_{B|Y,\Lambda}(b|y,\lambda)$. Here, $P(\lambda)$ represents the probability of observing the hidden state $\lambda$, and $P_{A|X,\Lambda}(a|x,\lambda)$ and $P_{B|Y,\Lambda}(b|y,\lambda)$ correspond to the conditional probabilities of obtaining results $a$ for Alice and $b$ for Bob, respectively, given their respective settings $x$ and $y$, and the hidden state $\lambda$. $P_{\Lambda}(\lambda)$ refers to the distribution of hidden internal states, where $\lambda$ represents a specific state and $\sum_{\lambda \in \Lambda} P_{\Lambda}(\lambda) = 1$ ensures normalization.
	
	Next, the non-signaling devices can be characterized as follows. The class $\mathfrak{N}$ denotes non-signaling distributions, which align with the principles of relativistic physics. $P_{A|X}(a|x)$ and $P_{B|Y}(b|y)$ represent the marginal distributions of Alice and Bob, respectively. These distributions are derived from the joint distribution $P(a,b|x,y)$, and they satisfy the conditions $\sum_{b \in B} P(a,b|x,y) = P_{A|X,Y}(a|x,y) = P_{A|X}(a|x)$ and $\sum_{a \in A} P(a,b|x,y) = P_{B|X,Y}(b|x,y) = P_{B|Y}(b|y)$. These conditions ensure the consistency of the marginal distributions regardless of the settings of the other party. Non-signaling property implies that the settings chosen by one party do not have any influence on the marginal distribution observed by the other party. By considering these elements and properties, we can analyze the behavior of non-signaling devices in the context of bipartite systems. Optimization over these sets can be performed using LP.
	
	The class $\mathfrak{Q}$ denotes quantum distributions, which adhere to the fundamental principles of quantum physics. It is noteworthy that the class of local distributions $\mathfrak{L}$, forms a subset of quantum distributions, $\mathfrak{L} \subset \mathfrak{Q}$, but $\mathfrak{Q} \subset \mathfrak{N}$. A Bell operator $\mathcal{I}$ can exhibit a characteristic where its maximum value allowed on the set $\mathfrak{Q}$, denoted as $I_Q$, is strictly greater than its maximum value on the set $\mathfrak{L}$, denoted as $I_L$. The existence of such operators is a consequence of Bell's theorem~\cite{bell2004speakable}. A \textit{Bell inequality} is a statement $\mathcal{I} \leq I_L$ that sets a limit on the value of this operator within the framework of local theories. We say that a Bell inequality is violated if, for a given quantum probability distribution $\{ P(a,b|x,y) \}$, we have $\mathcal{I} > I_L$. The task of optimization over $\mathfrak{Q}$, in particular Bell operators, is NP-hard, as shown by Kempe \textit{et al.} in 2008 at FOCS~\cite{kempe2011entangled}.
	
	The statement $\mathbb{P}(a,b|x,y) \in \mathfrak{Q}$ is true if and only if the following conditions are satisfied, involving the existence of a Hilbert space $\mathcal{H}$, a state (vector) $\ket{\psi}$, and a set of operators (measurements) $\{E^a_x, F^b_y\}_{a,b,x,y}$ such that:
	\begin{enumerate}
		\item The operators $E^a_x$ and $F^b_y$ are projectors. This property ensures that the operators correspond to observable quantities with non-negative probabilities.
		\item Different results with the same setting, represented by $E^a_x$ and $E^{a'}_x$, are orthogonal to each other, given by $E^a_x E^{a'}_x = 0$; similarly for Bob's measurements $F^b_y$ and $F^{b'}_y$. This orthogonality condition signifies that different measurement outcomes are mutually exclusive.
		\item The sum of all operators $E^a_x$ for a fixed $x$ equals the identity operator $\openone$, denoted as $\sum_a E^a_x = \openone$. Similarly, the sum of all operators $F^b_y$ for a fixed $y$ is equal to $\openone$, expressed as $\sum_b F^b_y = \openone$. These normalization conditions ensure that the probabilities of all possible outcomes sum up to 1.
		\item The operators representing measurements for Alice, $E^a_x$, and those for Bob, $F^b_y$, commute with each other, denoted as $[E^a_x, F^b_y] = 0$. This commutation property indicates that the order of measurements performed by Alice and Bob does not affect the results.
		\item The joint probability distribution $P(a,b|x,y)$ can be expressed as the expectation value of the operators $E^a_x$ and $F^b_y$ acting on the state $\ket{\psi}$, given by
		\begin{equation}
			\label{eq:PabxyPsiEF}
			P(a,b|x,y) = \bra{\psi} E^a_x F^b_y \ket{\psi}.
		\end{equation}
	This equation illustrates that the probabilities arise from performing measurements on the quantum state $\ket{\psi}$.
	\end{enumerate}

	Now, we provide a concise overview of the key aspects pertaining to dimension-bounded scenarios~\cite{li2011semi,brunner2008testing,gallego2010device,pawlowski2011semi}. We will discuss them in details in secs~\ref{sec:dimConstraint} and~\ref{sec:seesaw}. Alice and Bob are assigned random inputs, $x$ and $y$. Subsequently, Alice sends a message to Bob based on her input, where the message takes the form of a quantum state $\rho_x$ of a specific dimension $d$. Bob receives the quantum state and performs a measurement $\{ M^b_y \}_b$ on it, yielding a result $b$. This leads to a conditional probability distribution $\{P_d(b|x,y)\}$ within a prepare-and-measure scheme, $P_d(b|x,y) = \Tr \left(\rho_{x} M^b_y\right)$. We assume the absence of entanglement between Alice and Bob in this context. Let $\mathcal{P}_d \equiv \left\{ \{ P_d(b|x,y) \}_{\{\rho_x\}_x, \{\{ M^b_y \}_b\}_y}, \right\}$ be the set of all probabilities of the discussed for in the given dimension $d$. We note that we have $\mathcal{P}_d \subseteq \mathcal{P}_{d+1}$, since increasing the dimension of communicated state we can send at most the same amount of data. A dimension witness $W$ is a linear function of conditional probability distributions, \textit{i.e.} it has the following form:
	\begin{equation}
		\label{eq:DW}
		\sum_{b,x,y} \beta_{b,x,y} P(b|x,y).
	\end{equation}
	The key property of dimension witnesses is that they allow to distinguish the dimensions for which inclusion is strict. Using the definition of probability distributions in prepare-and-measure scheme, we may introduce a notion of dimension witnesses which is analogous to the concept of Bell operators.

	\section{Mathematical framework of optimization}
	\label{sec:frameworkConvex}
	
	% ...in fact, the great watershed in optimization isn't between linearity and nonlinearity, but convexity and nonconvexity - R. Tyrrell Rockafellar, in SIAM Review, 1993
	
	% https://www.solver.com/convex-optimization
	
	% opis w dużym stopniu według "MO19_ch8 - z Introduction to Nonlinear Optimization Theory, Algorithms, and Applications with MATLAB.pdf" p.1-2
	
	A general \textit{static optimization} problem~\cite{magnus2019matrix}, or optimization in finite-dimensional spaces, is a task of determining the values of a certain variable $x \in \mathcal{F} \subseteq \mathbb{R}^n$, called the \textit{decision variable}, for which a given function $f_0 : \mathcal{F} \rightarrow \mathbb{R}$, called \textit{target}, attains its minimum; these points are called \textit{minimizers}, and their set is called the \textit{optimal set}~\cite{beck2014introduction}. The whole set $\mathcal{F}$ is called the \textit{feasible set}\index{feasible set} (and other names like \textit{feasible region}\index{feasible region}, or \textit{solution space}\index{solution space}, or \textit{search space}\index{search space}, are often used)~\cite{beavis1990optimisation}.
	A point $x_0 \in \mathcal{F}$ with the property that for all $x_1 \in \mathcal{F}$ it holds $f(x_1) \geq f(x_0)$ is called a \textit{global minimum}. A point $x_0 \in \mathcal{F}$ for which there exists a neighborhood (in metric space sense) $\mathcal{N}$ such that $x_1 \in \mathcal{N} \cap \mathcal{F} \implies f_0(x_1) \geq f_0(x_0)$ is called a \textit{local minimum}. Every global minimum is also a local minimum.
	
	Further part of this section covers the crucial topic of duality in optimization. In sec.~\ref{sec:Banach} we discuss general optimization in Banach spaces, covering essential techniques and concepts. Next, in sec.~\ref{sec:FR} we explore the Fenchel-Rockafellar scheme, delving into strong duality and constraint qualification, which are fundamental principles in optimization. Then, in sec.~\ref{sec:LagrangianDuality} we discuss an alternative, but less general, way of construction of dual problems, \textit{viz.} the Lagrangian scheme. Lastly, in sec.~\ref{sec:convexCone} we delve into the more specific case of convex cone optimization and show how both dualization schemes apply to it.
	
	\subsection{Convex and linear programming}
	
	Now, let us specify what we mean by \textit{convex optimization} problems. In simple words, these are tasks of minimization of a convex function over a convex set~\cite{beck2014introduction,Boyd04}. To be more specific, the so-called \textit{functional form} of convex problems is defined as follows. Let $m, n \in \mathbb{N}$. The commonly used general form of convex problems is the following:
	\begin{align}
		\label{convex-primal}
		\begin{split}
			\text{minimize } &\null f_0(x) \\
			\text{subject to } &\null f_i(x) \leq b_i, i = 1, \dots, m,
		\end{split}
	\end{align}
	where $f_0, \dots, f_m: \mathbb{R}^n \rightarrow \mathbb{R}$ are convex function, i.e. for any $x_0,x_1 \in \mathbb{R}^n$ and $\lambda \in [0,1]$ they satisfy the so-called Jensen’s inequality: % Boyd04p91
	\begin{equation}
		\label{eq:convexFunctionJensen}
		f_i (\lambda x_0 + (1 - \lambda) x_1) \leq \lambda f_i(x_0) + (1-\lambda) f_i(x_1).
	\end{equation}
	If we replace $\leq$ with $<$ in~\eqref{eq:convexFunctionJensen}, we say that $f_i$ is \textit{strictly convex}\index{convexity!strict}.
	The set of points satisfying the constraints of~\eqref{convex-primal}, i.e. the feasible set\index{feasible set} $\mathcal{F} = \bigcap_{i=1}^m \{x \in \mathbb{R}^n : f_i(x) \leq b_i\} \subseteq \mathbb{R}^n$, is a convex set. A crucial property of convex programs is that all their local minimum points are also global minimum points. What is more, the optimal set for a convex problem is also a convex set. If $f_0$ is strictly convex then there exists at most one global minimum. The \textit{optimal value}\index{value of optimization problem} or, when there is no ambiguity, simply the \textit{value}, of an optimization problem corresponds to the optimal objective function value, representing the optimal outcome attainable for the objective function while adhering to all constraints. The \textit{solution}\index{solution of optimization problem} to an optimization problem refers to the collection of decision variables that yield the optimal objective function value. This solution must satisfy all constraints imposed by the problem. It is worth noting that in certain scenarios, multiple optimal solutions can exist, indicating that various sets of decision variables yield the same optimal objective function value.
	
	The terms \textit{optimization} and \textit{optimization problem} are often used interchangeably in the literature to refer to the task of finding the optimal (usually minimal, as above) value of an objective function. However, the terms have a subtle difference in their meaning. Optimization typically refers to the act of determining the optimal value itself. The optimization problem encompasses not only finding the optimal value but also the associated decision variables or parameters that achieve that optimal value, i.e. the solution. Thus, we distinguish the minimization problems where the task is to find both the value $\min_{x \in \mathcal{F}} [f_0(x)]$ and the solution $\argmin_{x \in \mathcal{F}} [f_0(x)]$, from the minimization, i.e. the task of finding the infimum $\inf_{x \in \mathcal{F}} [f_0(x)]$. Note that the infimum may not even be attained, whereas for the minimum there always exists a solution attaining it.
	
	Particular examples of convex optimization problems are LPs and SDPs, being the main topic of this Review. We start with the formulation of LP. Let $m, n \in \mathbb{N}$, and $m \leq n$. The so-called primal canonical form of LP is the following optimization task in variable $x$:
	\begin{align}
		\label{LP-primal}
		\begin{split}
			\text{minimize } &\null c^{T} \cdot x \\
			\text{subject to } &\null A x = b, x \geq 0,
		\end{split}
	\end{align}
	where $A \in \mathbb{R}^{n \times m}$, $b \in \mathbb{R}^m$, $x, c \in \mathbb{R}^n$. The dual problem of LP is
	\begin{align}
		\label{LP-dual}
		\begin{split}
			\text{maximize } &\null b^{T} \cdot y \\
			\text{subject to } &\null c - A^T y = z, \\
			&\null z \geq 0.
		\end{split}
	\end{align}
	In the above problems, the variable $x$ is called the \textit{primal variable}\index{variable!primal}, $y$ the \textit{dual variable}\index{variable!dual}, $z$ the \textit{dual slack variable}\index{variable!dual slack}, $A$ is the \textit{linear constraint matrix}\index{linear constraint matrix}, $b$ is the \textit{right-hand side of the linear constraint}, and $c$ is the \textit{linear coefficient}.
	
	% \subsection{Difficulty of classes of optimization problems}
	% uszczegółowić, że convex > conic > SDP > SCOP > QP > LP
	% dla SDP > SCOP patrz https://arxiv.org/abs/1610.04901
	% wspomnieć, że każdy z nich to jakiś convex cone
	% SCOP eqref{eq:SecondOrderCone}
	
	% ogólny overview wg. https://en.wikipedia.org/wiki/Convex_optimization
	% wspomnieć np. o https://link.springer.com/article/10.1007/BF00120662 (i dyskusji z tego), jak odrobina zmiany robi NP-trudnym
	
	% https://en.wikipedia.org/wiki/Conic_optimization
	% https://www.solver.com/linear-quadratic-programming
	% https://people.smp.uq.edu.au/YoniNazarathy/teaching_projects/studentWork/Duality.pdf

	\subsection{Sets and cones definitions}
	
	% sets
	Let $X$ be a linear space, and let $C \subseteq X$.
	\textit{Core} of a set $C \subset X$, for a Banach space $X$, is the set of all points in $C$ such that for any direction $d$ in $X$ there exist $T_d > 0 $ such that for all $t \in [0, T_d]$ we have $x + t d \in C$, \textit{viz.}: % BL06p34
	\begin{equation}
		\label{eq:coreOfSet}
		\core{C} \equiv \left\{ x \in C : \bigforall_{\substack{d \in X, \\ \Norm{d} = 1}} \bigexists_{T_d > 0} \bigforall_{t \in [0, T_d]} (x + t d) \in C \right\}.
	\end{equation}
	Contrast it with the \textit{interior} of a set $C$ defined as:
	\begin{equation}
		\interior{C} \equiv \left\{ x \in C : \bigexists_{T > 0} \bigforall_{\substack{d \in X, \\ \Norm{d} = 1}} \bigforall_{t \in [0, T]} (x + t d) \in C \right\}.
	\end{equation}
	It is easy to see that $\interior{C} \subseteq \core{C}$.
	% TODO: zintegrować z definicjami w następnej subsekcji
	$C$ is defined to be \textit{convex} if
	\begin{equation}
		\bigforall_{x_1, x_2 \in C} \bigforall{\lambda \in [0,1]} \lambda x_1 + (1 - \lambda) x_2 \in C.
	\end{equation}
	In particular, $\emptyset$ is a convex set\index{convex!set}.
	$C$ is defined to be \textit{absorbing}~\cite[p.244]{BL06} if it is convex and $X = \bigcup_{t \geq 0} t C$. Obviously, $0 \in \core{C}$ for absorbing $C$.
	
	% cones
	A subset $K$ of a vector space $\mathcal{X}$ over an ordered field $F$ is called a \textit{cone}, or \textit{nonnegative homogeneous}, if and only if~\cite{zalinescu2002convex}:
	\begin{equation}
		x \in K, \lambda \in F_+ \implies \lambda x \in K,
	\end{equation}
	where $F_+$ is the set of all non-negative scalars of $F$. Thus, the cone is simply the set invariant under multiplication by non-negative scalars. The cone $K$ is a \textit{convex cone} if and only if~\cite{Boyd04}
	\begin{equation}
		\label{eq:ConvexCone}
		x_1, x_2 \in K \implies x_1 + x_2 \in K,
	\end{equation}
	that can be intuitively understood as $K + K \subseteq K$.
	
	% dual cones
	For an arbitrary, \textit{not} necessarily being a cone, subset $K$ of a vector space $\mathcal{X}$ over an ordered field $F$, the \textit{dual cone}\index{dual cone} is defined as
	\begin{equation}
		\label{eq:DualCone}
		K^{*} \equiv \{ z \in \mathcal{X} : x \in K \implies \bk{z}{x} \geq 0 \},
	\end{equation}
	or, by the Riesz representation theorem, equivalently up to isomorphism, % https://math.stackexchange.com/questions/598392/what-is-the-relation-between-dual-spaces-and-inner-product
	\begin{equation}
		K^{*} \equiv \{ z \in \mathcal{X}^{*} : x \in K \implies z \cdot x \geq 0 \}.
	\end{equation}
	It can be shown~\cite{Boyd04} that the set $K^{*}$ defined by~\eqref{eq:DualCone} is always a convex cone. If $K = K^{*}$, then $K$ is called a \textit{self-dual cone}\index{self-dual cone}. Examples of a self-dual cone for $m \in \mathbb{N}_{+}$ are: the \textit{positive orthant cone} of $\mathcal{X} = \mathbb{R}^m$:
	\begin{equation}
		K_{+}^m \equiv \{ x \in \mathbb{R}^m : x_1 \geq 0, \cdots x_m \geq 0 \} = \mathbb{R}_{+}^m,
	\end{equation}
	the \textit{Lorentz} (or \textit{second order}, or \textit{quadratic}) cone:
	\begin{equation}
		\label{eq:SecondOrderCone}
		K_q^m \equiv \{ (x,t) \in \mathbb{R}^{m+1} : \SN{x} \leq t \},
	\end{equation}
	and the PSD cone $\mathbb{S}^n_+$ discussed further in sec.~\ref{sec:PSDdefinition}.
	% https://math.stackexchange.com/questions/2293218/proof-of-lorentzseconde-grade-cone-is-convex-and-self-dual
	
	% semialgebraic sets
	A set $\mathcal{S} \subseteq \mathbb{R}^n$ is called a basic closed semialgebraic set\index{semialgebraic set!basic} if and only if there exist a set of polynomials $\{f_i\}_{i \in [m]}$, $f_i: \mathbb{R}^n \rightarrow \mathbb{R}$, such that $\mathcal{S} = \left\{ x \in \mathbb{R}^n : \bigforall_{i \in [m]} f_i(x) \geq 0 \right\}$. Similarly, it is called a basic open semialgebraic set if and only if $\mathcal{S} = \left\{ x \in \mathbb{R}^n : \bigforall_{i \in [m]} f_i(x) > 0 \right\}$. Compare this with algebraic sets\index{algebraic set} which have the form $\mathcal{S} = \left\{ x \in \mathbb{R}^n : \bigforall_{i \in [m]} f_i(x) = 0 \right\}$. A set $\mathcal{S}$ is called semi-algebraic\index{semialgebraic set} if there exists a set $\{\mathcal{S}_{i,j}\}_{i \in [k], j \in [r_i]}$ for some $k, r_i \in \mathbb{N}_{+}$ such that each $\mathcal{S}_i$ is a basic closed semialgebraic set or a basic open semialgebraic set or an algebraic set and $\mathcal{S} = \bigcup_{i \in [k]} \bigcap_{j \in [r_i]} \mathcal{S}_{i,j}$. Any algebraic set is obviously semialgebraic. A consequence of the famous Tarski-Seidenberg principle~\cite{Tarski1949-TARADM-3,seidenberg1954new} is the fact that the set of semialgebraic sets is closed under projections, i.e. if $S \in \mathbb{R}^{n_1 + n_2}$ is a semialgebraic set, then also its projection onto the first $n_1$ coordinates is a semialgebraic set. We refer to Chapter~2 of~\cite{bochnak2013real} for a detailed discussion of semialgebraic sets. The positive orthant, Lorentz, and PSD cone are basic closed semialgebraic sets; the same holds for polyhedra and spectrahedra discussed in sec.~\ref{sec:SDPreps}. % TODO: https://mathworld.wolfram.com/CylindricalAlgebraicDecomposition.html

	\subsection{Convex analysis: Functions, convex conjugate, and Fenchel-Rockafellar theorem}
	
	Consider an arbitrary function $f: X \rightarrow \mathbb{R} \cup \{-\infty,+\infty\}$. The function is defined to be \textit{proper}\index{proper function} if it never takes the value $-\infty$, and is not identically equal to $+\infty$. The \textit{epigraph}\index{epigraph} of $f$ is defined as
	\begin{equation}
		\epi{f} \equiv \{ (x,r) \in X \times \mathbb{R}: r \geq f(x) \}, % L06p22
	\end{equation}
	so it is the set of all point above the graph of the function. The hypograph\index{hypograph} of $f$ is $\hyp{f} \equiv \{ (x,r) \in X \times \mathbb{R}: f(x) \geq r \}$.
	One usually formulates the definition of a convex function in terms of Jensen's inequality~\eqref{eq:convexFunctionJensen}; this is the convexity in \textit{analytical} sense. Alternatively, epigraph allows to provide a geometric sense of convexity, \textit{viz.} $f$ is defined to be \textit{convex} if $\epi{f}$ is convex, and \textit{concave} if $-f$ is convex; see~\cite[p.12]{L06} for a discussion.
	The \textit{effective domain}\index{effective domain} of $f$ is defined as
	\begin{equation}
		\dom{f} \equiv \{ x \in X: f(x) < +\infty \}. % L06p22
	\end{equation}
	$f$ is defined to be \textit{lower semi-continuous}\index{lower semi-continuous} ($\lsc$) if $\epi{f}$ is a closed subset of $X \times \mathbb{R}$. % L06p34
	The set $\cont{f}$ is the set of all points where $f$ is finite and continuous. % BZ05p128
	
	% Boyd04p101-102
	Let $F: X \times Y \rightarrow \mathbb{R} \cup \{-\infty,+\infty\}$, with $Y$ a linear spaces, be convex in both parameters. Let $C \subseteq X$ be a non-empty convex set. Then~\cite[p.88]{Boyd04}, the function
	\begin{equation}
		\label{eq:convexFromInf}
		\theta(y) = \inf_{x \in C} F(x, y)
	\end{equation}
	is convex in $y$, as long as
	\begin{equation}
		\label{eq:thetaNotMinusInf}
		\bigforall_{y \in Y} \theta(y) \neq -\infty.
	\end{equation}
	The epigraph of $\theta$ is $\epi{\theta} = \left\{ (x,t) : \bigexists_{y \in Y} (x,y,t) \in \epi{F} \right\}$, and is convex, as a projection of a convex set $\epi{F}$.
	Indeed, let $y_0, y_1 \in \dom{\theta}$. Then
	\begin{equation}
		\bigforall_{\epsilon > 0} \bigexists_{x_0, x_1 \in C} F(x_0, y_0) \leq \theta(y_0) + \epsilon \text{ and } F(x_1, y_1) \leq \theta(y_1) + \epsilon,
	\end{equation}
	and for any $\lambda \in [0,1]$ we have
	\begin{equation}
		\begin{aligned}
			\theta& (\lambda y_0 + (1-\lambda) y_1) = \inf_{x \in C} F(x, \lambda y_0 + (1-\lambda) y_1) \leq \\
			&F(\lambda x_0 + (1-\lambda) x_1, \lambda y_0 + (1-\lambda) y_1) \leq \lambda F(x_0,y_0) + (1-\lambda) F(x_1,y_1) \leq \\
			&\lambda \theta(y_0) + (1-\lambda) \theta(y_1) + \epsilon.
		\end{aligned}
	\end{equation}
	Since $\epsilon$ can be arbitrarily small, Jensen's inequality~\eqref{eq:convexFunctionJensen} for $\theta$ follows.
	
	A well known operation of the holomorphic functional calculus is the extension of a function defined on real values to Hermitian matrices. This extension allows for the evaluation of functions that are not originally defined on matrices, but can be extended to them through the use of complex analysis techniques. Any Hermitian matrix $H \in \mathbb{H}^n$ can be diagonalized by a unitary matrix $U$, so that $H = U \cdot \Diag[(d_i)_{i \in [n]}] \cdot U^\dagger$. A function $f : \mathbb{R} \rightarrow \mathbb{R}$ can be applied to the eigenvalues, and thus the function can then be extended to the entire matrix as $f(H) \equiv U \cdot \Diag[(f(d_i))_{i \in [n]}] \cdot U^\dagger$. We say that $f$ is an \textit{operator monotone}\index{operator!monotone} (or matrix monotone) when for any $M_1, M_2 \in \mathbb{H}^n$, from $M_1 \succeq M_2$ it follows that $f(M_1) \succeq f(M_2)$. Next, we say that $f$ is \textit{operator convex}~\index{operator!convex} (or matrix convex) if it satisfies Jensen's inequality~\eqref{eq:convexFunctionJensen} in L\"{o}wner's partial order, i.e. $\lambda f(M_1) + (1-\lambda) f(M_2) \succeq f (\lambda M_1 + (1 - \lambda) M_2)$ for all $\lambda \in [0,1]$~\cite{hansen2003jensen}. Finally, we say that $f$ is \textit{operator concave}\index{operator!concave} (or matrix concave) when $-f$ is operator convex~\cite{carlen2010trace,agarwal2022brief}. The \textit{matrix epigraph}\index{matrix epigraph} of $f$ is $\epiM{f} \equiv \left\{ (X, R) \in \mathbb{H}^n_{++} \times \mathbb{H}^n : R \succeq f(X) \right\}$, and the \textit{matrix hypograph} of $f$ is $\hypM{f} \equiv \left\{ (X, R) \in \mathbb{H}^n_{++} \times \mathbb{H}^n : f(X) \succeq R \right\}$.
	
	The (commutative) \textit{perspective} of a function\index{perspective of a function} $f : \mathbb{R}^n \rightarrow \mathbb{R}$ is the function $\Persp_f : \mathbb{R}^{n+1} \rightarrow \mathbb{R}$ defined as $\Persp_f(x,t) \equiv t f(x \cdot t^{-1})$ with $\dom{\Persp_f} = \left\{ (x,t) : x/t \in \dom{f}, t > 0 \right\}$. The operation of perspective preserves the convexity, i.e. if $f$ is convex then $\Persp_f$ is convex~\cite[p.89]{Boyd04}. If $M_1, M_2 \in \mathbb{H}^n$ commute, then $\Persp_f(M_1,M_2)$ is well-defined by extending $\Persp_f$ to matrices~\cite{effros2009matrix}. To cover also the non-commutative case, the \textit{non-commutative perspective} of an operator convex function is defined as the unique extension of the corresponding (commutative) perspective that preserves homogeneity and convexity~\cite{effros2014non}. The formula for non-commutative perspective is $\PerspM_f[M_1, M_2] \equiv M_2^{1/2} \cdot f \left( M_2^{-1/2} M_1 M_2^{-1/2} \right) \cdot M_2^{1/2}$, with $\dom{\PerspM_f} = \mathbb{H}^n \times \mathbb{H}^n_+$~\cite{ebadian2011perspectives}. For instance, the non-commutative perspective of the negative logarithm function is the operator relative entropy~\cite{fujii1989relative,fujii1992operator,fujii2018relative,fujii2022relative}, \textit{viz.}
	\begin{equation}
		\label{eq:relOpEntropy}
		\begin{aligned}
			S(M_1|M_2) &= M_2^{1/2} \cdot \eta[M_2^{-1/2} M_1 M_2^{-1/2}] \cdot  M_2^{1/2} = \PerspM_{(- x \log{x})}[M_1,M_2] \\
			&= -M_1^{1/2} \cdot \log \left( M_1^{-1/2} M_2 M_1^{-1/2} \right) \cdot M_1^{1/2} = \PerspM_{(-\log)}[M_2,M_1]
		\end{aligned}
	\end{equation}
	for $\eta(x) \equiv - x \log{x}$ and invertible $M_1$ and $M_2$~\cite{fujii2018relative}.
	
	In~\cite{pusz1975functional,ando1978topics} the notion of the \textit{matrix geometric mean}\index{matrix geometric mean} $M_1 \# M_2 \equiv M_1^{1/2} \cdot [ M_1^{-1/2} M_2 M_1^{-1/2} ]^{1/2} \cdot M_1^{1/2}$, which satisfy certain general properties~\cite{kubo1980means}, was introduced for PD $M_1$ and $M_2$. Its direct generalization is the so-called $t$-weighted matrix geometric mean
	\begin{equation}
		\label{eq:wGMean}
		M_1 \#_t M_2 \equiv \PerspM_{(x \mapsto x^t)}[M_1,M_2] = M_1^{1/2} \cdot [ M_1^{-1/2} M_2 M_1^{-1/2} ]^{t} \cdot M_1^{1/2},
	\end{equation}
	and thus $M_1 \# M_2 = M_1 \#_{1/2} M_2$~\cite{bhatia2009positive,sagnol2013semidefinite}. It can be shown that $t$-weighted matrix geometric mean is operator concave for $t \in [0,1]$, and operator convex for $t \in [-1,0] \cup [1,2]$~\cite{bhatia2009positive}.
	
	The \textit{indicator}\index{indicator function} function of $C$ is defined as
	\begin{equation}
		\label{eq:indicatorFunction}
		I_C[x] \equiv 
		\begin{cases}
			0 & \text{ if } x \in C, \\
			+\infty & \text{ otherwise}.
		\end{cases}
	\end{equation}
	The indicator function $I_C[x]$ is convex if and only if, the set $C$ is convex.
	It can also be shown that the indicator function is lower (upper) semi-continuous if and only if, $C$ is closed (open). % https://math.stackexchange.com/questions/706671/lower-semicontinuity-of-indicator-function and https://www.math.uh.edu/~rohop/fall_06/Chapter5.pdf page 5
	
	The mean value of a random variable $x$ we denote as $\mean{x}$, and the standard deviation as $\standardDeviation{x}$. The covariance\index{covariance} between random variables $x_i$ and $x_j$ is defined as $\cov[x_i, x_j] \equiv \mean{(x_i - \mean{x_i}) \cdot (x_i - \mean{x_i})}$, and their correlation\index{correlation} is defined as $\text{corr}[x_i, x_j] \equiv \cov[x_i, x_j] / (\standardDeviation{x_i} \cdot \standardDeviation{x_j})$. The variance of $x$ is the covariance of the variable with itself, $\var[x] \equiv \cov[x,x] \geq 0$.

	\subsubsection{Fenchel conjugate}~\\
	% TODO: to co bieżemy z L06 jest z X będącym Banach space - napisać tak, żeby było jasne, że X wszędzie jest Banach space
	
	For an arbitrary function $f: X \rightarrow \mathbb{R} \cup \{+\infty\}$, where $X$ is a Banach space, the \textit{Fenchel conjugate}~\cite{fenchel_1949}, or \textit{convex conjugate}, or \textit{Legendre transform}, or \textit{Legendre-Fenchel transform}, or simply the conjugate, being the basic operation in convex analysis, is defined as % L06p88
	\begin{equation}
		\label{eq:convexConjugateDef}
		f^{*}: X^{*} \rightarrow \mathbb{R} \cup \{-\infty,+\infty\} : x^{*} \mapsto \sup_{x \in X} \{ \bkc{x^{*}}{x} - f(x) \}.
	\end{equation}
	The conjugate operation can be applied multiple times. For instance the function $(f^{*})^{*}$ is defined on $X^{**}$. With an abuse of notation by $f^{**}$ we denote the restriction of $(f^{*})^{*}$ to $X$, so that~\cite[p.83]{L06}:
	\begin{equation} % L06p92
		f^{**}: X \rightarrow \mathbb{R} \cup \{-\infty,+\infty\} : x \mapsto \sup_{x^{*} \in X^{*}} \{ \bkc{x^{*}}{x} - f^{*}(x^{*}) \}.
	\end{equation}
	It holds
	\begin{equation}
		\epi{f^{*}} = \bigcap_{x \in X} \epi{ \{ \bkc{\cdot}{x} - f(x) \} }.
	\end{equation}
	Since for any $x \in X$ the set $\epi{ \{ \bkc{\cdot}{x} - f(x) \} }$ is convex and closed in weak\textsuperscript{*} topology on $X^{*} \times \mathbb{R}$ (and the natural topology on $\mathbb{R}$), we have that $f^{*}$ is always a convex and $\lsc$ function in that topology, no matter on the form of $f$. % L06p89
	It is trivial to see that $f_1 \leq f_2 \implies f_1^{*} \geq f_2^{*}$ and $f^{*}(0) = -\inf_{x \in X} f(x)$~\cite[p.82]{L06}.
	
	A direct consequence of the definition~\eqref{eq:convexConjugateDef} is for any $f: X \rightarrow \mathbb{R} \cup \{-\infty,+\infty\}$ it holds~\cite[p.184]{BC10}:
	\begin{equation} % TODO: sprawdzić
		f \equiv +\infty \iff f^{*} \equiv -\infty \iff -\infty \in f^{*}(X), % BC10p184
	\end{equation}
	and, in particular if $f^{*}$ is proper, then also $f$ is proper.
	From $\bigforall_{x \in X} \bigforall_{x^{*} \in X^{*}} f(x) \geq \bkc{x^{*}}{x} - f^{*}(x^{*})$ taking supremum over $x^{*}$ we get $\bigforall_{x \in X} f^{**}(x) \leq f(x)$, i.e.
	\begin{equation} % L06p92
		\label{eq:biconIneq}
		f^{**} \leq f.
	\end{equation}
	The inequality~\eqref{eq:biconIneq}, together with the convexity and $\lsc$ properties of the Fenchel conjugate, motivate to call $f^{**}$ a \textit{regularization}, or a \textit{convex lower semicontinuous relaxation}. % L06p88
	If $f$ itself is convex and $\lsc$, and there exist $x^{*} \in X^{*}$ and $\alpha \in \mathbb{R}$ such that $f \geq \bkc{x^{*}}{\cdot} + \alpha$ then the \textit{Fenchel-Moreau theorem}\index{Fenchel-Moreau theorem} states the equality, % TODO: jak dokładnie brzmi F-M? W L06p93 nie potrzeba convex lsc tylko pewną nierówność; a w BC10p190 nie potrzeba nierówności, tylko proper convex lsc. Rozstrzygnąć i i tych kilku miejscach, gdzie używam M-F (czyli de facto tam gdzie f=f^{**}) , sprawdzić, czy używam właściwie.
	\begin{equation}
		\label{eq:FenchelMoreauTheorem}
		f = f^{**},
	\end{equation}
	see~\cite[p.84]{L06} for the proof.

	\subsubsection{Subgradient ad Fenchel-Rockafellar theorem}~\\
	\label{sec:withFenchelRockafellarTheorem}
	
	Consider an arbitrary function $f: X \rightarrow \mathbb{R} \cup \{ +\infty \}$. An element $x^{*} \in X^{*}$ is called a \textit{subgradient} of $f$ at $x \in \dom{f}$ when
	\begin{equation}
		\bigforall_{x' \in X} f(x') - f(x) \geq \bkc{x^{*}}{x' - x}. % L06p44, BZ05p125
	\end{equation}
	The (possibly empty) set of all subgradients of $f$ at $x \in \dom{f}$ is called \textit{subdifferential}\index{subdifferential} and denoted by $\partial f(x)$.
	Directly from the definitions of subgradient and convex conjugate, it follows that
	\begin{equation}
		\label{eq:ffc} % TODO: czy funkcja musi być np. convex lub x \in dom f? w L06p96 nie ma takich założeń, ale w Prop.4.4.1 BZ05p143 są, a w BC10p185 jest wymaganie proper
		\bigforall_{x \in \dom f \subseteq X} \bigforall_{x^{*} \in X^{*}} x^{*} \in \partial f(x) \iff f(x) + f^{*}(x^{*}) = \bkc{x^{*}}{x} \implies x \in \partial f^{*}(x^{*}). % L06p96-97
	\end{equation}
	If $f(x) = f^{**}(x)$, e.g. for $f$ proper convex $\lsc$ by the Fenchel-Moreau theorem, then the implication in~\eqref{eq:ffc} becomes equivalence.
	Note that a part of~\eqref{eq:ffc}, the so-called \textit{Fenchel-Young inequality}\index{Fenchel-Young inequality},
	\begin{equation}
		\label{eq:FenchelYoungIneq}
		\bigforall_{x \in X} \bigforall_{x^{*} \in X^{*}} f(x) + f^{*}(x^{*}) \geq \bkc{x^{*}}{x}, 
	\end{equation}
	always holds~\cite[p.51]{BL06}. % BL06p51
	The notion of the subgradient does not require the function $f$ to be convex. Nonetheless, it uses global properties of $f$ and is most useful in the context of convex functions. % L06p44
	
	The Fenchel-Rockafellar theorem\index{Fenchel-Rockafellar theorem} provides sufficient conditions for the subdifferential of a convex function $f : X \rightarrow \mathbb{R} \cup \{ +\infty \}$ to be non-empty at $x$, i.e. $\partial f (x) \neq \emptyset$~\cite[p.121]{BZ05}, \textit{viz.}:
	\begin{enumerate}
		\item $f$ is $\lsc$ and $x \in \core (\dom{f})$, or
		\item $x \in \cont{f}$.
	\end{enumerate}

	% TODO: może uzupełnić np. na podstawie abstraktu z https://doi.org/10.1016/B978-044450550-7/50005-3
	% bardziej zilustrować, że Lagrange jest tylko szczególnym przykładem, a innym jest np. tighter z Appendix C \cite{brown2019framework}
	
	\subsection{Optimization in Banach spaces}
	\label{sec:Banach}
	% np. według
	% https://en.wikipedia.org/wiki/Dual_cone_and_polar_cone
	% https://people.smp.uq.edu.au/YoniNazarathy/teaching_projects/studentWork/Duality.pdf
	% https://www.ams.org/journals/qam/1961-19-03/S0033-569X-1961-0135625-6/ (Philip Wolfe, A duality theorem for non-linear programming 1961)
	% https://math.stackexchange.com/questions/329501/the-dual-of-the-dual-is-the-primal
	% https://en.wikipedia.org/wiki/Duality_(optimization)
	% https://www.net.in.tum.de/fileadmin/TUM/NET/NET-2011-07-2/NET-2011-07-2_20.pdf
	% Borwein-Lewis2006_Book_ConvexAnalysisAndNonlinearOpti.pdf zwłaszcza chap. 5.3
	% https://en.wikipedia.org/wiki/Fenchel%27s_duality_theorem
	% https://en.wikipedia.org/wiki/Wolfe_duality
	
	For the rest of this section, let $X$ and $Y$ be Banach spaces. The duality between product space $X \times Y$ and $X^{*} \times Y^{*}$ is given by $\bkc{(x^{*},y^{*})}{(x,y)} \equiv \bkc{x^{*}}{x} + \bkc{y^{*}}{y}$.
	
	Now, we provide a discussion of the basic concept of duality in optimization. The treatment we provide is relatively extensive, yet we find this topic to be of particular importance, and, additionally, often quite confusing. We discuss the two major duality schemes of Fenchel-Rockafellar and Lagrangian. For SDP the two schemes lead to the same results, but since most of the works refer to either of those two, it is useful to recognize and understand both. We consider the general approach, not limited to convex optimization unless explicitly stated.
	As we will show, for an optimization problem formulated as a minimization task, we can formulate a related maximization task with the property of the so-called \textit{duality} meaning that any of feasible solution of the former its value, is at least as large as the value of any feasible solution of the latter. % TODO: gdzieś podać jawnie rozróżnienie między feasible solution oraz optimal solution oraz the value of a feasible solution oraz the optimal value
	The former optimization task is called a \textit{primal} problem, and the latter a \textit{dual} problem.
	
	For both Fenchel-Rockafellar and Lagrangian schemes, we consider a single, possibly non-convex, function $F: X \times Y \rightarrow \mathbb{R} \cup \{ +\infty \}$.
	The \textit{primal problem} is defined as
	\begin{align}
		\label{eq:L-primal}
		\begin{split}
			\text{minimize } &\null F(x, 0) \\
			\text{subject to } &\null x \in X,
		\end{split}
	\end{align}
	and the primal optimization is $\inf_{x \in X} \{ F(x,0) \}$; its value is called the \textit{primal value}\index{primal value} and denoted by $p$. A value of $x \in X$ for that the primal value $p$ is attained, if exist, is called a \textit{primal solution}\index{primal solution}.
	The \textit{primal function}\index{primal function}, or the \textit{target function}\index{target function} is
	\begin{equation}
		\label{eq:targetFunctionAndPerturbation}
		f(x) \equiv F(x, 0).
	\end{equation}
	Actually, in most of the cases, one is interested in some, domain-specific, target function $f$, and in this sense, $F$ is secondary to $f$. For a given $f$ there exist many different possible functions $F$ that satisfy~\eqref{eq:targetFunctionAndPerturbation}. For this reason $F$ is called a \textit{perturbation function}\index{perturbation function} of $f$.
	
	The \textit{dual problem} is defined as
	\begin{align}
		\label{eq:L-dual}
		\begin{split}
			\text{maximize } &\null -F^{*}(0, y^{*}) \\
			\text{subject to } &\null y^{*} \in Y^{*},
		\end{split}
	\end{align}
	and the dual optimization is $\sup_{y^{*} \in Y^{*}} \{ -F^{*}(0, y^{*}) \}$, where $F^{*} : X^{*} \times Y^{*} \rightarrow \mathbb{R} \cup \{-\infty,+\infty\}$ is the Fenchel conjugate with respect to both variables, i.e.
	\begin{equation}
		F^{*}(x^{*}, y^{*}) = \sup_{(x,y) \in X \times Y} \left\{ \bkc{(x^{*},y^{*})}{(x,y)} - F(x, y) \right\}.
	\end{equation}
	Its value is called the \textit{dual value} and denoted by $d$; a value of $y \in Y^{*}$ for that the value $d$ is attained, if exists, is called the \textit{dual solution}.
	The \textit{dual function}\index{dual function} is~\cite[p.216]{Boyd04}:
	\begin{equation}
		\label{eq:dualFunction}
		g(y^{*}) \equiv -F^{*}(0, y^{*}).
	\end{equation}
	
	The \textit{value function} is defined as
	\begin{equation}
		\label{eq:valueFunction}
		\theta: Y \rightarrow \mathbb{R} \cup \{-\infty,+\infty\}: y \mapsto \inf_{x \in X} F(x,y). % BC10p279, L06p108
	\end{equation}
	The duality property is called \textit{weak}\index{duality!weak} when the value $p$ of the optimal solution of the primal problem is greater or equal to the value $d$ of the optimal solution of the dual problem, i.e. $p \geq d$.
	This property can be obtained directly from the definition of the Fenchel conjugate giving:
	\begin{equation}
		\label{eq:LweakDuality}
		F(x, 0) + F^{*}(0, y^{*}) \geq \bkc{(0,y^{*})}{(x,0)} = 0. % L06p109
	\end{equation}
	The duality is called \textit{strong}\index{duality!strong}, if the equality of the optimal primal and dual solution holds, i.e. $p = d$.
	The difference $p - d \geq 0$ is called the \textit{duality gap}\index{duality gap}. % TODO czy to się nie powtarza już któryś raz zdefiniowane?
	The \textit{duality gap} $\Delta$ is defined as
	\begin{equation}
		\label{eq:dualityGap}
		\Delta \equiv 
		\begin{cases}
			0 & \text{ if } p = d \in \{-\infty, +\infty\} \\
			p - d & \text{ otherwise}.
		\end{cases}
	\end{equation}
	For the strong duality to hold, we need equality in~\eqref{eq:LweakDuality}. From~\eqref{eq:ffc} since $\bkc{(0,y^{*})}{(x,0)} = 0$ we see that the strong duality\index{duality!strong} for $(\bar{x},\bar{y}^{*}) \in X \times Y^{*}$ is equivalent to~\cite[pp.101-102]{L06}:
	\begin{equation}
		(0,\bar{y}^{*}) \in \partial F(\bar{x},0), % L06p109-110
	\end{equation}
	and if $F(\bar{x},0) = F^{**}(\bar{x},0)$, then also to~\cite[p.103]{L06}:
	\begin{equation}
		(\bar{x},0) \in \partial F^{*}(0,\bar{y}^{*}). % L06p111
	\end{equation}

	If the perturbation functions $F$\index{perturbation function} is proper and is both convex and $\lsc$ in the second parameter, then it is called a \textit{dualizing parametrization}~\cite{RW09}\index{dualizing parametrization} of the minimization $\inf_{x \in X} \{ F(x, 0) \}$, resp. of the primal problem~\eqref{eq:L-primal}.
	Thus, $F$ provides a family of optimizations $\inf_{x \in X} \{ F(x, y) \}$, resp. problems, parametrized by the so-called \textit{parameter} variable $y$~\cite[pp.100-101]{L06}.
	We stress that the same primal problem~\eqref{eq:L-primal} is obtained with any other function $F': X \times Y \rightarrow \mathbb{R} \cup \{ +\infty \}$ satisfying $F(\cdot,0) \equiv F'(\cdot,0)$, but a different function $F' \neq F$ may lead to a different dual problem~\eqref{eq:L-dual}.

	\subsection{Fenchel-Rockafellar dualization scheme}
	\label{sec:FR}
	
	First, for the Fenchel-Rockafellar duality~\cite{Boyd04,BZ05,BL06,BC10}, consider a bounded linear map $A: X \rightarrow Y$, and two, possibly non-convex, functions, $f: X \rightarrow \mathbb{R} \cup \{+\infty\}$ and $g: Y \rightarrow \mathbb{R} \cup \{+\infty\}$. For the triple $(A, f, g)$ define
	\begin{equation}
		\label{eq:FxyFR}
		F(x, y) \equiv f(x) + g(Ax + y).
	\end{equation}
	The primal problem~\eqref{eq:L-primal} is thus
	\begin{align} % TODO: minimize of empty is +inf, and maximize is -inf - podać to w definicjach (i sprawdzić, czy tak)
		\label{eq:FR-primal}
		\begin{split}
			\text{minimize } &\null f(x) + g(Ax) \\
			\text{subject to } &\null x \in X.
		\end{split}
	\end{align}
	The \textit{value function}~\eqref{eq:valueFunction} is equal
	\begin{equation}
		\label{eq:valueFunctionFR}
		\theta: Y \rightarrow \mathbb{R} \cup \{-\infty,+\infty\}: y \mapsto \inf_{x \in X} \left( f(x) + g(Ax + y) \right) % BZ05p135
	\end{equation}
	with $\theta(0) = p$.
	We have~\cite[p.106]{L06}:
	\begin{equation}
		\begin{aligned} %BC10p282
			F^{*}(x^{*}, y^{*}) &= \sup_{(x,y) \in X \times Y} \left\{ \bkc{(x^{*},y^{*})}{(x,y)} - f(x) - g(Ax + y) \right\} \\
			&= \sup_{(x,y) \in X \times Y} \left\{ \bkc{x^{*} + A^{*} y^{*}}{x} + \bkc{-A^{*} y^{*}}{x} + \bkc{y^{*}}{y} - f(x) - g(Ax + y) \right\} \\
			&= \sup_{(x,y) \in X \times Y} \left\{ \bkc{x^{*} + A^{*} y^{*}}{x} + \bkc{-y^{*}}{Ax + y} - f(x) - g(Ax + y) \right\} \\
			&= \sup_{x \in X} \left\{ \bkc{x^{*} + A^{*} y^{*}}{x} - f(x) + \sup_{y \in Y} \left( \bkc{-y^{*}}{Ax + y} - g(Ax + y) \right) \right\} \\
			&= \sup_{x \in X} \left\{ \bkc{x^{*} + A^{*} y^{*}}{x} - f(x) + g^{*}(-y^{*}) \right\} \\
			&= f^{*}(x^{*} + A^{*} y^{*}) + g^{*}(-y^{*}).
		\end{aligned}
	\end{equation}
	
	The dual problem~\eqref{eq:L-dual} is thus equal
	\begin{align}
		\label{eq:FR-dual}
		\begin{split}
			\text{maximize } &\null -f^{*}(A^{*} y^{*}) - g^{*}(-y^{*}) \\ % TODO: czy \dagger dla Banacha?; zdefiniować, co to \dagger i co * - jest ok, coś podałem z L06p294 - czy na pewno ok? czy raczej \dagger nie jest typowo Hilberta, a tu wszędzie powinien być Banach i *?
			\text{subject to } &\null y^{*} \in Y^{*}, % zdefiniować przestrzeń sprzężoną *
		\end{split}
	\end{align}
	The weak duality\index{duality!weak}, \textit{viz.} $p \geq d$, can be derived directly from the Fenchel-Young inequality~\eqref{eq:FenchelYoungIneq}. % BZ05p143
	Indeed from the adjoint operator definition~\eqref{eq:adjointOperator} we get
	\begin{equation}
		\begin{aligned}
			\bigforall_{x \in X} & \bigforall_{y^{*} \in Y^{*}} \left( f(x) + f^{*}(A^{*} y^{*}) \right) + \left( g(Ax)  + g^{*}(-y^{*})\right) \\
			& \geq \bkc{A^{*} y^{*}}{x} + \bkc{-y^{*}}{Ax} = \bkc{y^{*}}{Ax} - \bkc{y^{*}}{Ax} = 0.
		\end{aligned}
	\end{equation}
	
	\subsubsection{Strong Duality}~\\
	\label{sec:strongDualityFR}
	\index{duality!strong}
	
	Knowing that the weak duality $p \geq d$ holds, to show the strong duality, we need to establish when $p \leq d$.
	Suppose that the subdifferential $\partial \theta$ is non-empty at $0$. We will show that this suffices for the strong duality, and in sec.~\ref{sec:decouplingLemma} provide the so-called constraint qualification conditions that ensure this. Let $y^{*} \in \partial \theta(0) \subseteq Y^{*}$. From the definition of subgradient for any $x \in X$ we have $\bigforall_{y \in Y} \theta(y - Ax) - \theta(0) \geq \bkc{y^{*}}{y - Ax}$, and thus, from the definition of the value function~\eqref{eq:valueFunctionFR} we get
	\begin{equation}
		\begin{aligned} % BZ05p136
			\bigexists_{y^{*} \in Y^{*}} & \bigforall_{x \in X} \bigforall_{y \in Y} \theta(0) \leq \theta(y - Ax) - \bkc{y^{*}}{y - Ax} \\
			& \leq f(x) + g(y) - \bkc{y^{*}}{y - Ax} = \left( f(x) + \bkc{A^{*} y^{*}}{x} \right) + \left( g(y) - \bkc{y^{*}}{y} \right).
		\end{aligned}
	\end{equation}
	Since the inequality holds for all $x \in X$ and $y \in Y$, taking the infimum of these variables by the definition of the convex conjugate~\eqref{eq:convexConjugateDef} we get $\bigexists_{y^{*} \in Y^{*}} \theta(0) \leq -f^{*}(-A^{*} y^{*}) - g^{*}(y^{*})$ or, equivalently by negating the sign of $y^{*}$ (as also $-y^{*} \in Y^{*}$), we get % BZ05p144
	\begin{equation}
		\bigexists_{y^{*} \in Y^{*}} \theta(0) \leq -f^{*}(A^{*} y^{*}) - g^{*}(-y^{*}).
	\end{equation}
	This, by~\eqref{eq:FR-dual}, shows that $p = \theta(0) \leq d$, and thus
	\begin{equation}
		\partial \theta(0) \neq \emptyset \implies p \leq d,
	\end{equation}
	and the strong duality holds.

	\subsubsection{The Decoupling Lemma}~\\
	\label{sec:decouplingLemma}
	
	Now, we will discuss the so-called Decouping Lemma~\cite{BZ05,borwein2006variational} providing a sufficient criterion, \textit{viz.} certain \textit{constraint qualification}\index{constraint qualification} condition, for the strong duality of the Fenchel-Rockafellar scheme. Since this topic is crucial for duality in convex optimization, for the sake of completeness, we provide in this work in~\ref{sec:decouplingProof} a summary of the proof of the theorem, following~\cite[p.127nn]{BZ05}.
	
	% TODO: tu zdaje się podajemy tylko jedno constraint qualification, (4.3.1) z BZ05; a jest jeszcze drugie, (4.3.2) - opisać je też
	Let us assume that $f: X \rightarrow \mathbb{R}$ and $g: Y \rightarrow \mathbb{R}$ are both convex and that the map $A: X \rightarrow Y$ is linear and bounded. We provide sufficient conditions for $\theta$ as defined in~\eqref{eq:valueFunctionFR} to have non-empty subdifferential at $0$, i.e. $\partial \theta(0) \neq \emptyset$. We use the Fenchel-Rockafellar theorem, see sec.~\ref{sec:withFenchelRockafellarTheorem}.
	One can directly check that $\theta$ when $f$ and $g$ are convex, then $\theta$ given by~\eqref{eq:valueFunctionFR} is a convex function with $\dom{\theta} = \dom{g} - A \dom{f}$, % BZ05p135
	where the set difference is the Minkowski difference.
	Indeed, $F(x,y)$ defined as~\eqref{eq:FxyFR} is convex as a sum of convex $f$ and $g$ is of the form~\eqref{eq:convexFromInf}, and satisfies the condition~\eqref{eq:thetaNotMinusInf}. % TODO: czy rzeczywiście spełnia {eq:thetaNotMinusInf}? upewnić się
	The Decoupling Lemma states that under the condition that both $f$ and $g$ are $\lsc$ and
	\begin{equation}
		\label{eq:decouplingLemmaCond1}
		0 \in \core \left( \dom{\theta} \right),
	\end{equation}
	the function $\theta$ defined as in~\eqref{eq:valueFunctionFR} is continuous at $0$.
	Then, from the Fenchel-Rockafellar theorem\index{Fenchel-Rockafellar theorem} stated in sec.~\ref{sec:withFenchelRockafellarTheorem}, it follows that $\partial \theta$ is non-empty at $0$, as required for the proof of strong duality as given in sec.~\ref{sec:strongDualityFR}.
	
	% TODO: podać przykłady innych constraint qualification, może jest jakiś artykuł przeglądowy?

	\subsection{Lagrangian dualization scheme}
	\label{sec:LagrangianDuality}

	Next, for the Lagrangian duality~\cite{BC10,L06}, consider a single, again possibly non-convex, function $F$.
	Nonetheless, again, if $F$ is convex in both parameters, then $\theta$ is of the form~\eqref{eq:convexFromInf} and satisfies \eqref{eq:thetaNotMinusInf}, thus $\theta$ is convex in this case. % TODO: upewnić się, że spełnia {eq:thetaNotMinusInf}
	The Lagrangian of $F$ is defined as~\cite{L06,RW09,BC10}
	\begin{equation}
		\label{eq:Lagrangian}
		\mathcal{L}: X \times Y^{*} \rightarrow \mathbb{R} \cup \{-\infty,+\infty\}: (x,y^{*}) \mapsto -\sup_{y \in Y} \{ \bkc{y^{*}}{y} - F(x,y) \},
	\end{equation}
	where one can easily recognize the Fenchel conjugate with respect to the parameter (i.e. the second) variable. % L06p117
	
	The Lagrangian allows reformulating the primal problem~\eqref{eq:L-primal}. From the definition~\eqref{eq:Lagrangian} it follows that
	\begin{equation}
		\label{eq:LtoPrimal_step}
		\bigforall_{x \in X} \bigforall_{y^{*} \in Y^{*}} \bigforall_{y \in Y} F(x,y) \geq \mathcal{L}(x,y^{*}) + \bkc{y^{*}}{y}.
	\end{equation}
	When $F$ is a dualizing parametrization, then by the Fenchel-Moreau theorem, see~\eqref{eq:FenchelMoreauTheorem}, the equality in~\eqref{eq:LtoPrimal_step} holds. From~\eqref{eq:LtoPrimal_step} we have for the primal optimization, cf.~\eqref{eq:L-primal}:
	\begin{equation}
		\label{eq:LtoPrimal}
		\inf_{x \in X} \{ F(x,0) \} \geq \inf_{x \in X} \sup_{y^{*} \in Y^{*}} \mathcal{L}(x,y^{*}).
	\end{equation}
	Also the dual optimization, cf.~\eqref{eq:L-dual}, can be easily expressed with the Lagrangian, \textit{viz.}~\cite[p.109]{L06}:
	\begin{equation}
		\label{eq:LtoDual}
		\begin{aligned} %L06p117
			\sup_{y^{*} \in Y^{*}}& \{ -F^{*}(0, y^{*}) \} = \sup_{y^{*} \in Y^{*}} \left\{ -\sup_{x \in X, y \in Y} \{ \bkc{(0,y^{*})}{(x,y)} - F(x,y)\} \right\} = \\
			& = \sup_{y^{*} \in Y^{*}} \left\{ -\sup_{x \in X} \sup_{y \in Y} \{ \bkc{y^{*}}{y} - F(x,y)\} \right\} = \sup_{y^{*} \in Y^{*}} \inf_{x \in X} \mathcal{L}(x,y^{*}).
		\end{aligned}
	\end{equation}
	The inner infimum of the last expression is sometimes used as an alternative definition~\cite[p.216]{Boyd04} of the dual function\index{dual function}~\eqref{eq:dualFunction}:
	\begin{equation}
		\label{eq:dualFunctionL}
		g(y^{*}) \equiv \inf_{x \in X} \mathcal{L}(x,y^{*}).
	\end{equation} 
	
	A direct consequence of~\eqref{eq:LtoPrimal} and~\eqref{eq:LtoDual} is another proof of the weak duality\index{duality!weak}:
	\begin{equation}
		\label{eq:saddleWeakDuality}
		\inf_{x \in X} \{ F(x,0) \} \geq \inf_{x \in X} \sup_{y^{*} \in Y^{*}} \mathcal{L}(x,y^{*}) \geq \sup_{y^{*} \in Y^{*}} \inf_{x \in X} \mathcal{L}(x,y^{*}) = \sup_{y^{*} \in Y^{*}} \{ -F^{*}(0, y^{*}) \}.
	\end{equation}
	The second inequality follows from the well-known max-min inequality~\cite{v1928theorie}. A value $(\bar{x},\bar{y}^{*}) \in X \times Y^{*}$ is defined to be a \textit{saddle point} of $\mathcal{L}$ when
	\begin{equation}
		\bigforall_{x \in X} \bigforall_{y^{*} \in Y^{*}} \mathcal{L}(\bar{x},y^{*}) \leq \mathcal{L}(\bar{x},\bar{y}^{*}) \leq \mathcal{L}(x,\bar{y}^{*}),
	\end{equation}
	i.e., in other words, when
	\begin{equation}
		\sup_{y^{*} \in Y^{*}} \mathcal{L}(\bar{x},y^{*}) = \mathcal{L}(\bar{x},\bar{y}^{*}) = \inf_{x \in X} \mathcal{L}(x,\bar{y}^{*}).
	\end{equation}
	From~\eqref{eq:saddleWeakDuality} we also directly get that for $F$ being a dualizing parametrization the strong duality\index{duality!strong} for $(\bar{x},\bar{y}^{*})$ is equivalent to saying that $(\bar{x},\bar{y}^{*})$ is a saddle point of $\mathcal{L}$, see e.g.~\cite[p.110]{L06} for a proof.

	\subsection{Convex cone optimization and duality}
	\label{sec:convexCone}
	
	We have introduced the general framework for optimization and discussed its duality. Now, we concentrate on a particular problem of convex cone programming, i.e. the optimization over variables belonging to a convex cone~\cite{NT97,Boyd04,dattorro2010convex,Mittelmann12}. We write convex cone optimization problems as:
	\begin{align}
		\label{convex-primal-cone}
		\begin{split}
			\text{minimize } &\null c^{\dagger} x \\
			\text{subject to } &\null \mathcal{A} x = b, \\
			&\null x \in K \subseteq \mathcal{X},
		\end{split}
	\end{align}
	in the primal form, cf.~\eqref{eq:L-primal}, and in the dual form, cf.~\eqref{eq:L-dual}, as:
	\begin{align}
		\label{convex-dual-cone}
		\begin{split}
			\text{maximize } &\null \bkc{y^{*}}{b} \\
			\text{subject to } &\null c^{\dagger} - \mathcal{A}^\dagger y^{*} = z^{*}, \\
			&\null z^{*} \in K^{*} \subseteq \mathcal{X}^{*},
		\end{split}
	\end{align}
	where $K$ is a nonempty, closed convex cone in an Euclidean space $\mathcal{X}$, cf.~\eqref{eq:ConvexCone}, $\mathcal{A}: \mathcal{X} \rightarrow \mathbb{R}^m$ is a linear operator, the operator $\mathcal{A}^\dagger : \left( \mathbb{R}^m \right)^{*} \rightarrow \mathcal{X}^{*}$ is its adjoint, $b \in \mathbb{R}^m$, $y^{*} \in \left( \mathbb{R}^m \right)^{*}$ and $c \in \mathcal{X}$. Note, that the spaces $\mathbb{R}^m$ and $\left( \mathbb{R}^m \right)^{*}$ are isomorphic with the transposition operation as the isomorphism.
	
	To derive~\eqref{convex-dual-cone} from~\eqref{convex-primal-cone} we need a parametrization of the family of problems~\cite[p.111-112]{L06}. \textit{One} of the possibilities is to introduce a variable $y$ used as the parameter for the linear constraints and take
	\begin{equation}
		\label{eq:FxyConic}
		F(x, y) =
		\begin{cases}
			c^{\dagger} x + I_{ \{ x : \mathcal{A} x - b = y \} }[x] & \text{ if } x \in K, \\
			+\infty & \text{ otherwise}.
		\end{cases}
	\end{equation}
	where we used the indicator function~\eqref{eq:indicatorFunction}. We stress that this is only one of multiple examples of a dualizing parametrization\index{dualizing parametrization} (note that the indicator is over a convex closed set, and thus is convex and $\lsc$): the most direct and simple, but yet arbitrary.
	To get the dual problem~\eqref{eq:L-dual} we calculate % L06p120
	\begin{equation}
		\label{eq:Fstar0yConic}
		\begin{aligned}
			-F^{*}(0, y^{*}) &= -\sup_{x \in X, y \in Y} \left\{ \bkc{y^{*}}{y} - F(x, y) \right\} = \inf_{x \in K, y \in Y} \left\{ c^{\dagger} x + I_{ \{ x : \mathcal{A} x - b = y \} }[x] - \bkc{y^{*}}{y} \right\} \\
			&= \inf_{x \in K} \left\{ c^{\dagger} x - \bkc{y^{*}}{\mathcal{A} x - b} \right\} = \inf_{x \in K} \left\{ \bkc{c^{\dagger} - \mathcal{A}^\dagger y^{*}}{x} + \bkc{y^{*}}{b} \right\}.
		\end{aligned}
	\end{equation}
	The term $\bkc{c^{\dagger} - \mathcal{A}^\dagger y^{*}}{x}$ is non-negative for all $x \in K$ if and only if $c^{\dagger} - \mathcal{A}^\dagger y^{*}$ belongs to the dual cone $K^{*}$; and if a negative value can be attained for some $x \in K$, then the infimum is $-\infty$. Thus
	\begin{equation}
		\label{eq:dualF0yConic}
		-F^{*}(0, y^{*}) =
		\begin{cases}
			\bkc{y^{*}}{b} & \text{ if } c^{\dagger} - \mathcal{A}^\dagger y^{*} \in K^{*}, \\
			-\infty & \text{ otherwise}.
		\end{cases}
	\end{equation}
	The problem~\eqref{convex-dual-cone} is derived as $\sup_{y^{*} \in Y^{*}} \{ -F^{*}(0, y^{*}) \}$, see~\eqref{eq:L-dual}, by introducing $z^{*} = c^{\dagger} - \mathcal{A}^\dagger y^{*}$.
	
	The same result can be equivalently achieved, but more step by step, with the approach using the Lagrangian~\eqref{eq:Lagrangian}, which if $x \in K$ for $F$ given by~\eqref{eq:FxyConic} is~\cite[p.266]{Boyd04}:
	\begin{equation}
		\begin{aligned}
			\mathcal{L}(x,y^{*}) &= -\sup_{y \in Y} \{ \bkc{y^{*}}{y} - F(x,y) \} = \inf_{y \in Y} \{ F(x,y) - \bkc{y^{*}}{y} \} \\
			&= \inf_{y : \mathcal{A} x - b = y, y \in Y} \{ c^{\dagger} x - \bkc{y^{*}}{y} \} = \bkc{c^{\dagger} - \mathcal{A}^\dagger y^{*}}{x} + \bkc{y^{*}}{b}
		\end{aligned}
	\end{equation}
	and $\mathcal{L}(x,y^{*}) = +\infty$ if $x \notin K$.
	The dual is derived as, \textit{cf.}~\eqref{eq:LtoDual} and~\eqref{eq:Fstar0yConic}:
	\begin{equation}
		\sup_{y^{*} \in Y^{*}} \inf_{x \in X} \mathcal{L}(x,y^{*}) = \sup_{y^{*} \in Y^{*}} \inf_{x \in K} \mathcal{L}(x,y^{*}).
	\end{equation}
	In particular, we see that the dual function~\eqref{eq:dualFunctionL} is the same as in~\eqref{eq:dualF0yConic}:
	\begin{equation}
		g(y^{*}) = \inf_{x \in X} \mathcal{L}(x,y^{*}) = \inf_{x \in K} \mathcal{L}(x,y^{*}) = -F^{*}(0, y^{*}) =
		\begin{cases}
			\bkc{y^{*}}{b} & \text{ if } c^{\dagger} - \mathcal{A}^\dagger y^{*} \in K^{*}, \\
			-\infty & \text{ otherwise}.
		\end{cases}
	\end{equation}

%TODO: policzyć to i odkomentować
%	Now, let us consider the following form of (primal) conic problems:
%	\begin{align}
%		\label{convex-primal-multicones}
%		\begin{split}
%			\text{minimize } &\null f_0(x) \\
%			\text{subject to } &\null \bigforall_{i \in [m]} \left( f_i(x) - y_i \right) \in K_i,\\
%			%&\null \bigforall_{i \in [m]} y_i \in K_i,
%		\end{split}
%	\end{align}
%	where $K_i$ are nonempty, closed convex cones.
%	Again, we choose the following dualizing parametrization\index{dualizing parametrization}:
%	\begin{equation}
%		F(x, y_1, \cdots, y_m) = f_0(x) + \sum_{i \in [m]} I_{K_i}[f_i(x) - y_i].
%	\end{equation}
%	The Lagrangian is
%	\begin{equation}
%		\begin{aligned}
%			\mathcal{L}(x, y_1^{*}, \cdots, y_m^{*}) &= \inf_{(y_1, \cdots, y_m) \in \left( \bigotimes_{i \in [m]} K_i \right)} \left\{ f_0(x) + \sum_{i \in [m]} \left( I_{K_i}[f_i(x) - y_i] - \bkc{y_i^{*}}{y_i} \right) \right\} \\
%			&= % TODO
%		\end{aligned}
%	\end{equation}
%	One can look at the problem~\eqref{convex-primal-cone} as a particular case of the problems~\eqref{convex-primal-multicones} with $m = 2$, $f_0(x) = c^{\dagger} x$, $f_1(x) = \mathcal{A} x - b$, and $f_2(x) = \Mat(x)$ with $K_1 = \{ 0 \} \subset \mathbb{R}^m$ and $K_2 = K \subseteq \mathcal{X}$.

	% TODO: \subsection{Physical and mathematical intuitions behind duality}
%	o mechanice Lagrange'a i np. \url{https://math.stackexchange.com/questions/223235/please-explain-the-intuition-behind-the-dual-problem-in-optimization}

	\section{Theory of semi-definite programming}
	\label{sec:theorySDP}

	In this section we delve into the foundational concepts and principles underlying the field of SDP. The sec.~\ref{sec:PSDdefinition} elucidates the fundamental properties and criteria for positive semi-definiteness of matrices, which form the basis for semidefinite optimization problems. In sec.~\ref{sec:formulations} we investigate various primal and dual formulations present in the literature. Next, sec.~\ref{sec:dualitySDP} explores the duality theory associated with SDP, highlighting the relationships between primal and dual problems. Finally, secs~\ref{sec:complexSDP} and~\ref{sec:slackAndEqual} cover specialized topics, shedding light on the utilization of complex variables, the incorporation of slack and surplus variables, and the treatment of mixed problems and equalities in the context of SDP.  In sec.~\ref{ssec:Schur} we discuss simple tricks related to the Schur complement. Then, in sec.~\ref{sec:IPM} we briefly discuss implementations of SDP solvers, and in sec.~\ref{sec:PredictorCorrector} we outline selected internal solver mechanisms that may impact the performance.
	
	\subsection{Definition and characterization of positive semidefiniteness}
	\label{sec:PSDdefinition}
	
	Discussion of SDP requires us to introduce the concept of \textit{positive-definite} (PD) or simply \textit{positive} matrices, as well as \textit{positive semi-definite} (PSD) or simply \textit{semi-definite} matrices. For a symmetric or Hermitian matrix $M$, we denote its positivity (semi-positivity) as $M \succ 0$ ($M \succeq 0$). Positive semi-definite matrices are also referred to as \textit{non-negative definite} or simply \textit{non-negative}. Several equivalent definitions or characterizations of such matrices can be found in the literature, and here we present three of them. Thus, a symmetric matrix $M \in \mathbb{R}^{n \times n}$ is considered positive (or non-negative) if all its eigenvalues are positive (or non-negative). Alternatively, $M$ is positive (or non-negative) if and only if for all $x \in \mathbb{R}^n$ with $x \neq 0$, it holds that $x^T M x > 0$ (or $x^T M x \geq 0$). Similarly, for a Hermitian matrix $M \in \mathbb{C}^{n \times n}$, it is PD (or PSD) if and only if for all $x \in \mathbb{C}^n$ with $x \neq 0$, we have $x^\dagger M x > 0$ (or $x^\dagger M x \geq 0$). The former definition based on eigenvalues seem to offer a greater intuitive understanding, while the latter is more prevalent in the existing literature on the subject. A more comprehensive exploration of the properties of positive-definite and positive semi-definite matrices can be found in references~\cite{matrixAnalysis,matrixAnalysis2}. It should be noted that a real PD (PSD) matrix satisfies the conditions for Hermitian matrices, making it a complex PD (PSD) matrix as well. Conversely, for a complex PD (PSD) matrix $M = M^R + i M^I$ (where $M^R$ and $M^I$ are real symmetric and antisymmetric matrices), we observe that $\forall_{\substack{x \in \mathbb{C}^n \ x \neq 0}} x^\dagger \frac{1}{2} (M + M^{\dagger}) x = x^\dagger M^R x \geq 0$, implying that the matrix $M^R$ is a real PD (PSD) matrix. It is evident that PSD matrices of size $n$ form a convex cone $\mathbb{S}^n_+$, as indicated in equation \eqref{eq:ConvexCone}, and this cone is self-dual.
	
	Now we state two very important properties characterizing PSD matrices by their possible decompositions~\cite{V13,matrixAnalysis2}. 
	It can be shown that for a Hermitian (symmetric) matrix $M \in \mathbb{C}^{n \times n}$ ($M \in \mathbb{R}^{n \times n}$) we have that $M \succeq 0$ is equivalent to each of the following statements:
	\begin{enumerate}
		\item There exists $L \in \mathbb{C}^{n \times n}$ ($L \in \mathbb{C}^{n \times n}$) such, that $M = L^{\dagger} L$ ($M = L^{T} L$), and $L$ is a lower triangular matrix.
		\item There exists a set of vectors $\{v_i\}_{i \in [n]}$, $v_{i} \in \mathbb{C}^n$ ($v_{i} \in \mathbb{R}^n$), such that $M_{i, j} = v_{i}^{\dagger} \cdot v_{j}$ ($M_{i, j} = v_{i}^{T} \cdot v_{j}$).
	\end{enumerate}
	In the first of these characterizations, a non-unique matrix $L$ is called the \textit{Cholesky decomposition} of $M$. The second characterization is equivalent to the existence of a matrix $B \in \mathbb{C}^{n \times n}$ ($B \in \mathbb{R}^{n \times n}$), such that $M = B^\dagger B$ ($M = B^T B$), so, in other words, $M$ is a multiplication table of vectors $\{v_{i}\}_i$ being columns of $B$. Some authors~\cite{Watrous11,watrous2018theory} existence of such matrix $B$ use as the definition of positive semi-definiteness. We say that $M$ is a Gram matrix\index{Gram matrix}, or a Gramian\index{Gramian}. The relation between the existence of a set of vectors and PSD property can be generalized to infinite-dimensional spaces~\cite{Mercer1909}. Trivially, the set of vectors is linearly independent if and only if the determinant of its corresponding Gram matrix is non-zero.
	
	The notion of PD and PSD is also characterized by the Sylvester criteria. Recall that the minor of a matrix is the determinant of a square submatrix obtained by removing some rows and columns of a larger matrix. If $I = J$ minor is \textit{principal}. If $I = J = \{1, \cdots k\}$ (for $k \leq n,m$) principal minor is \textit{leading}. Let $M$ be an $m \times n$ matrix, $I \subseteq \{1, \cdots, m\}$ and $J = \{1, \cdots, n\}$. Let $(M)_{I,J}$ be a submatrix with elements contained in rows from $I$ and columns from $J$. Determinant of $(M)_{I,J}$ is a minor of $M$.
	Sylvester's criteria\index{Sylvester's criteria} provide necessary and sufficient conditions for positive definiteness and semi-definiteness of a Hermitian matrix~\cite{gilbert1991positive}. %TODO: w tym cytowania tylko dla real nie Hermitian; może tu coś szukać: https://arxiv.org/pdf/0709.2458.pdf
	Sylvester's criterion for positive definiteness states that a Hermitian matrix is positive definite if and only if, all its leading principal minors are positive.
	Sylvester's criterion for positive semi-definiteness states that a Hermitian matrix is positive semi-definite if and only if, all principal minors are non-negative.
	For example an SDP constraint$\begin{bmatrix}
		1 & x \\
		x & 1
	\end{bmatrix} \succeq 0$
	implies by the second Sylvester's criterion that $\Ab{x} \leq 1$.
	
	For the L\"{o}wner's partial order $\succeq$ it can be easily shown that if $A, B \succeq 0$, then $A + B \succeq 0$. If we multiply a PSD matrix by a non-negative constant, we get another PSD matrix. Thus the set of PSD matrices forms a pointed convex cone. It also follows for $A, B \succeq 0$ that $\Tr(A B) \geq 0$ and $A^{\frac{1}{2}}$ exists and is PSD.

	\subsection{Formulations of semidefinite optimization problems}
	\label{sec:formulations}
	
	In the literature there exist a couple of equivalent formulations of SDPs, each has both primal and dual form.
	The author prefers the so-called standard or canonical form of SDP given below in~\eqref{SDP-primal} and~\eqref{SDP-dual} in sec.~\ref{sec:CanonicalSDP}, and used in many of the classical textbooks~\cite{BL06,anjos2011handbook,gartner2012approximation}, reviews~\cite{todd2001semidefinite,nemirovski2007advances}, SDP fundamental papers~\cite{M97,AHO98,TTT98,T99,Sturm99} and implementations~\cite{TTT99,Sturm02,SDPT3b,TTT12,Mittelmann12}, sometimes with slight changes in labeling~\cite{ben2011lectures}, different notations for the Frobenius product~\eqref{eq:Frob}, and more general form of conic formulations~\cite{NT98,cheng2006implementation}.
	Another important formulation is the one used by Vandenberghe and Boyd~\cite{sdp96,Boyd04}, which we provide in sec.~\ref{sec:BoydSDP}. This form seems to be preferred in many quantum information papers~\cite{jevzek2002finding,eldar2003semidefinite,brandao2004robust,NPA08} with direct influence of~\cite{sdp96}, which is apparently the default reference to the SDPs.
	The third important formulation was given by Watrous in his lecture notes~\cite{Watrous11} and textbook~\cite{watrous2018theory}, see sec.~\ref{sec:WatrousSDP} below. It has an elegant symmetric form and also is used in many quantum information books and papers~\cite{tomamichel2015quantum,cavalcanti2016quantum}, especially involving quantum channels~\cite{magesan2012characterizing,merkel2013self,piani2016robustness,kueng2016comparing,bu2017maximum}.
	
	The paradigmatic part of all the formulations are linear matrix inequalities (LMI), i.e. expressions of the form~\cite{boyd1994linear}:
	\begin{subequations}
		\begin{equation}
			\label{eq:LMI}
			F(x) \succeq 0, \text{ where}
		\end{equation}
		\begin{equation}
			\label{eq:LMexpression}
			F(x) \equiv F_0 + \sum_{i \in [m]} x_i F_i,
		\end{equation}
	\end{subequations}
	$x \in \mathbb{R}^m$ is a variable, and $F_i$, for $i = 0, \cdots, m$, are symmetric constant matrices $\mathbb{R}^{n \times n}$. The origin of LMIs is in control theory including solving Lyapunov stability problems, and their interconnection with convex optimization has been noted e.g. by Pyatnitskii and Skorodinskii~\cite{pyatnitskiy1982numerical,boyd1994linear}. In fact, SDPs can be intuitively viewed as optimization problems with linear target functions and LMIs as constraints. Any SDP can be formulated as either a primal or dual problem of the formulations given below. From the form of constraints~\eqref{eq:LMI} it directly follows that they are convex:
	\begin{equation}
		F \left( \lambda x + (1-\lambda) x' \right) = \lambda F(x) + (1-\lambda) F(x') \succ 0
	\end{equation}
	for $\lambda \in [0,1]$. The linear target function is obviously also convex. Thus the SDP problems are convex, so we can use the methods of sec.~\ref{sec:frameworkConvex}. We also note that any number of LMIs can be reformulated as a single LMI involving block-diagonal matrices, with each block referring to a relevant LMI. We refer to~\cite{chesi2010lmi} for an overview of applications of LMIs.

	\subsubsection{The canonical or standard form}~\\
	\label{sec:CanonicalSDP}
	
	Again, let $m, n \in \mathbb{N}$, $m \leq \frac{n(n+1)}{2}$. An SDP problem in a \textit{canonical}, or \textit{standard}, \textit{primal} form is the following optimization task in a variable $X \in \mathbb{S}^n$:
	\begin{align}
		\label{SDP-primal}
		\begin{split}
			\text{minimize } &\null \Frob{C}{X} \\
			\text{subject to } &\null \Frob{A_i}{X} = b_i, \text{ for } i \in [m] \\
			&\null X \succeq 0, \\
		\end{split}
	\end{align}
	where $C \in \mathbb{S}^n$ and $A_1, \cdots A_m \in \mathbb{S}^n$ are symmetric matrices. The matrices $A_i$, $C$, and vector $b \in \mathbb{R}^m$ define the SDP problem. Note that the fact that these matrices are symmetric is not restrictive. For a symmetric matrix $X$ and a matrix $C$ we have $\Frob{C}{X} = \Tr \left( \frac{1}{2}(C + C^T) X \right)$, and thus we may always take a symmetric matrix $\frac{1}{2}(C + C^T)$ instead of $C$. We assume that $A_1, \cdots A_m$ are linearly independent (otherwise we can reduce this set).
	Obviously, LP problem~\eqref{LP-primal} may be written in the form of SDP~\eqref{SDP-primal}, if $X$ is constrained to be a diagonal matrix, with the diagonal entries used as the $x$ variable. Thus, LP can be considered as a particular case of SDP. The goal expression $\Frob{C}{X} = \Tr(C X)$, but the former notation is more often used; similarly $\Frob{A_i}{X} = \Tr(A_i X)$.
	
	A canonical \textit{dual} SDP problem for \eqref{SDP-primal} is the optimization task in variables $y \in \mathbb{R}^m$ and $Z \in \mathbb{S}^n$ of the following form
	\begin{align}
		\label{SDP-dual}
		\begin{split}
			\text{maximize } &\null b^{T} \cdot y \\
			\text{subject to } &\null C - \sum_{i \in [m]} y_i A_i = Z. \\
			&\null Z \succeq 0,
		\end{split}
	\end{align}
	Some authors~\cite[p.39-40]{brown2019constructions} rewrite the canonical form~\eqref{SDP-primal} and~\eqref{SDP-dual} with substitutions $F_i$ instead of $A_i$, $-C$ instead of $C$, and $-\lambda_i$ instead of $y_i$, turning the primal problem to maximization, and the dual problem to minimization.
	
	Similarly as in LP, $X$ is called the \textit{primal variable}\index{variable!primal}, $y$ the \textit{dual variable}\index{variable!dual}, $Z$ the \textit{dual slack variable}\index{variable!dual slack}, $\{A_i\}$ are \textit{linear constraint matrices}\index{linear constraint matrix}, $b$ is the \textit{right hand side of the linear constraint}, and $C$ is the \textit{linear coefficient}.
	If $X, Z \in \mathbb{R}^{n \times n}$ and $y \in \mathbb{R}^{m}$ satisfies conditions specified by~\eqref{SDP-primal} and~\eqref{SDP-dual}, then they are called a \textit{feasible solution}. A feasible variable $X$ is called a \textit{primal solution}\index{primal solution}, and feasible variables $Z$ and $y$ constitute the \textit{dual solution}\index{dual solution}. An optimal solution is required to be feasible. The values of $\Frob{C}{X}$ and $b^{T} \cdot y$ are called the values of the primal and dual solutions, or \textit{values of the problem}\index{value of the problem}, respectively. We have $\Frob{C}{X} \geq b^{T} \cdot y$. Usually, an SDP solver is expected to find both primal and dual solutions.
	If either $C = 0 \in \mathbb{R}^{n \times n}$ or $b = 0$, then such a problem is called \textit{feasibility problem}\index{feasibility problem} and refers to finding whether \textit{any} solution of given, the primal or dual, problem exists.
	As the dual form is often delivered from the Lagrange duality~\ref{sec:LagrangianDuality}, and $y$, in that case, plays the role of Lagrange multipliers\index{Lagrange multipliers}, this name is also usually attributed to the dual variable $y$~\cite{Boyd04}.
	
	% TODO: odkomentować, gdy będzie gotowe: In~\ref{App:PrimalDualExample} we provide an explicit example of how to formulate a problem in primal and dual forms.
	The fact that primal formulation refers to minimization and dual to maximization problems, is not restrictive. We can always change the sign of the matrix $C$ or the vector $b$ to get the desired optimization problem fitting into the standard form in~\eqref{SDP-primal} and~\eqref{SDP-dual}.
	What is more, a problem formulated in one of the forms given by~\eqref{SDP-primal} and~\eqref{SDP-dual} may be reformulated in the other one. The issue of choosing the proper formulation is not always obvious and can have a very significant impact on the difficulty of the problem to a solver \cite{dualizeIt}. This can be illustrated by the example in Tab.~\ref{tab:NPAsize} showing the sizes of some SDP problems in dual and primal formulations. Generally, one should choose the formulation which leads to a smaller number of constraints, given by the number $m$ unless some special properties of the structure of the formulation can be used to further simplify the process of solving of the problem, see e.g. sec.~\ref{sec:PredictorCorrector}.

	\subsubsection{The Vandenberghe and Boyd and the SDPA forms}~\\
	\label{sec:BoydSDP}
	
	In the formulation popularized by~\cite{sdp96,Boyd04} the primal optimization task is:
	\begin{align}
		\label{SDP-primal-Boyd}
		\begin{split}
			\text{minimize } &\null c^{T} \cdot x \\
			\text{subject to } &\null F(x) \succeq 0, \\
		\end{split}
	\end{align}
	where $F(x)$ is given by~\eqref{eq:LMexpression}. As stated in~\cite{sdp96} the aim of this formulation is to make the primal formulation ``as \textit{explicit} as possible''. The dual of~\eqref{SDP-primal-Boyd} is
	\begin{align}
		\label{SDP-dual-Boyd}
		\begin{split}
			\text{maximize } &\null - \Tr \left[ F_0 Z \right] \\
			\text{subject to } &\null \Tr \left[ F_i Z \right] = c_i, i \in [m],\\
			&\null Z \succeq 0,
		\end{split}
	\end{align}
	where the variable is a symmetric matrix $Z \in \mathbb{R}^{n \times n}$.
	
	The form where $F_0$ takes an opposite sign is often referred to as the SDPA (Semidefinite Programming Algorithm), see sec.~\ref{sec:softwareAndIPM} for a discussion. Stated explicitly, keeping the original notation and naming of the variables, the SDPA primal form is~\cite{fujisawa2002sdpa}:
	\begin{align}
		\label{SDP-primal-SDPA}
		\begin{split}
			\text{minimize } &\null \sum_{i \in [m]} c_i x_i, \\
			\text{subject to } &\null X \equiv -F_0 + \sum_{i \in [m]} F_i x_i \succeq 0, \\
		\end{split}
	\end{align}
	and the SDPA dual form is:
	\begin{align}
		\label{SDP-dual-SDPA}
		\begin{split}
			\text{maximize } &\null \Tr \left[ F_0 Y \right] \\
			\text{subject to } &\null \Tr \left[ F_i Z \right] = c_i, i \in [m],\\
			&\null Z \succeq 0.
		\end{split}
	\end{align}

	\subsubsection{The Watrous symmetric form}~\\
	\label{sec:WatrousSDP}
	
	The third common form of SDPs is given by John Watrous~\cite{Watrous11,watrous2018theory}. This form is designed to show the symmetry between the primal and dual problems and is particularly convenient for quantum channel analysis.
	For two complex Euclidean spaces $\mathcal{X}$ and $\mathcal{Y}$, a semidefinite program in the Watrous form is defined as a triple $(\Phi, A, B)$, where $\Phi: \mathsf{L}\left[ \mathcal{X}, \mathcal{X} \right] \rightarrow \mathsf{L}\left[ \mathcal{Y}, \mathcal{Y} \right]$ is a Hermitian and trace-preserving map, $A \in \Herm{\mathcal{X}}$, and $B \in \Herm(\mathcal{Y})$.
	The primal problem in the Watrous form is:
	\begin{align}
		\label{SDP-primal-Watrous}
		\begin{split}
			\text{maximize } &\null \bkc{A}{X} \\
			\text{subject to } &\null \Phi(X) = Y,\\
			&\null X \succeq 0,
		\end{split}
	\end{align}
	and the dual is:
	\begin{align}
		\label{SDP-dual-Watrous}
		\begin{split}
			\text{minimize } &\null \bkc{B}{Y} \\
			\text{subject to } &\null \Phi^\dagger(Y) \succeq A,\\
			&\null Y \in \Herm(\mathcal{Y}),
		\end{split}
	\end{align}
	with a remark that in the orignal notation of~\cite{Watrous11,watrous2018theory} Watrous uses $^*$ instead of $^\dagger$ to denote the Hermitian conjugate, $\geq$ instead of $\succeq$ to denote the L\"{o}wner's partial order, and $X \in \text{Pos}(\mathcal{X})$ instead of $X \succeq 0$.

	\subsubsection{The Kronecker-canonical form for convex cones}~\\
	\label{sec:KroneckerSDP}
	
	Here we briefly show the methodology that offers an alternative way of expressing the canonical formulation of~\eqref{SDP-primal} and~\eqref{SDP-dual}, resembling the LP formulations in~\eqref{LP-primal} and~\eqref{LP-dual}. This formalism, in fact, is more encompassing and applicable to a wide range of conic optimization problems, as discussed in sec.~\ref{sec:convexCone}. The general formulation~\eqref{convex-primal-cone} and~\eqref{convex-dual-cone} is highly convenient and valuable, as it facilitates seamless transition between primal and dual formulations for any convex cone by establishing the corresponding dual cone.
	
	It can be easily verified that for any real matrices $A$, $B$, and $C$, the relationship $\left( A \otimes B \right) \Vector(C) = \Vector{\left( B C A^T \right)}$ holds. We define a matrix $\mathcal{A} \equiv [a_1; \dots; a_m] \in \mathbb{R}^{n^2 \times m}$, where $a_i = \Vector(A_i)$, referring to the matrices in~\eqref{SDP-primal}. Hence, $a_i$ represents the $i$-th column of $\mathcal{A}$. Consequently, we have $\Vector \left( \sum_{i \in [m]} y_i A_i \right) = \mathcal{A} y$ and $\Frob{A_i}{X} = (\mathcal{A}^T x)_i$, where $(\mathcal{A}^T x)_i$ denotes the $i$-th element of the vector $\mathcal{A}^T x$, $c = \Vector(C)$, and $x = \Vector(X)$. When $K$ represents the self-dual cone of real or convex PSD $n$ by $n$ matrices, substituting these expressions into~\eqref{convex-primal-cone} and~\eqref{convex-dual-cone} yields an alternative and equivalent formulation for the canonical SDP, commonly used, for instance, in~\cite{NT98,SeDuMi,TTT99,Mittelmann12}, as illustrated in~\eqref{eq:primalFeas} and~\eqref{eq:dualFeas}.

	\subsection{Duality of semi-definite programming}
	\label{sec:dualitySDP}
	
	% zobaczyć Gärtner-Matousek2012_Book_ApproximationAlgorithmsAndSemi.pdf rozdział 4
	
	% Weak duality proof: Tr(XZ) = 0
	Recall that an important property of primal and dual formulations is that any feasible solution to a primal problem provides an upper bound on all feasible solutions to the dual problem. The \textit{weak duality} property of SDP, \textit{viz.} $\Frob{C}{X} \geq b^T \cdot y$ is derived as follows:
	\begin{equation}
		\begin{aligned}
			\Frob{C}{X} &- b^T \cdot y = \Frob{\left( Z + \sum_{i \in [m]} y_i A_i \right)}{X} - b^T \cdot y \\
			&= \Tr (ZX) + \sum_{i \in [m]} y_i \cdot \Tr (A_i X) - b^T \cdot y = \Tr(X Z) \geq 0.
		\end{aligned}
	\end{equation}
	In LP, value of the primal and dual problems are always equalmeaning the strong duality\index{duality!strong}. Now, we provide a sufficient condition for strong duality to occur in SDP, as it is observed in many cases. Let $p^{*}$ be the optimal value of the primal SDP problem, and $d^{*}$ be the optimal value of the dual SDP problem. One can show that it holds $p^{*} = d^{*}$ if at least one of the conditions is satisfied~\cite{A95,B00}:
	\begin{enumerate}
		\item There exist $y \in \mathbb{R}^{m}$, such that $C - \sum_{i \in [m]} y_i A_i \succ 0$, i.e. the dual problem is strictly feasible (then also the value $d^{*}$ is attained).
		\item There exists $X \succ 0$, such that $\Frob{A_i}{X} = b_i$, i.e. the primal problem is strictly feasible (then also the value $p^{*}$ is attained).
	\end{enumerate}
	These statements are called the Slater conditions~\cite{slater2013lagrange}\index{Slater conditions}.
	
	One of the most confusing and intriguing questions in the theory of SDP is asking whether the dual of the dual form is the primal form. The affirmative answer can be derived in the following way:
	\begin{equation}
		\begin{aligned}
			\sup_y & \left\{ b^T \cdot y : C - \sum_i (A_i y_i) \succeq 0 \right\} = \sup_y \inf_{X \succeq 0} \left\{ b^T \cdot y + \Frob{\left( C - \sum_i (A_i y_i) \right)}{X} \right\} \leq \\
			& \leq 
			%\inf_{X \succeq 0} \sup_y \left\{ b^T \cdot y + \Frob{\left( C - \sum_i (A_i y_i) \right)}{X} \right\} = 
			\inf_{X \succeq 0} \sup_y \left\{ \Frob{C}{X} + \sum_i y_i \cdot (b_i - \Frob{A_i}{X}) \right\} = \inf_{\substack{X \succeq 0, \\ \Frob{A_i}{X} = b_i}} \left\{ \Frob{C}{X} \right\}.
		\end{aligned}
	\end{equation}
	Indeed:
	\begin{subequations}
		\begin{equation}
			\inf_{X \succeq 0} \left\{ b^T \cdot y + \Frob{X}{Z} \right\} = \begin{cases} b^T \cdot y & \text{if } Z \succeq 0 \\ -\infty & \text{otherwise} \end{cases}, \text{ and}
		\end{equation}
		\begin{equation}
			\sup_{y} \left\{ \Frob{C}{X} + \sum_i y_i \cdot (b_i - \Frob{A_i}{X}) \right\} = \begin{cases} \Frob{C}{X} & \text{if } \bigforall_{i} \Frob{A_i}{X} = b_i \\ +\infty & \text{otherwise} \end{cases}.
		\end{equation}
	\end{subequations}
	%TODO: podać ogólnie jakie są warunki dual dual = primal (czy już są gdzieś podane?)

% TODO: zrobić coś z tym poniżej
%	%\subsection{Farkas' lemma}
%	
%	%według https://en.wikipedia.org/wiki/Farkas%27_lemma
%	wspomnieć o Garg, Anupam; Mermin, N. D. (1984), "Farkas's Lemma and the Nature of Reality: Statistical Implications of Quantum Correlations", Foundations of Physics, 14: 1–39, %doi:10.1007/BF00741645
%	
%	%\subsection{Slater's conditons}
%	% https://en.wikipedia.org/wiki/Slater%27s_condition
%	\cite[p.240,248-250]{Boyd04}
%	978-3-0348-0439-4 Traces and Emergence of Nonlinear Programming.pdf page 293nn

	% TODO: \subsection{Lagrange multipliers and Karush-Kuhn-Tucker conditions}
	%Rockafellar_LagrangeMultAndOptimality.pdf

	% TODO: \subsection{Properties of semi-definite matrices}
	%TODO: inne, np. Gram matrix, S-lemma (S-procedure na wikipedii), iloczyn Frobeniusa (i ewentualnie inne), LMI
	% https://books.google.pl/books?id=AllToPtYGGgC&lpg=PA62&dq=The%20rank%20of%20the%20Gram%20matrix%20of%20vectors%20in&hl=pl&pg=PA62#v=onepage&q=The%20rank%20of%20the%20Gram%20matrix%20of%20vectors%20in&f=false
	% https://en.wikipedia.org/wiki/Gramian_matrix
	% do Gram matrix, ciekawe jest to z 2012.00554 strona 5-6
	% czy może to zadziała w SDP nie tylko w LP? https://en.wikipedia.org/wiki/Linear-fractional_programming

	\subsection{Complex variables in semidefinite problems}
	\label{sec:complexSDP}
	
	In sec.~\ref{sec:CanonicalSDP} we considered SDPs where the problems were defined by real matrices and vectors, and the optimization was carried over real-valued variables. One of the first works showing how to reduce a problem where some of the elements of the problem involve complex numbers was~\cite{goemans2001approximation}.
	
	Let $B \in \mathbb{C}^{n \times n}$ be a Hermitian matrix, $B^R$ and $B^I$ its real and imaginary parts, respectively, i.e. $B = B^R + i B^I$ and $B^I = -B^I$. Then $B \succeq 0$ if and only if
	\begin{equation}
		\label{eq:HermitianToSymmetric}
		\begin{bmatrix} B^R & -B^I \\ B^I & B^R \end{bmatrix} \succeq 0.
	\end{equation}
	Indeed, for any complex vector $w = u + i v \in \mathbb{C}^n$ we have
	\begin{equation}
		\begin{aligned}
			w^{\dagger} B w &= \left( u^T - i v^T \right) \left( B^R + i B^I \right) (u + i v) = \left( u^T B^R u + v^T B^R v - u^T B^I v + v^T B^I u \right) + \\
			& + i (u^T B^R v - v^T B^R u + u^T B^I u + v^T B^I v) \geq 0
		\end{aligned}
	\end{equation}
	if and only if $\begin{bmatrix} u^T & v^T \end{bmatrix} \begin{bmatrix} B^R & -B^I \\ B^I & B^R \end{bmatrix} \begin{bmatrix} u \\ v \end{bmatrix} \succeq 0$. This is because $u^T B^I u = v^T B^I v = 0$ and $u^T B^R v = v^T B^R u$. Thus any SDP problem defined in terms of complex vectors and Hermitian matrices can be stated as a problem involving only real vectors with symmetric matrices. For instance, let us consider the case when both the linear coefficient $C$ and the primal variable $X$ are complex matrices, $C = C^R + i C^I$, with $C^R, C^I \in \mathbb{R}^{n \times n}$ and $X \in \mathbb{C}^{n \times n}$. When we reframe this as a real-valued SDP, then the target function takes the form of $\begin{bmatrix} C^R & -C^I \\ C^I & C^R \end{bmatrix}$, cf.~\eqref{eq:HermitianToSymmetric}, and the primal real-valued variable $X^{(R)} = \begin{bmatrix} X_{11} & X_{12} \\ X_{21} & X_{22} \end{bmatrix}$ does not require explicit constraints. Instead, the resulting complex variable $X$ is retrieved as $X \equiv 2 X_{11} + i \left( X_{12} - X_{12}^T \right)$. % TODO: dowód, że to nie ogranicza ogólności
%% from solvesdp.m in YALMIP
%n = size(complexInfo.replaced{i},1);
%re = 2*double(complexInfo.new{i}(1:n,1:n));            
%im = 2*double(complexInfo.new{i}(1:n,n+1:end));
%im=triu((im-im')/2)-(triu((im-im')/2))';
%assign(complexInfo.replaced{i},re + sqrt(-1)*im);

	We now briefly discuss the formulation of complex SDP problems for which the target is given by real linear coefficient $C$.
	Consider $C \in \mathbb{S}^n$, and $X \in \mathbb{H}^n$. Let $X = X^R + i X^I$, where $X^R \in \mathbb{S}^n$ and $X^I \in \mathbb{R}^{n \times n}$. We have:
	\begin{equation}
		\Tr (C X) = \Tr (C X^R) + i \Tr(C X^I) = \Tr (C X^R).
	\end{equation}
	Since $X^I$ is antisymmetric, the Frobenius product of symmetric and antisymmetric matrix is always equal to $0$.
	Thus if $C$ is real and we are interested only in finding the value of the solution, then we can ignore the imaginary part occurring in the problem.

	\subsection{Slack and surplus variables, mixed problems and equalities}
	\label{sec:slackAndEqual}
	
	Slackness and complementary slackness are both concepts used in optimization theory, particularly in convex optimization. Slackness refers to the idea that in an optimal solution, some of the inequality constraints are satisfied with equality, i.e. there is no \textit{slack} or excess capacity in the system. The extent to which they diverge from the equality can be expressed as a new PSD variable, which then can be introduced to convert inequality constraints to equality constraints, as elucidated below. The concept of slack variables is commonly used in LP and SDP. Recall that in LP and SDP formulations, the objective function is optimized subject to a set of constraints, where the constraints can be in the form of equalities or inequalities. In the case of linear constraints of the form $A x \leq b$, where $A$ is an $m \times n$ matrix and $b$ is a column vector of length $m$, introducing a slack variable $\tilde{x}_i$ for each constraint $i$ allows us to convert the inequality constraint into an equality constraint. The idea is to add a non-negative variable $\tilde{x}_i$ to the left-hand side of the $i$-th constraint so that the resulting expression becomes equality. Specifically, if the $i$-th constraint is:
	\begin{equation}
		a_{i1} x_1 + \cdots a_{in} x_n \leq b_i
	\end{equation}
	then we can add a slack variable $\tilde{x}_i \geq 0$ to obtain an expression that is equivalent to the previous constraint:
	\begin{equation}
		a_{i1} x_1 + \cdots a_{in} x_n + \tilde{x}_i = b_i.
	\end{equation}
	The new variable $\tilde{x}_i$ is called a slack variable because it measures the amount by which the left-hand side of the $i$-th constraint falls short of the right-hand side $b_i$. If the left-hand side is already equal to $b_i$, then $\tilde{x}_i$ is zero.
	
	On the other hand, if we have inequality constraints of the form $A x \geq b$, then we introduce a surplus variable $\tilde{x}_i$ that measures the amount by which the left-hand side of the $i$-th constraint exceeds the right-hand side $b_i$. Specifically, we add a non-negative variable $\tilde{x}_i$ to the left-hand side of the $i$-th constraint, so that the resulting expression becomes an equality:
	\begin{equation}
		a_{i1} x_1 + \cdots a_{in} x_n - \tilde{x}_i = b_i.
	\end{equation}
	Therefore, we can see that the use of slack and surplus variables allows us to convert any inequality constraint into an equality constraint, which makes it easier to fit to the canonical form.
	
	% TODO: complementary slackness
	On the contrary, complementary slackness\index{complementary slackness}, refers to the idea that for an optimal solution, the primal variables and the corresponding dual variables are \textit{complementary}, in the sense that their product is zero. In other words, slackness is a condition that holds between the primal variables (e.g., the decision variables in a LP) and the primal constraints, while complementary slackness is a condition that holds between the primal solution and the dual solution in a convex optimization problem. For instance, when the strong duality of SDP holds, we have $p^{*} = d^{*}$, i.e. $\Frob{C}{X} = b^{T} \cdot y = \Frob{A_i}{X} \cdot y$. Thus $0 = \Frob{\left( C - \sum_{i \in [m]} y_i A_i \right)}{X} = \Tr \left[ Z X \right]$. This means that if the solutions of primal and dual SDP problems exist, then strong duality is equivalent to complementary slackness.
	Summing up, whereas slackness conditions tell us which constraints are active in the optimal solution (i.e., satisfied with equality), the complementary slackness conditions tell us which constraints are binding (i.e., have nonzero dual variables) and which are nonbinding (i.e., have zero dual variables). Both conditions are important for understanding and characterizing optimal solutions in optimization problems.
	
	One often considers the so-called mixed LP-SDP cone. The primal mixed problem in variables $(x_L,X_S) \in \mathbb{R}^{n_L} \times \mathbb{S}^{n_S \times n_S}$ is of the following form:
	\begin{align}
		\label{eq:SDP-primal-lpVariables}
		\begin{split}
			\text{minimize } &\null c_L^{T} \cdot x_L + \Frob{C_S}{X_S} \\
			\text{subject to } &\null (A_L^T)_{i,:} \cdot x_L + \Frob{{A_S}_i}{X_S} = b_i, \text{ for } i \in [m] \\
			&\null x_L \geq 0, X_S \succeq 0.
		\end{split}
	\end{align}
	where $A_L \in \mathbb{R}^{n_L \times m}$, $b \in \mathbb{R}^m$, $x_L, c_L \in \mathbb{R}^{n_L}$, $C_S \in \mathbb{S}^{n_S \times n_S}$ and ${A_S}_1, \cdots {A_S}_m \in \mathbb{S}^{n_S \times n_S}$.
	The dual mixed problem in variables $(y,z_L,Z_S) \in \mathbb{R}^m \times \mathbb{R}^{n_L} \times \mathbb{S}^{n_S \times n_S}$ is given by the formula:
	\begin{align}
		\label{eq:SDP-dual-lpVariables}
		\begin{split}
			\text{maximize } &\null b^{T} \cdot y \\
			\text{subject to } &\null c_L - A_L y = z_L, C_S - \sum_{i \in [m]} y_i {A_S}_i = Z_S, \\
			&\null z_L \geq 0, Z_S \succeq 0.
		\end{split}
	\end{align}
	We note that as any LP can be reformulated as SDP, the mixed problems are not more general than the SDP problems. One can see that to embed an LP in SDP it is sufficient to place the $n_L$ linear variables on the diagonal of a new SDP variable of size $n_L + n_S$, where $n_S$ is the size of the SDP original variable. On the other hand, mixed problems are useful for efficient solver implementations, as the numerical methods needed to solve SDP are more expensive in terms of computational effort than LP. Thus, when stating the problem in the mixed form, then one may obtain a reduction of the computational cost of the solver.
	% TODO: wspomnieć o conic formulation, jak w Mittelman12
	
	We now discuss, how equality constraints are expressed in the canonical form of SDP. Recall that in the primal canonical SDP~\eqref{SDP-primal}, equalities have the form $\Frob{A_i}{X} = b_i$ for $i \in [m]$, where $m$ is one of the two parameters describing the size of the problem. Thus, to add a linear constraint on variables within $X$ we add a new matrix $A_i$ \textit{increasing} size of the problem to $m+1$.
	On the other hand, if we add a linear constraint in the dual form~\eqref{SDP-dual}, we can do one of the following.
	The first possibility is to eliminate one of the variables $y_i$, e.g. with LU or QR decomposition, and thus \textit{reduce} size of the problem to $m-1$. This simplifies the SDP but requires an additional effort of variable elimination, which itself requires some computational resources and is not always implemented. For instance, the elimination is performed in YALMIP when the user passes an option \Minline{'removeequalities'} to the model, as discussed in~\ref{sec:matlab:1stYALMIP}.
	The second possibility is to reframe the optimization problem over $(x_F,X_S) \in \mathbb{R}^{n_F} \times \mathbb{S}^{n_S \times n_S}$ for some $n_F, n_S \in \mathbb{N}$ in a form different than the canonical form, cf.~\eqref{eq:SDP-primal-lpVariables}, \textit{viz.}
	\begin{align}
		\label{eq:SDP-primal-freeVariables}
		\begin{split}
			\text{minimize } &\null c_F^{T} \cdot x_F + \Frob{C_S}{X_S} \\
			\text{subject to } &\null (A_F^T)_{i,:} \cdot x_F + \Frob{{A_S}_i}{X_S} = b_i, \text{ for } i \in [m] \\
			&\null X_S \succeq 0.
		\end{split}
	\end{align}
	where $A_F \in \mathbb{R}^{n_F \times m}$, $b \in \mathbb{R}^m$, $x_F, c_F \in \mathbb{R}^{n_F}$, $C_S \in \mathbb{S}^{n_S \times n_S}$ and ${A_S}_1, \cdots {A_S}_m \in \mathbb{S}^{n_S \times n_S}$, and there is no constraint on $x_F$. The dual problem of~\eqref{eq:SDP-primal-freeVariables} in variables $(y,z_F,Z_S) \in \mathbb{R}^m \times \{0\}^{n_F} \times \mathbb{S}^{n_S \times n_S}$ is of the following form:
	\begin{align}
		\label{eq:SDP-dual-freeVariables}
		\begin{split}
			\text{maximize } &\null b^{T} \cdot y \\
			\text{subject to } &\null c_F - A_F y = z_F, \\
			&\null C_S - \sum_{i \in [m]} y_i {A_S}_i = Z_S, Z_S \succeq 0. \\
		\end{split}
	\end{align}
	The forms~\eqref{eq:SDP-primal-freeVariables} and~\eqref{eq:SDP-dual-freeVariables} are sometimes called the standard
	form with free variables~\cite{hansson2014sampling}. Note that the trivial cone $\{0\}^{n_F}$ appearing in~\eqref{eq:SDP-dual-freeVariables} is the dual cone of $\mathbb{R}^{n_F}$ appearing in~\eqref{eq:SDP-primal-freeVariables}. Since this implies $z_F = 0$, the first condition in~\eqref{eq:SDP-dual-freeVariables} is equivalent to $A_F y = c_F$, providing a way to express equality constraints on the dual variable $y$. This possibility of treating the equality constraint requires the solver to be able to deal with free variables in the primal problem. This is beyond the capabilities of the usual interior point methods as discussed in sec.~\ref{sec:IPM}, see also~\cite{meszaros1998free}.
	The standard way of dealing with free variables, implemented in SeDuMi and SDPT3, is to represent a vector $x_F \in \mathbb{R}^{n_F}$ as a difference of two vectors $x_{(+)}, x_{(-)} \in \mathbb{R}^{n_F}_+$ as $x_F = x_{(+)} - x_{(-)}$. This guides us to the third possibility of representing equality constrain $(A_F)_{i,:} \cdot y = {c_F}_i$, as two inequalities:
	\begin{equation}
		\begin{aligned}
			& (A_F)_{i,:} \cdot y \geq {c_F}_i - \epsilon, \\
			& (A_F)_{i,:} \cdot y \leq {c_F}_i + \epsilon,
		\end{aligned}
	\end{equation}
	for some small $\epsilon$. In consequence, equalities even in dual form increase complexity. For instance, YALMIP allows the user to specify, how to treat the explicitly stated equality constraints with the mentioned option \Minline{'removeequalities'}, or, in case the chosen solver requires it, YALMIP does it automatically with the function \Minline{solveequalities}. A detailed disussion of conversion the problems~\eqref{eq:SDP-primal-freeVariables} and~\eqref{eq:SDP-dual-freeVariables} to the canonical form~\eqref{SDP-primal} and~\eqref{SDP-dual} is provided in~\cite{kobayashi2007conversion}.

	\subsection{Schur complement and submatrices}
	\label{ssec:Schur}
	
	% Definition of Schur complement
	% https://www.quora.com/Linear-Algebra-What-is-the-motivation-behind-defining-the-Schur-complement-of-a-matrix
	Consider a partition of a square matrix $M \in \mathbb{C}^{(n_1 + n_2) \times (n_1 + n_2)}$ to submatrices, \textit{viz.} $	M = \begin{bmatrix} A & B \\ C & D \end{bmatrix}$, where $A \in \mathbb{C}^{n_1 \times n_1}$ and $D \in \mathbb{C}^{n_2 \times n_2}$ are square matrices. Recall that the principal submatrix of a square matrix\index{principal submatrix} is the special case of a submatrix where the same rows and columns are removed. Thus $A$ and $D$ are principal submatrices. It is easy to see that a principal submatrix of a PSD matrix is also PSD. For instance consider an arbitrary vector $v \in \mathbb{C}^{n_1 + n_2}$ with non-zero entries only in the first $n_1$ positions, and a vector $v' \in \mathbb{C}^{n_1}$ with entries equal to the first $n_1$ entries of $v$. Obviously, since $v^\dagger M v \geq 0$ it also holds that $v'^\dagger A v' \geq 0$.
	
	Assume the submatrix $D$ to be invertible. For numerical implementations, it would also be desirable for $D$ to be well-conditioned to give accurate results under finite-precision arithmetic upon inverting it. The Schur complement of block $D$ is given by $A - B D^{-1} C$ and denoted $M / D$. Similarly, $M / A \equiv D - C A^{-1} B$~\cite{schur1917potenzreihen,jbilou2004some,Boyd04,gallier2020schur}.
	% Properties of Schur complement
	% https://en.wikipedia.org/wiki/Schur_complement
	We have $M \succeq 0 \implies M / D \succeq 0$; and for symmetric $M$ (i.e. $B = C^T$): %https://en.wikipedia.org/wiki/Schur_complement#Schur_complement_condition_for_positive_definiteness_and_positive_semi-definiteness
	\begin{subequations}
		\begin{equation}
			M \succ 0 \iff A, M / A \succ 0, \text{ and}
		\end{equation}
		\begin{equation}
			\label{eq:SchurIff2}
			A \succ 0 \implies \left[ M \succeq 0 \iff M / A \succeq 0 \right].
		\end{equation}
	\end{subequations}
	Obviously the same holds for $D$.
	% Schur complement for solving linear equations
	The Schur complement method~\cite{Schur06} allows reducing a large system of equations to a smaller one, involving only a subset of variables, e.g. $M \begin{bmatrix} x_1 \\ x_2 \end{bmatrix} = \begin{bmatrix} y_1 \\ y_2 \end{bmatrix}$, by solving two simpler equations, namely $(A - B D^{-1} C) x_1 = y_1 - B D^{-1} y_2$, and then $C x_1 + D x_2 = y_2$. The reason why this method is useful is that one usually needs $O(n^3)$ operations to solve linear equations with $n$ variables, and thus it is profitable to decompose the initial equation into two smaller equations. % TODO: zrobić to: This we will use later, in implementation discussion.
	
	% Schur trick
	Schur complement is also a tool for introducing the following useful trick\index{Schur trick}. Consider an LMI:
	\begin{equation}
		\begin{bmatrix}
			t & c(x) \\
			c^T(x) & F(x)
		\end{bmatrix} \succeq 0,
	\end{equation}
	where $x \in \mathbb{R}^k$ for some $k$, $F(x) \in \mathbb{R}^{m \times m}$ is a linear matrix expression of the form~\eqref{eq:LMexpression} for some $m$, $c: \mathbb{R}^k \rightarrow \mathbb{R}^m$ is a linear function, and $t$ is a positive number. From~\eqref{eq:SchurIff2} it follows
	\begin{equation}
		\label{eq:SchurTrick}
		t \geq c(x) \cdot F^{-1}(x) \cdot c^T(x).
	\end{equation}
	The expression~\eqref{eq:SchurTrick} allows for expression fairy generic non-linear constraints as LMIs, e.g. taking $k = 2$, $m = 1$ with $c(x) = x_1$ and $F(x) = x_2$ we get the constraint $t \geq \frac{x_1^2}{x_2}$.
	Similarly, from Schur lemma it follows that for $A, B, R \in \mathbb{H}^n$:
	\begin{equation} % https://www.sciencedirect.com/science/article/pii/S0024379516304852 App.2.1
		\label{eq:SchurMGMean}
		\begin{bmatrix}
			A & R \\ R & B
		\end{bmatrix} \succeq 0
		\Leftrightarrow
		B \succeq R A^{-1} R
		\Leftrightarrow
		[A^{-1/2} B A^{-1/2}]^{1/2} \succeq A^{-1/2} R A^{-1/2}
		\Leftrightarrow
		A \# B - R \succeq 0,
	\end{equation}
	where the $1/2$-weighted matrix geometric mean is given by~\eqref{eq:wGMean}.
	
	% TODO: odkomentować o nullity, rozbudować i podać jakieś zastosowanie - lub usunąć
%	A convenient tool for formulating such for of LMIs is the nullity theorem. Recall that the nullity\index{nullity} of a linear operator (matrix) is the dimension of its kernel.
%	The nullity theorem~\cite{gustafson1984note,fiedler1986completing} states, in particular, that the nullity of a block in a matrix equals the nullity of the complementary block in its inverse matrix.
%	Consider a partition to submatrices of a matrix and its inverse, e.g.:
%	\begin{equation}
%		\begin{bmatrix}
%			A & B \\
%			C & D
%		\end{bmatrix}^{-1}
%		=
%		\begin{bmatrix}
%			X & Y\\
%			W & Z
%		\end{bmatrix},
%	\end{equation}
%	where the shape of the one partition is the transpose of the shape of the other partition. The theorem states that the nullity of $A$ equals the nullity of $Z$ and that the nullity of $D$ equals the nullity of $X$.

	% TODO: \subsection{Useful lemmas}
	%np. wg. https://ocw.mit.edu/courses/sloan-school-of-management/15-094j-systems-optimization-models-and-computation-sma-5223-spring-2004/lecture-notes/sdp094_digest.pdf
	
		\subsection{How does a solver use Interior Point Methods?} % jak działa IPM (Newton step, predictor-corrector, search directions)
	\label{sec:IPM}
	
	%  implementations/1-s2.0-S0377042700004337-main.pdf wygląda na fajny przegląd, łącznie ze złożonością obliczeniową (stąd https://www.sciencedirect.com/science/article/pii/S0377042700004337)
	
	Even though the topic of implementation of IPM is not specific to quantum information, from our experience, it is useful to have at least a general understanding of how actually a solver is deriving its results. This helps to identify potential problems, estimate the difficulties, and interpret the outputs of the solver. An important concept in path-following IPMs is the \textit{central path}\index{central path}, which consists of a set of feasible solutions $(X,y,Z)$ parameterized by a non-negative variable $\nu$~\cite{monteiro1989interiorI,monteiro1989interiorII,kojima1989primal,zhang1992superlinear,kojima1993primal,goldfarb1998interior,potra2000interior}. The central path is defined by the following conditions:
	\begin{subequations}
		\label{eq:centralPathCanonical}
		\begin{equation}
			X, Z \succeq 0 \quad \text{(PSD matrices)},
		\end{equation}
		\begin{equation}
			\Frob{A_i}{X} = b_i \quad \text{for } i \in [m] \quad \text{(linear constraints)},
		\end{equation}
		\begin{equation}
			C - \sum_{i \in [m]} y_i A_i = Z \quad \text{(dual feasibility)},
		\end{equation}
		\begin{equation}
			X Z = \nu \openone \quad \text{(approx. complementary slackness)},
		\end{equation}
	\end{subequations}
	where $\{A_i\}$, $b_i$, and $C$ are problem-specific matrices for the canonical formulation described in sec.~\ref{sec:CanonicalSDP}. The central path captures a family of solutions that gradually approaches the optimal solution as $\nu$ increases. By following this path, usually employing the concept of the Newton steps, IPMs efficiently navigate the feasible region of the SDP problem toward the optimal solution. Equivalently~\eqref{eq:centralPathCanonical} can be written in Kronecker-canonical form, see sec.~\ref{sec:KroneckerSDP}:
	\begin{subequations}
		\label{eq:optimalConds}
		\begin{equation}
			\begin{aligned}
				\label{eq:primalFeas}
				\mathcal{A}^T x = b \\
			\end{aligned}
		\end{equation}
		\begin{equation}
			\begin{aligned}
				\label{eq:dualFeas}
				c - \mathcal{A} y = z, \\
			\end{aligned}
		\end{equation}
		\begin{equation}
			\begin{aligned}
				\label{eq:centralPath}
				X Z = \nu \openone,
			\end{aligned}
		\end{equation}
	\end{subequations}
	together with $X, Z \succeq 0$. Further the expression $\Tr(Z X)$ is referred to as the \textit{gap}\index{duality gap}. Note that strong duality of an SDP problem implies the complementary slackness, see sec.~\ref{sec:slackAndEqual}, stating that the optimal primal and dual variables are orthogonal, \textit{i.e.}
	\begin{equation}
		\label{eq:gapTrZX}
		\Tr(Z X) = 0,
	\end{equation}
	meaning that the gap is equal to $0$. On the other hand, it should be stressed that the gap~\eqref{eq:gapTrZX} and the duality gap are closely related but different terms. The former is defined in terms of the primal and dual variables $X$ and $Z$, even if they do not provide a feasible solution, i.e. even if they do not satisfy the conditions in~\eqref{SDP-primal} and~\eqref{SDP-dual}. The latter is defined as~\eqref{eq:dualityGap} and expresses the difference between the optimal primal and dual solutions of the problem. We note here that practical implementations of SDP solvers usually find solutions that are not feasible in a strict sense. Instead, the solutions satisfy the condition from~\eqref{SDP-primal} and~\eqref{SDP-dual} only with some accuracy. Here we discuss the expressions we use further in this work to evaluate primal and dual infeasibility. See \cite{Mittelmann12} for more details on the issue of infeasibility norms. Let $c \equiv \Vector(C)$, $x \equiv \Vector(X)$ and $z \equiv \Vector(Z)$, as in the Kronecker-canonical form, see sec.~\ref{sec:KroneckerSDP}. Let us define the following terms, \textit{viz.} the residuals for feasibility conditions in~\eqref{SDP-primal} and~\eqref{SDP-dual}, \textit{cf.}~\eqref{eq:primalFeas} and~\eqref{eq:dualFeas}:
	\begin{subequations}
		\begin{equation}
			\label{eq:rp}
			r_p \equiv b - \mathcal{A}^T x \in \mathbb{R}^m,
		\end{equation}
		\begin{equation}
			\label{eq:rd}
			r_d \equiv c - \mathcal{A} y - z \in \mathbb{R}^{n^2}.
		\end{equation}
	\end{subequations}
	Using the above conditions~\eqref{eq:optimalConds} we get that the Newton step $(\Delta X, \Delta y, \Delta Z)$ is supposed to satisfy the following formulae:
	\begin{equation}
		\label{eq:rprd}
		\begin{aligned}
			\mathcal{A}^T \Delta x = r_p, \\
			\mathcal{A} \Delta y + \Delta z = r_d.
		\end{aligned}
	\end{equation}
	If the method assumes that both $r_p$ and $r_d$ are zero, $r_p=r_d=0$, it is referred to as a feasible IPM\index{interior point methods!feasible}. Otherwise, it is considered an infeasible IPM\index{interior point methods!infeasible}. The conditions~\eqref{eq:rprd} are supposed to iteratively bring the variables $(X,y,Z)$ closer to the feasibility constraints, \eqref{eq:primalFeas} and~\eqref{eq:dualFeas}. On the other hand, note that those two equations do not determine fully the solution; yet we still need to consider~\eqref{eq:centralPath} somehow. At the same time, from the strong duality, we get that the optimal solution, $(X^{*}, y^{*}, Z^{*})$ has the property that it is on the central path at the point with $\nu = 0$, i.e. the gap\index{duality gap} is $0$, see sec.~\ref{sec:dualitySDP}. Thus, the Newton step should not only ensure feasibility but also reduce the gap between primal and dual solutions. Actually, the condition imposed on the Newton step from the condition~\eqref{eq:centralPath} is most problematic for another reason. The problem arises from the fact that the matrices $X^{(i)}$ and $Z^{(i)}$, where $i$ stands for the current iteration, possibly do not commute. For this reason, the following form of conditions on the target of the Newton step is imposed:
	\begin{equation}
		\label{eq:NewtonSymmetrization}
		\Theta_{\nu}(X, Z) = \nu \openone \in \mathbb{R}^{n \times n},
	\end{equation}
	where $\Theta_{\nu}(X, Z)$ is a symmetrization of $X Z$; we discuss a couple of symmetrizations below. The algorithm stops when residual norms and the gap\index{duality gap} $\Ab{\Tr(X Z)}$, are all less than the specified threshold. Examples of the primal\index{primal residual norm} and dual residual\index{dual residual norm} norms used e.g. in~\cite{myThesis} are $\frac{1}{1 + |b|_F} \Ab{b - \mathcal{A}^T x}_F$, and $\frac{1}{1 + |c|_F} \Ab{c - \mathcal{A} y - z}_F$, respectively, where we recognize normalized norms of the expressions~\eqref{eq:rp} and~\eqref{eq:rd}.
	
	The problem of efficiently solving the Newton system~\eqref{eq:optimalConds} numerically is discussed in detail in Chapter 5 of~\cite{myThesis}. It is worth noting that by employing the Schur complement method, see sec.~\ref{ssec:Schur}, we can reduce the system of equations to a smaller size. Solving a linear system of $k$ equations typically requires $O(k^3)$ floating point operations (FlOps). In the initial system, we have $k=2n^2 + m$ equations, where $n > m$. Consequently, obtaining the solution requires $O(n^6)$ FlOps. However, the Schur complement equation has a size of $m$ and only requires $O(m^3)$ FlOps.
	
	Now let's examine the search directions discussed earlier. Consider the condition stated in~\eqref{eq:centralPath}, namely $X Z = \nu \openone$. When we take a step, we have the equation $(X + \Delta X)(Z + \Delta Z) = \nu \openone$, or, neglecting the second-order term $\Delta X \Delta Z$, we have $\Delta X Z + X \Delta Z = \nu \openone - X Z$. It is required that the steps $\Delta X$ and $\Delta Z$ be symmetric. The second equation in~\eqref{eq:rprd} reveals that $\Delta Z$ is always symmetric, given the numerical precision. However, the same cannot be said for $\Delta X$, which may not be symmetric. Consequently, there is a necessity to symmetrize~\eqref{eq:centralPath}. For instance, in 1998 the following natural symmetrization of~\eqref{eq:centralPath}, called AHO\index{search direction!AHO}, was introduced by Alizadeh, Haeberly and Overton~\cite{AHO98}:
	\begin{equation}
		\Theta_{\nu}^{AHO}(X, Z) \equiv \left( \frac{1}{2} (X Z + Z X) = \nu \openone \right).
	\end{equation}
	While the search direction defined by this symmetrization holds historical significance, it is no longer widely utilized by the majority of SDP solvers. For instance, recent implementations of SDPT3~\cite{TTT12} have omitted this particular search direction.
	Another search direction, called HKM\index{search direction!HKM}, has been introduced independently by Helmberg \textit{et al.} (1996)~\cite{HRVW96}, Kojima \textit{et al.} (1997)~\cite{KSH97} and Monteiro (1997)~\cite{M97}, and is used by many modern solvers. The HKM has the following primal form:
	\begin{equation}
		\label{eq:HKM_direction}
		\Theta_{\nu}^{HKM,primal}(X, Z) \equiv \left( Z^{\frac{1}{2}} X Z^{\frac{1}{2}} = \nu \openone \right),
	\end{equation}
	and the dual form $\Theta_{\nu}^{HKM,dual}(X, Z) \equiv \left( X^{\frac{1}{2}} Z X^{\frac{1}{2}} = \nu \openone \right)$. A third important search direction is the Nesterov and Tod~\cite{NT97,NT98}, or NT, direction\index{search direction!NT}. Its definition is more involved than in the cases of AHO and HKM. To define it, we consider a matrix $W$ satisfying $W^{-1} X W^{-1} = Z$. With the aid of the matrix $W$, the symmetrization of~\eqref{eq:centralPath} can be formulated as $\frac{1}{2} \left( W^{-\frac{1}{2}} X Z + Z X W^{-\frac{1}{2}} \right) = \nu W^{-1}$.
	
	The work of Monteiro and Zhang (MZ) from 1998~\cite{Z98,MZ98,M98}, introduced the following family of search directions, which includes the three mentioned directions, \textit{viz.} AHO, HKM and NT. The linear transformation of MZ\index{search direction!MZ} is given by the following formula
	\begin{equation}
		\label{eq:MZoperator}
		H_P(M) \equiv \frac{1}{2} \left( P M P^{-1} + P^{-T} M^T P^T \right),
	\end{equation}
	with $P$ being an invertible matrix. Next, we define $\Theta_{\nu}(X,Z) \equiv \left( H_P(X Z) = \nu \openone \right)$. By selecting different invertible matrices $P$, we can observe that each of the search directions discussed can be obtained. Specifically, when $P = \openone$, we obtain the AHO search direction. On the other hand, choosing $P = Z^{\frac{1}{2}}$ and $P = X^{-\frac{1}{2}}$ leads to the primal and dual HKM search directions, respectively. Furthermore, if we consider an invertible matrix $P$ satisfying the condition $P^T P = W^{-1}$, we obtain the NT search direction.
	
	Once the search direction has been determined, the next step in an IPM is to select the appropriate step length in that direction. This step length is crucial for ensuring the feasibility of the iterates. Specifically, the IPM chooses a step-length, represented by a pair of constants $\alpha$ and $\beta$ in the range $(0,1]$, such that the following conditions are satisfied:
	\begin{subequations}
		\label{eq:stepPSDcond}
		\begin{equation}
			X + \alpha \Delta X \succeq 0,
		\end{equation}
		\begin{equation}
			Z + \beta \Delta Z \succeq 0.
		\end{equation}
	\end{subequations}
	It is important to note that while Newton's method is used to determine the direction, the iterative solver does not necessarily take the full Newton step. To provide further clarity, it is worth mentioning the following points. According to~\cite{Toh02}, the value of $t$ that solves the optimization problem defined by $M$ and $\Delta M$ can be obtained in the following way. The objective is to maximize $t$ subject to the constraint $M + t \Delta M \succeq 0$. The formula to compute this step length is given by $\max \left( \text{eig}(C^{-T} \Delta M C^{-1}) \right)$, where $C$ represents the Cholesky decomposition of $M$. By using this formula, we can determine the appropriate step lengths $\alpha$ and $\beta$ for the IPM. It is important to note that convergence proofs of IPM algorithms often necessitate that at each step, the current solution is in close proximity (in some defined sense) to the central path. This requirement imposes additional constraints on the step length. However, for the sake of efficiency, these constraints may be relaxed, at the cost of losing the guarantee of convergence. Furthermore, it is worth highlighting that a common cause of failure in IPM algorithms arises during the Schur complement matrix decomposition. This issue tends to occur when the iterates approach the optimal solution and the primal variable and dual slack variables become nearly orthogonal, resulting in $\Tr(XZ) \approx 0$. In such cases, the Schur complement matrix may become ill-conditioned, leading to numerical instability or inaccurate results.

	\subsection{Solver internal mechanisms: predictor-corrector, warm start, problem structure}
	\label{sec:PredictorCorrector}
	
	Now, we will discuss several internal mechanisms employed by solvers to provide a deeper understanding of their functionality and potential challenges or performance gains. We begin by introducing the concept of predictor-corrector, shedding light on its significance and usage within solvers. Additionally, we explore the application of perturbations to iterates, which can prove beneficial when tackling numerical issues while solving complex problems with stringent constraints. Furthermore, we emphasize the importance of leveraging the structure of specific problems to unlock potential solver optimizations and fine-tune performance. By highlighting these aspects, we aim to provide insights into the diverse approaches and strategies that can be employed in solver implementations.
	
	% predictor-corrector
	
	As we mentioned above, when discussing the constrain~\eqref{eq:centralPath}, we expect an SDP solver to approach the value $\nu = 0$. But here a question araises: Should this be done immediately or gradually? And in the latter case: \textit{How} gradually? One of the most popular answers has been given by Mehrotra~\cite{Mehrotra92}, who introduced the so-called predictor-corrector mechanism\index{predictor-corrector}. The predictor-corrector method is a powerful numerical technique widely used in solvers for solving complex mathematical problems efficiently. This iterative algorithm combines two essential steps: prediction and correction. In the prediction step, an approximate solution is computed based on the available information. This predicted solution is then refined in the correction step by incorporating additional calculations or adjustments to improve its accuracy and convergence towards the proper solution. 
	
	The first step of the predictor-corrector mechanism is the calculation of the predictor direction, sometimes referred to as an \textit{affine scaling direction}~\cite{Sturm02}\index{affine scaling direction}, which employs an aggressive strategy to advance along the central path. In this approach, the target for the Newton step is set to $\nu = 0$ in~\eqref{eq:NewtonSymmetrization}, indicating that the predictor step aims to approach the optimal solution. Let $(\delta X, \delta y, \delta Z)$ represent the predictor step, with the step-length determined by $\alpha_P$ and $\beta_P$. These quantities are subsequently used to compute the value of $\nu$ that will serve as the objective for the Newton step direction in the subsequent iteration of the SDP solver. The predictor step itself is not taken but rather employed to derive a second-order correction for the corrector, or \textit{centering}, direction. The actual procedure of calculation of the new value of $\nu$ is quite complicated. We refer Reader to sec.~6.5 of~\cite{myThesis} for more details. Just to provide some taste of the method we mention that e.g. in SDPT3 solver the value $\alpha_P$ is upper bounded by a certain user-specified parameter \Minline{gam}, and the following formula for the corrector step~\cite{TTT98,TTT12}:
	\begin{equation}
		\label{eq:PredCorr_SDPT3}
		\frac{1}{n} \Tr(X Z) \cdot \left( \frac{\Tr \left( (X + \alpha_P \delta X) (Z + \beta_P \delta Z) \right)}{\Tr(X Z)} \right)^{\text{expon\_used}},
	\end{equation}
	where $\text{expon\_used}$ is a variable whose value is either a constant or is determined with some other algorithm basing on another user-specified parameter \Minline{expon}. Next, in the corrector part of the iteration one sets the new value of $\nu$ and calculates the Newton step $(\Delta X, \Delta y, \Delta Z)$ for the second time, with a different right-hand-side (RHS). Often, to the goal $\nu \openone$ in~\eqref{eq:centralPath}, an additional term $F(X, Z, \delta X, \delta Z)$ with a second order correction is added. Finally, the step-lengths $\alpha_C$ and $\beta_C$ for the corrector step are computed to ensure the preservation of positivity. In the subsequent iteration, the IPM algorithm sets the following:
	\begin{subequations}
		\begin{equation}
			X \equiv X + \alpha_C \Delta X,
		\end{equation}
		\begin{equation}
			y \equiv y + \beta_C \Delta y,
		\end{equation}
		\begin{equation}
			Z \equiv Z + \beta_C \Delta Z.
		\end{equation}	
	\end{subequations}

	% warm start
	
	The concept of warm start in solvers plays a crucial role in optimizing computational efficiency and reducing solution times for various optimization problems. Warm start refers to the technique of providing an initial feasible solution to a solver, obtained as a guess, or some generic scheme, or has been calculated from an already solved similar or related problem. Rather than starting the solver from scratch, the warm start approach utilizes the information from a previously solved problem to accelerate the convergence of subsequent iterations. By leveraging this initial solution, warm start techniques can significantly improve the overall performance of solvers. We will only briefly overview the warm start in solvers, and refer to sec.~6.3 of~\cite{myThesis} for a detailed discussion. It has been observed that it is desirable for the initial iterate to have the magnitude of at least the same order as the optimal solution. The following method of cold-start is used in SDPT3~\cite{TTT12}: $X^{(0)} \equiv \xi \openone$, $Z^{(0)} \equiv  \eta \openone$, and $y^{(0)}$ is the zero vector of the relevant dimension, where $\xi \equiv \max \left( 10, \sqrt{n}, n \max_{i \in [m]} \frac{1+|b_i|_F}{1+|A_i|_F} \right)$ and $\eta \equiv \max \left( 10, \sqrt{n}, |C|_F, \max_{i \in [m]} |A_i|_F \right)$. In~\cite{myThesis} we proposed and analysed warm start strategies for NPA problems.	
	
	% perturbations in iterations
	
	Another mechanism used in SDP solvers, which proved to be advantageous to certain scenarios, is the utilization of perturbations during iterations. Perturbations involve introducing slight modifications to the current iterates under specific conditions. The primary purpose of employing perturbation strategies is to prevent solvers from becoming stuck near the boundary of e.g. the PSD cone. While the primary objective of perturbations is not to reduce the number of iterations or CPU time, but rather to circumvent failures, it has been observed that in instances where the solver does not encounter failures, the most efficient solution tends to be one without perturbations~\cite{myThesis}. By incorporating perturbations into the iterative process, the solver can navigate away from critical regions and explore a wider solution space, potentially avoiding numerical instabilities or convergence issues. While we will not delve into the details of perturbation strategies, we present a simple example in Algorithm~\ref{alg:perturbation} to illustrate their form. The purpose of this strategy is to strike a balance between improving the duality gap and ensuring feasibility. In our observations of SDPs occurring in NPA~\cite{myThesis}, we have found that the reduction of the size of the gap often presents the greatest challenge. Specifically, the iterates tend to reach the feasibility threshold relatively quickly, and the majority of iterations are dedicated to reducing the gap.
	\begin{algorithm}
		\begin{algorithmic}
			\If {$\text{gap} > 100 \cdot \epsilon_P $}
			\State $X \gets X + 0.01 \cdot t_p \cdot \openone$
			\EndIf
			\If {$\epsilon_P > 100 \cdot \epsilon_D$}
			\State $Z \gets Z + 0.1 \cdot t_p \cdot \openone$
			\EndIf
		\end{algorithmic}
		\caption{\label{alg:perturbation}Example of a perturbation of the iteratively improved solution in an SDP solver.}
	\end{algorithm}		
	It is important to note that the efficiency and effectiveness of perturbation strategies may vary depending on the problem and solver being employed. In cases where the solver encounters failures, perturbations can provide a crucial mechanism to overcome such issues and continue the iterative process. However, it is worth noting that in scenarios where the solver operates smoothly without failures, perturbations may introduce additional computational overhead without providing substantial benefits in terms of solution quality or convergence speed. Overall, the use of perturbations in iterations offers a valuable approach to enhance the robustness and reliability of solvers when tackling problems involving the PSD cone. Nevertheless, the decision to incorporate perturbations should be made considering the specific problem characteristics, solver behavior, and desired trade-offs between reliability and computational efficiency.

	% exploiting structure of problems
	
	The last mechanism we mention about is exploiting the specific structure of the problem by a solver to enhace its performance. It allows for tailored solver optimizations that can significantly enhance performance. By exploiting the problem structure, solvers can take advantage of inherent characteristics such as symmetry, sparsity, or specific constraints to reduce computational complexity. This approach enables the solver to focus computational resources on the most relevant parts of the problem, leading to faster convergence and more efficient solutions. Additionally, by understanding the problem structure, solvers can employ specialized algorithms and techniques that are specifically designed to leverage the problem's unique properties, further improving solution quality and computational efficiency. For instance, in~\cite{myThesis} we have proposed a special data format to improve performance of operations taken by a solver dedicated the problems occuring in the dual formulation of NPA problems, where such properties as the sparsity and entries pattern was taken into account. 
	
	One of the first generic approaches exploiting sparsity patterns\index{sparsity patterns} was given by Fujisawa \textit{et al.}~\cite{Fujisawa97}, where the methods to leverage the sparsity of the problem matrices to improve efficiency were explored. The computation of the Schur complement matrix, as discussed in Section~\ref{ssec:Schur}, is often recognized as the most computationally intensive step in solving an SDP problem. One strategy they employ to support the Schur complement calculations is reordering the matrices $\{A_i\}_i$. They demonstrate that the most effective reordering is one that arranges the matrices in a non-increasing order based on the number of nonzero elements $f_i$. Additionally, due to the symmetry of the Schur complement $B$, only entries $B_{i,j}$ with $j \geq i$ need to be directly evaluated. Fujisawa \textit{et al.} presented three different methods for computing the element $B_{i,j}$, with the choice depending on the sparsity patterns of matrices $A_i$ and $A_j$. For highly sparse matrices, they utilize the formula $B_{i,j} = \sum_{a,b,c,d \in [n]} (A_i)_{c,d} W_{c,a} W_{d,b} (A_j)_{a,b}$ to compute the Schur complement. The calculation of this quantity requires $(2 f_i + 1) f_j$ multiplications.
	The work~\cite{waki2006sums} introduces the so-called \textit{correlative sparsity pattern graph}, which relates to a certain sparse structure in the objective and constraint polynomials of unconstrained and inequality-constrained sparse polynomial optimization problems. The graph is used to obtain sets of the supports for sums-of-squares polynomials and get the improved performance of relevant SDPs.
	Sparsity patterns that are more specific to non-commutative polynomial optimization, particularly relevant to quantum information problems, were investigated in~\cite{klep2021sparse}. The method suggested partitioning the input variables into cliques based on the so-called \textit{correlative sparsity pattern}\index{sparsity patterns!correlative sparsity} exhibited by the polynomials present in the objective function and constraints. In~\cite{wang2021exploiting} another particular type of sparsity occurring in the input data for large-scale sparse noncommutative polynomial optimization problems, called \textit{term sparsity}\index{sparsity patterns!term sparsity} is introduced.

	\section{Constructions of semi-definite programming useful for quantum information}
	\label{sec:basicTools}
	
	In this section, we provide an overview of popular constructions which are used as building blocks for more complicated optimization problems used in quantum information. In sec.~\ref{sec:SDPreps} we discuss semidefinite representations of semialgebraic functions which allow e.g. to express approximations of various types of quantum entropies as SDPs. In sec.~\ref{sec:PPT} we provide an overview of the separability criteria of quantum states originating from the famous PPT criterion. Next, in sec.~\ref{sec:C-J} we describe the Choi-Jamiołkowski isomorphism and highlight its relation to PSD constraints. Then, in sec.~\ref{sec:SoS} the SoSs decomposition important for polynomial optimization is discussed, and in sec.~\ref{sec:Lovasz} we describe the famous Lov{\'a}sz $\theta$ function. Afterwards, we will discuss the application of moment matrices in the realm of quantum information and explore different aspects of their use. These include sec.~\ref{sec:NPA} which discusses the relationship between correlation matrices and the so-called moment matrices, together with the Navascués-Pironio-Acin hierarchy which is a method for optimization over non-commuting variables; secs~\ref{sec:dimConstraint} and~\ref{sec:seesaw}, which investigates three distinct methods for optimizing probability distributions subject to dimension constraints.
	
	% TODO: lmibook.pdf rozdziały 2.1 i 2.2
	
	% TODO: może z tego tekstu coś się przyda?
	%A more comprehensive overview of applications of SDP may be found, e.g. in \cite{sdp96,Boyd04}. These include a famous MAX-CUT and MAX-k-SAT relaxations by Goemans and Williamson \cite{maxcut}, maximum eigenvalue, matrix norm minimization, and combinatorial optimization problems \cite{GLS84,A91,Overton92,MoharPoljak93,A95,Goemans97,B00,BYZ00}.

% TODO: napisać tę sekcję	
%	\subsection{Polynomial optimization}
%	
%	\subsubsection{Positivstellensatz}~\\
%	http://www.mit.edu/~parrilo/pubs/files/SDPrelaxations.pdf
%   POSITIVITY AND SUMS OF SQUARES A GUIDE FOR RECENT RESULTS.pdf całe zawiera przegląd
%	
%	to cytuje NTV: [J.W. Helton and S. A. McCullough, A Positivstellensatz for Non-Commutative Polynomials, Trans. Am. Math. Soc. 356, 3721 (2004).]
%\subsubsection{Bilevel polynomial optimization}
%\label{sec:bilevel}
%\cite{jeyakumar2016convergent}
%	
%	

%	\subsection{Lasserre’s hierarchy for commutative polynomials}
%	\label{sec:Lasserre}
%	\cite{lasserre2001global}

	\subsection{Semidefinite representations of semialgebraic sets}
	\label{sec:SDPreps}
	
%	\cite{fawzi2019semidefinite}
%	
%	Parrilo2003\_Article\_SemidefiniteProgrammingRelaxat.pdf
%	
%	Parrilo-SDPLog.pdf
%	
%	Semidefinite Approximations of the Matrix Logarithm (wydrukowane)
%	
%	tu jest co to \textbf{operator concave}: https://www.hindawi.com/journals/ijanal/2015/649839/
%	
%	\textbf{operator perspective} jest opisane tu: https://projecteuclid.org/journals/annals-of-functional-analysis/volume-5/issue-2/Non-commutative-perspectives/10.15352/afa/1396833504.full
	
	% 10_positivstellensatz_2003_12_07_02_screen
	
	Spectrahedron\index{sec:SDPrepresentation} is defined as a geometric object that can be characterized as a solution set of an LMI~\eqref{eq:LMI} or, in other words, it is an intersection of the PSD cone with a linear affine subspace. This representation allows spectrahedra to capture the feasible regions of SDPs. When spectrahedra are subjected to linear or affine transformations, the resulting shapes are referred to as \textit{projected spectrahedra}, or \textit{spectrahedral shadows}, or \textit{SDP representable sets}~\cite{helton2009sufficient}. These projected spectrahedra retain the convexity property and also belong to the class of semialgebraic sets\index{semialgebraic set}. It is worth noting that while every spectrahedral shadow is a convex semialgebraic set, but the converse statement, which was posed as a question by Nemirovski in~\cite{nemirovski2007advances}, previously conjectured~\cite{helton2009sufficient} to be true until 2017~\cite{scheiderer2018spectrahedral}, does not hold in general. This means that not all convex semialgebraic sets can be represented as spectrahedra. The notion of spectrahedra was introduced in~\cite{ramana1995some}; see~\cite{vinnikov2012lmi} for an overview, and~\cite{bengtsson2017geometry} for a detailed discussion of their relation to the geometry of quantum states. An example of a spectrahedron of particular interest to quantum information is the spectraplex\index{spectraplex}, which is the set of all PSD matrices in a given dimension with trace $1$, i.e. the normalized quantum states.
	
	To be more specific, we say that a set $\mathcal{S} \subseteq \mathbb{R}^n$ is LMI representable\index{LMI representable set} or has an LMI representation\index{LMI representation}~\cite{helton2007linear} if there exists a set of $n+1$ symmetric $n \times n$ matrices $\{A_i\}_{i \in \{0\} \cup [n]} \subset \mathbb{S}^n$ such that
	\begin{equation}
		\mathcal{S} = \left\{ x \in \mathbb{R}^n : A_0 + \sum_{i \in [n]} A_i x_i \succeq 0 \right\}.
	\end{equation}
	The question of which closed convex sets can be SDP represented for $n = 2$ was first posed by Parrilo in~\cite{parrilo2003minimizing}. Suppose that for some $N \in \mathbb{N}_{+}$ there exist the following two sets of symmetric $n \times n$ matrices: $\{A_i\}_{i \in \{0\} \cup [n]} \subset \mathbb{S}^n$ and $\{B_j\}_{j \in [N]} \subset \mathbb{S}^n$, such that $\mathcal{S}$ is a projection to $\mathbb{R}^n$ of the set
	\begin{equation}
		\label{eq:liftedLMI}
		\hat{\mathcal{S}} = \left\{ (x,u) \in \mathbb{R}^{n+N} : A_0 + \sum_{i \in [n]} A_i x_i + \sum_{j \in [N]} B_j u_j \succeq 0 \right\},
	\end{equation}
	so that $\mathcal{S} = \left\{ x \in \mathbb{R}^n : \bigexists_{u \in \mathbb{R}^N} (x,u) \in \hat{\mathcal{S}} \right\}$. Then, we say that $\mathcal{S}$ is \textit{SDP representable}\index{SDP representable set}, or has an \textit{SDP representation}\index{SDP representation} or a \textit{lifted LMI representation}\index{LMI representation!lifted}, or is a spectrahedral shadow~\cite{scheiderer2018spectrahedral}.
	For instance, from~\eqref{eq:SchurMGMean} it follows that the hypograph of the matrix geometric mean, \textit{viz.} $\hypM{[\#]} = \left\{ (X, Y, R) \in \mathbb{H}^n_{++} \times \mathbb{H}^n_{++} \times \mathbb{H}^n : X \# Y - R \succeq 0 \right\}$, has an SDP representation. As shown in~\cite[Prop.~1]{fawzi2017lieb}, for a odd $p \in \mathbb{N}_+$, and $l \in \mathbb{N}_+$, with $p < 2^l$, it holds that
	\begin{equation}
		\label{eq:hypMMGMean}
		\hypM{[\#_{p / 2^l}]} = \left\{ (X, Y, R_l) : \bigexists_{(R_i)_{i \in [l-1]} \subset \mathbb{H}^n},
		\bigforall_{i \in [l]}
		\begin{bmatrix}
			X \#_{m_i} Y & R_i \\ R_i & R_{i-1}
		\end{bmatrix} \succeq 0,
		R_0 = Y
		\right\},
	\end{equation}
	where $(m_i)_{i \in [l]}$ is the binary expansion of $p / 2^l$, i.e. $p / 2^l = \sum_{i \in [l]} m_i / 2^{l - i + 1}$ and $m_0 = 0$. Thus, there exist an SDP representation of $\hypM{[\#_{p / 2^l}]}$ consisting of $l$ LMI, each of size $2n$ by $2n$.
	
	A necessary condition for a set to be LMI representable is to be convex and basic closed semialgebraic. If a set is SDP representable, then it might not be basic closed semialgebraic, but it must be convex semialgebraic. In~\cite{helton2009sufficient,helton2010semidefinite} sufficient conditions for SDP representability were given. It was also shown that the set of all SDP representable sets is closed under taking linear images or preimages, finite intersections, or convex hulls of finite unions~\cite{helton2009sufficient,netzer2009note}. In~\cite{netzer2010semidefinite} it was shown that the interior of an SDP representable set is again an SDP representable set. The result of~\cite{scheiderer2018semidefinite} was that closed convex hulls of one-dimensional semialgebraic sets are also SDP representable. The seminal work~\cite{parrilo2003semidefinite} showed how to construct a complete family of SDPs of polynomial size, which can be used to prove the infeasibility of a finite set of polynomial constraints.
	
	Since the feasible set\index{feasible set} of SDP is a semialgebraic set, SDPs cannot be directly used to model non-semialgebraic sets and functions, even if they are convex. The work~\cite{fawzi2019semidefinite} provided a method to approximate certain useful non-semialgebraic sets with SDP representations of relatively small size. Consider a non-semialgebraic concave function $g : \mathbb{R} \rightarrow \mathbb{R}$. Suppose it has an integral representation $g(x) = \int_0^1 f_t(x) dt$, and that the integral can be approximated, e.g. using one of the Gauss quadratures~\cite{quarteroni2007foundations}, and then $g(x) \approx r_m(x) \equiv \sum_{j \in [m]} w_j f_{t_j}(x)$, where the weights $w_j$ and nodes $t_j$ depend on the quadrature. The quantity $m$ that defines how many terms occur in the quadrature is called the order of the quadrature. The order of the quadrature determines the accuracy of the approximation, with higher orders resulting in more accurate approximations. For a function of particular interest, the logarithm, we have $\log(x) = \int_0^1 \frac{x - 1}{t \cdot (x - 1) + 1} dt$. For any fixed $t$, the hypograph of the concave function $f_t(x) \equiv \frac{x - 1}{t \cdot (x - 1) + 1}$ has an SDP representation, \textit{viz.} $f_t(x) \geq r$ if and only if $\begin{bmatrix} x - 1 - r & -\sqrt{t} r \\ -\sqrt{t} r & 1 - r t \end{bmatrix} \succeq 0$. One can show that the matrix hypograph of $f_t$ for any $t \in [0,1]$ has the following SDP representation~\cite[Prop.~2]{fawzi2019semidefinite}:
	\begin{equation}
		\begin{aligned}
			\hypM{f_t} &= \left\{ (X, R) \in \mathbb{H}^n_{++} \times \mathbb{H}^n : (X - \openone) \cdot [t \cdot (X - \openone) + \openone]^{-1} \succeq R \right\} \\
			&= \left\{ (X, R) \in \mathbb{H}^n_{++} \times \mathbb{H}^n : 
			\begin{bmatrix}
				X - \openone - R & -\sqrt{t} R \\
				-\sqrt{t} R & \openone - t R
			\end{bmatrix} \succeq 0
			\right\}.
		\end{aligned}
	\end{equation}

	The approximation $\log{x} \approx r_m(x)$ is most accurate for small values of $x$. Since we have $\log(x) = \frac{1}{h} \log(x^h)$, it is often beneficial to use another approximation, \textit{viz.} $r_{m,k}(x) \equiv 2^k r_m(x^{1/{2^k}})$. It can be shown that $r_{m,k}$ is operator concave~\cite[Prop.~3]{fawzi2019semidefinite} and $\PerspM_{r_{m,k}}[Y,X] = 2^k \PerspM_{r_m} [X \#_{2^{-k}} Y, X]$~\cite[eq.~18]{fawzi2019semidefinite}. What is more, consider $V, X \in \mathbb{H}^n_{++}$ and $R \in \mathbb{H}^n$.
	It holds~\cite[p.~272]{fawzi2019semidefinite} $R' \succeq \PerspM_{-r_m}[Y,X]$ if and only if there exist $(R'_i)_{i \in [m]} \subset \mathbb{H}^n$ such that $R' = \sum_{i \in [m]} R'_i$ and $\bigforall_{i \in [m]}
	\begin{bmatrix}
		Y - X + R'_i & \sqrt{t_i} R'_i \\ \sqrt{t_i} R'_i & X + t_i R'_i
	\end{bmatrix} \succeq 0$.

	From~\eqref{eq:relOpEntropy} and $\log{x} \approx r_{m,k}(x)$ we can expect $S(X|Y) \approx \PerspM_{(-r_{m,k})}[Y,X]$ and thus is some sense:
	\begin{equation}
		\begin{aligned}
			\epiM{S(X|Y)} \approx K^n_{m,k} &\equiv \left\{ (X,Y,R') \in \mathbb{H}^n_+ \times \mathbb{H}^n_+ \times \mathbb{H}^n : R' \succeq \PerspM_{(-r_{m,k})}[Y,X] \right\} \\
			&= \left\{ (X,Y,R') \in \mathbb{H}^n_+ \times \mathbb{H}^n_+ \times \mathbb{H}^n : R' \succeq -2^k \PerspM_{r_m} [X \#_{2^{-k}} Y, X] \right\}.
		\end{aligned}
	\end{equation}
	This, together with the SDP representation~\eqref{eq:hypMMGMean}, shows that the set $K^n_{m,k}$ has an SDP representation~\cite[Theorem~3]{fawzi2019semidefinite}. This, in consequence, allows optimizations over quantum entropies~\cite{fawzi2018efficient,brown2021computing} using SDP.

	\subsection{Doherty-Parillo-Spedalieri conditions of separability}
	\label{sec:PPT}

	% TODO:
	%	PPT, DPS
	%	zdaje się, że quantum de Finetti jest w pewnym sensie uzupełnieniem DPS, w sensie, że pokazuje że DPS nie jest dokładnie separowalne, ale że jest blisko, coraz bliżej (łagodna postać wyekstrahowana jest w 2012.00554 s 13 lemma 7)
	%	w linii 87 pliku IsSeparable.m z pakietu QETLAB jest dużo referencji
	
	The Doherty-Parillo-Spedalieri (DPS)~\cite{DPS02,DPS04} method is a powerful technique used to determine the separability of multipartite quantum states. It provides a hierarchy of SDP relaxations to approximate the optimal value of the separability problem. DPS introduces a hierarchy of conditions involving partial transpositions allowing for a stronger test of separability and is a reliable approach for studying the separability of multipartite quantum systems with more than two parties.
	The DPS method utilizes a series of SDP relaxations, where each relaxation introduces additional variables and constraints. At each level of the hierarchy, the DPS method formulates an SDP. The PSD conditions play a crucial role in these relaxations, as they express the constraint that the obtained solutions are physically valid quantum states.
	Roughly speaking, in the DPS method for a given state $\rho_{\mathcal{A}\mathcal{B}}$ one asks whether there exist a hierarchy of symmetric extensions, i.e., a family of states $\rho_{\mathcal{A} \mathcal{B}_1 \cdots \mathcal{B}_N}$ defined for any $N$, such that $\bigforall_{i \in [N}] \rho_{\mathcal{A}\mathcal{B}} = \Tr_{\mathcal{B}_j : j \neq i} [\rho_{\mathcal{A} \mathcal{B}_1 \cdots \mathcal{B}_N}]$. It happens that the state $\rho_{\mathcal{A}\mathcal{B}}$ is separable if and only if such a hierarchy exists for each natural $N$. The state $\rho_{\mathcal{A}\mathcal{B}}$ is separable if and only if a hierarchy of symmetric extensions exists for every natural number $N$. For a given fixed value of $N$, the task of verifying the existence of a symmetric extension is equivalent to an SDP. Consequently, an algorithm can be devised by incrementally examining the extendability condition for increasing values of $N$. This algorithm is guaranteed to terminate if the initial state $\rho_{\mathcal{A}\mathcal{B}}$ is entangled, thus it detects the non-separability. However, if the state is separable, the algorithm will continue indefinitely without termination.
	
	Let us provide a more detailed overview of the concept of DPS. Consider a state $\rho_{\mathcal{A}\mathcal{B}}$ residing in the composite Hilbert space $\mathcal{A} \otimes \mathcal{B}$, which exhibits separability, meaning that it can be expressed as a convex combination of pure product states:
	\begin{equation}
		\rho_{\mathcal{A}\mathcal{B}} = \sum_{i} \lambda_i \proj{\phi}_\mathcal{A} \otimes \proj{\varphi}_\mathcal{B},
	\end{equation}
	where the coefficients $\lambda_i$ satisfy the conditions $\sum_{i} \lambda_i = 1$ and $\lambda_i \geq 0$.
	Now, let $\tilde{\rho}$ denote a state on ${A^k} \otimes {\mathcal{B}^l}$, where $\mathcal{A}^k$ represents a product of $k$ spaces $\mathcal{A}$, and the same applies to $\mathcal{B}$. The state $\tilde{\rho}$ is referred to as an \textit{extension}~\cite{fannes1988symmetric,raggio1988quantum} of $\rho$ if it satisfies the condition:
	\begin{equation}
		\rho_{\mathcal{A}\mathcal{B}} = \Tr_{\mathcal{A}^{k-1} \mathcal{B}^{l-1}} \left[ \tilde{\rho} \right],
	\end{equation}
	where the partial trace is taken over all but the first copy of each space.
	Let $\mathcal{S}_\mathcal{A}$ represent the set of all permutation operators among copies of the space $\mathcal{A}$, and the same applies to $B$. The state extension $\tilde{\rho}$ is considered symmetric if, for every $P \in \mathcal{S}_\mathcal{A} \otimes \mathcal{S}_B$, the following condition holds:
	\begin{equation}
		\tilde{\rho} = P \tilde{\rho} P.
	\end{equation}
	On the other hand, the state extension $\tilde{\rho}$ is classified as PPT\index{Positive Partial Transpose} if $\tilde{\rho}$ remains positive after applying any partial transposition on the subsystems. When $\rho_{\mathcal{A}\mathcal{B}}$ is separable, it is guaranteed that for any values of $k$ and $l$, there exists an extension $\tilde{\rho}$ that is a PPT symmetric extension of $\rho_{\mathcal{A}\mathcal{B}}$. The fundamental principle behind the DPS hierarchy is to examine whether a PPT symmetric extension of $\rho_{\mathcal{A}\mathcal{B}}$ exists for fixed values of $k$ and $l$, and if not, this implies that $\rho_{\mathcal{A}\mathcal{B}}$ is not separable.
	As the constraints of PPT symmetric extension can be expressed as SDPs, the DPS method enables optimization over a relaxation of the set of separable states on the given spaces. As the values of $k$ and $l$ increase, the relaxation approaches the actual set of all separable states more closely. Therefore, starting from a PPT symmetric extension state $\tilde{\rho}$, it is possible to construct a state $\varrho$ on $\mathcal{A} \otimes \mathcal{B}$ that is, in a certain sense, \textit{close} to being separable.
	
	The method can be applied in analogous way to more than to parties, as in the following example involving three subsystems. We will now demonstrate an application of DPS, which involves a method for representing quantum states and measurements using a single SDP variable. Consider three unit vectors: $\ket{\Phi^{\lambda}} = \sum_{i \in [d_\mathcal{A}]} \sum_{j \in [d_\mathcal{B}]} \phi_{i j} \ket{ij}_{\mathcal{A}\mathcal{B}}$, $\ket{u^{\lambda}} = \sum_{i \in [d_\mathcal{A}]} \mu_i^{\lambda} \ket{i}_{\mathcal{A}}$, and $\ket{v^{\lambda}} = \sum_{j \in [d_\mathcal{B}]} \nu_j^{\lambda} \ket{j}_{\mathcal{B}}$. Here, $\lambda$ represents a global hidden variable with an arbitrary probability distribution ${p^{\lambda}}$. Next, we define the following operator
	\begin{equation}
		\label{eq:WABApBp}
		W_{\mathcal{A} \mathcal{B} \mathcal{A}' \mathcal{B}'} \equiv \sum_{\lambda} p^{\lambda} \cdot \left[ \proj{\Phi^{\lambda}}_{\mathcal{A}B} \otimes \proj{u^{\lambda}}_{A'} \otimes \proj{v^{\lambda}}_{\mathcal{B}'} \right].
	\end{equation}
	For two subsystems $S_1$, $S_2$, both of dimension $d_S$, the SWAP operator\index{SWAP operator} is defined in the following way:
	\begin{equation}
		\label{eq:SWAP}
		SWAP(S_1, S_2) \equiv \sum_{i,j \in [d_S]} \ket{i j} \bra{j i}_{S_1 S_2}.
	\end{equation}
	The idea that a tensor product of the density matrix $\rho$ and measurements $\{M\}$ contain all elements necessary to express the probabilities of the form $\Tr (\rho M)$ was formulated in~\cite{werner1989quantum}. In~\cite{navascues2014characterization}, where the so-called Navascués-de la Torre-Vértesi (NTV) SDP hierarchy was introduced, it was recognized that this idea in combination with DPS hierarchy allows approximating the probabilities $\Tr (\rho M)$ as entries in SDP variables. NTV provided one of the methods to approximate the set of quantum correlations with dimension constraints; other such methods are discussed in sec.~\ref{sec:dimConstraint}. A similar approach was used in~\cite[eq. (4)]{yu2021complete} to express a constraint of purity of the state, and in consequence to develop a method of providing rank constraints on the considered operators.
	Direct calculations show that $\Tr \left[ W_{\mathcal{A} \mathcal{B} \mathcal{A}' \mathcal{B}'} \left( SWAP(\mathcal{A}, \mathcal{A}') \otimes SWAP(\mathcal{B}, \mathcal{B}') \right) \right]$ is equal to: % -> zeszyt 30.03.2021
	\begin{equation}
		\begin{aligned}
			\sum_{\lambda} p^{\lambda} \cdot \Tr \left[ \tilde{W}_{\mathcal{A} \mathcal{B} \mathcal{A}' \mathcal{B}'}^\lambda \right] &= \sum_{\lambda} p^{\lambda} \cdot \left[ \sum_{i_1, i_2 \in [d_\mathcal{A}]} \sum_{j_1, j_2 \in [d_\mathcal{B}]} \phi_{i_1 j_1}^{\lambda} \phi_{i_2 j_2}^{\lambda*} \mu_{i_2}^{\lambda} \mu_{i_1}^{\lambda*} \nu_{j_2}^{\lambda} \nu_{j_1}^{\lambda*} \right] \\
			&= \sum_{\lambda} p^{\lambda} \cdot \Tr \left[ \proj{\Phi^{\lambda}}_{\mathcal{A}\mathcal{B}} \left( \proj{u^{\lambda}}_\mathcal{A} \otimes \proj{v^{\lambda}}_\mathcal{B} \right) \right],
		\end{aligned}
	\end{equation}
	where $\tilde{W}_{\mathcal{A} \mathcal{B} \mathcal{A}' \mathcal{B}'}^\lambda$ is defined as:
	\begin{equation}
		 \sum_{\substack{i_1, i_2, i_3, \\ i_4, i_5, i_6 \in [d_\mathcal{A}]}} \sum_{\substack{j_1, j_2, j_3, \\ j_4, j_5, j_6 \in [d_\mathcal{B}]}} \phi_{i_1 j_1}^{\lambda} \phi_{i_2 j_2}^{\lambda*} \mu_{i_3}^{\lambda} \mu_{i_4}^{\lambda*} \nu_{j_3}^{\lambda} \nu_{j_4}^{\lambda*} \ket{i_1 j_1 i_3 j_3} \bk{i_2 j_2 i_4 j_4}{i_5 j_5 i_6 j_6} \bra{i_6 j_6 i_5 j_5}_{\mathcal{A} \mathcal{B} \mathcal{A}' \mathcal{B}'}.
	\end{equation}
	The resulting expression is the probability of projection of the state on some projective measurements. % TODO: jaka jest jawna formuła na stan i pomiary sformułowana za pomocą wektorów w tych równaniach?
	With similar calculations, we also obtain:
	\begin{equation}
		\Tr \left[ W_{\mathcal{A} \mathcal{B} \mathcal{A}' \mathcal{B}'} \left( SWAP(\mathcal{A}, \mathcal{A}') \otimes \openone_{\mathcal{B} \mathcal{B}'} \right) \right] = \sum_{\lambda} p^{\lambda} \cdot \Tr \left[ \proj{\Phi^{\lambda}}_{\mathcal{A}\mathcal{B}} \left( \proj{u^{\lambda}}_\mathcal{A} \otimes \openone_\mathcal{B} \right) \right].
	\end{equation}
	We see that operators created in a similar way like~\eqref{eq:WABApBp} contain entries expressing Frobenius products~\eqref{eq:Frob} of a state and measurements, and SWAP operators provide a tool to extract them to obtain quantum probabilities.	
	It is easy to generalize the above formulae to cover cases involving e.g. more projective measurements and more parties.
	Obviously, the operator~\eqref{eq:WABApBp} is separable, and this constraint is imposed with the discussed DPS method.

	\subsection{Choi-Jamiołkowski isomorphism and quantum channels}
	\label{sec:C-J}
	
	The Choi-Jamiołkowski isomorphism introduced in 1972 by Jamiołkowski in~\cite{jamiolkowski1972linear} and, independently in 1975 by Choi in~\cite{choi1975completely} is a fundamental concept in quantum information theory that establishes a correspondence between quantum states and quantum channels, see~\cite{jiang2013channel} for a detailed discussion and historical remarks. The isomorphism also called a \textit{state-channel duality}, provides a mathematical framework to represent quantum channels as density matrices, enabling the study and manipulation of quantum processes using tools from quantum state theory. We say that a map is PSD when it is transforming PSD matrices to PSD matrices; it is \textit{completely PSD} if $\mathcal{E} \otimes \openone$ is PSD for $\openone$ acting over arbitrary space; a linear map is a \textit{trace preserving} when the trace of the input matrix is equal to the trace of the output matrix. The Choi-Jamiołkowski isomorphism is defined as follows. Given a linear PSD map $\mathcal{E}: \mathbb{C}^{d \times d} \rightarrow \mathbb{C}^{d' \times d'}$, i.e. $\mathcal{E} \in \mathsf{L}\left[ \mathbb{C}^{d \times d} , \mathbb{C}^{d' \times d'} \right]$, that transforms input states on $\mathbb{C}^{d \times d}$ to output states on $\mathbb{C}^{d' \times d'}$, its corresponding Choi matrix $J(\mathcal{E}) \in \mathbb{C}^{d' \times d'} \otimes \mathbb{C}^{d \times d}$ is the PSD matrix defined in the following way:
	\begin{equation}
		\label{eq:ChoiMatrix}
		J(\mathcal{E}) \equiv \sum_{i,j \in [d]} \mathcal{E} \left[ \kb{i}{j} \right] \otimes \kb{i}{j} = d \cdot \left( \mathcal{E} \otimes \openone_{d} \right) \left[ \proj{\Phi^{+}} \right],
	\end{equation}
	where $\ket{\Phi^{+}} \equiv \frac{1}{\sqrt{d}} \sum_{i \in [d]} \ket{i} \otimes \ket{i}$ is the maximally entangled state.
	
	Given a PSD matrix, it is possible to reconstruct the corresponding quantum channel. This reconstruction process allows us to extract useful information about the properties and behavior of the quantum channel. Let us consider a state $\rho = \sum_{k,l \in [d]} \rho_{k,l} \kb{k}{l} \in \mathbb{C}^{d \times d}$. Then
	\begin{equation}
		\openone_{d'} \otimes \rho^T = \left( \sum_{m \in [d']} \proj{m} \right) \otimes \left( \sum_{k,l \in [d]} \rho_{k,l} \kb{l}{k} \right) = \sum_{m \in [d']} \sum_{k,l \in [d]} \rho_{k,l} \proj{m} \otimes \kb{l}{k},
	\end{equation}
	and we have
	\begin{equation}
		\begin{aligned}
			J(\mathcal{E}) \cdot \left( \openone_{d'} \otimes \rho^T \right) &= \sum_{i,j \in [d]} \sum_{k,l \in [d]} \sum_{m \in [d']} \rho_{k,l} \cdot \left( \mathcal{E} \left[ \kb{i}{j} \right] \otimes \kb{i}{j} \right) \cdot \left( \proj{m} \otimes \kb{l}{k} \right) \\
			&= \sum_{i,j,k,l \in [d]} \sum_{m \in [d']} \rho_{k,l} \cdot \left( \mathcal{E} \left[ \kb{i}{j} \right] \cdot \proj{m} \right) \otimes \left( \kb{i}{j} \cdot \kb{l}{k} \right) \\
			&= \sum_{i,j,k \in [d]} \sum_{m \in [d']} \rho_{k,l} \cdot \left( \mathcal{E} \left[ \kb{i}{j} \right] \cdot \proj{m} \right) \otimes \kb{i}{k} \\
			&= \sum_{i,j,k \in [d]} \rho_{k,l} \cdot \mathcal{E} \left[ \kb{i}{j} \right] \otimes \kb{i}{k} \equiv Y.
		\end{aligned}
	\end{equation}
	Then, we take the partial trace of $Y$ over the second subspace, removing the input space:
	\begin{equation}
		\begin{aligned}
			\Tr_2 Y &= \sum_{l \in [d]} \bra{l} Y \ket{l} = \sum_{i,j,k,l \in [d]} \rho_{k,l} \cdot \mathcal{E} \left[ \kb{i}{j} \right] \bk{l}{i} \bk{k}{l} \\
			&= \sum_{i,j \in [d]} \rho_{l,j} \mathcal{E} \left[ \kb{l}{j} \right] = \mathcal{E} \left[ \sum_{i,j \in [d]} \rho_{l,j} \kb{l}{j} \right] = \mathcal{E} \left[ \rho \right].
		\end{aligned}
	\end{equation}
	Thus we have the following crucial property:
	\begin{equation}
		\label{eq:ChoiResult}
		\Tr_2 \left[ J(\mathcal{E}) \cdot \left( \openone_{d'} \otimes \rho^T \right) \right] = \mathcal{E} \left[ \rho \right].
	\end{equation}
	A direct consequence of~\eqref{eq:ChoiResult} is that for a POVM $\{M^b\}_b$ on $\mathbb{C}^{d' \times d'}$, we have $\Tr \left[ J(\mathcal{E}) \cdot \left( M^b \otimes \rho^T \right) \right] = \Tr \left[ \mathcal{E} \left[ \rho \right] M^b \right]$, which is the probability of the outcome $b$ of the POVM applied to the output state of the channel.

	It can be shown, that any linear map $\mathcal{E}$ is completely PSD if and only if its Choi matrix~\eqref{eq:ChoiMatrix} $J(\mathcal{E})$ is PSD.
	Similarly, the Choi-Jamiołkowski isomorphism captures the property of trace preserving with the constraint that the Choi matrix after tracing out the first subsystem is equal to $d \cdot \openone_{d}$. Trivially, for $J(\mathcal{E})$ defined in~\eqref{eq:ChoiMatrix} and trace preserving $\mathcal{E}$ we have $\Tr \left[ \mathcal{E} \left[ \kb{i}{j} \right] \right] = \Tr [\kb{i}{j}] = \delta_{i,j}$ and thus $\Tr_1 [J(\mathcal{E})] = \openone_{d}$. Conversely, consider a matrix $X' \in \mathbb{C}^{d' \times d'} \otimes \mathbb{C}^{d \times d}$ satisfying $\Tr_1 [X] = \openone_{d}$. Let $X' = \sum_{i,j \in [d]} X'_{i,j} \otimes \kb{i}{j}$. Then since $\Tr[\kb{i}{j}] = \delta_{i,j}$, we have $\bigforall_{i,j} \Tr [X'_{i,j}] = \delta_{i,j}$, and thus for any $\rho$ it holds that:
	\begin{equation}
		\Tr \left[ X' \cdot \left( \openone_{d'} \otimes \rho^T \right) \right] = \Tr_2 \left[ \sum_{i,j \in [d]} \Tr_1 \left[ X'_{i,j} \otimes \left( \kb{i}{j} \cdot \rho^T \right) \right] \right] = \Tr_2 \left[ \sum_{i \in [d]} \proj{i} \cdot \rho^T \right] = \Tr[\rho].
	\end{equation}
	
	It is possible to employ the Choi-Jamiołkowski isomorphism to express other properties of quantum channels. Consider a channel $\mathcal{E}_{A'B' \leftarrow AB}$ transforming states states on $\mathbb{C}^{n_A \times n_A} \otimes \mathbb{C}^{n_B \times n_B}$ to states on $\mathbb{C}^{n_{A'} \times n_{A'}} \otimes \mathbb{C}^{n_{B'} \times n_{B'}}$, for some $n_A$, $n_{A'}$, $n_B$, and $n_{B'}$, with the Choi matrix $J(\mathcal{E}_{A'B' \leftarrow AB}) \in \mathbb{C}^{n_{A'} \times n_{A'}} \times \mathbb{C}^{n_{B'} \times n_{B'}} \otimes \mathbb{C}^{n_A \times n_A} \otimes \mathbb{C}^{n_B \times n_B}$. One usually interprets $A$ and $B$ as inputs and $A'$ and $B'$ as outputs of Alice and Bob, respectively. For instance, one can show that the marginal output state of Alice is a result of a fixed operation on the marginal input state of Alice, or, in other words, the channel is non-signaling from Bob to Alice~\cite{cubitt2011zero} if and only if $\Tr_{B'} \left[ J(\mathcal{E}_{A'B' \leftarrow AB}) \right] = \Tr_{BB'} \left[ J(\mathcal{E}_{A'B' \leftarrow AB}) \right] \otimes \openone_B$
	\cite[eq.(22)]{leung2015power}; similar relation holds for channels non-signaling from Alice to Bob.
	Suppose that Alice and Bob are controlling ancillary subsystems, $\tilde{A}$ and $\tilde{B}$, respectively. A channel is called PPT-preserving\index{Positive Partial Transpose} when it transforms a bipartite PPT state $\rho_{A \tilde{A} B \tilde{B}}$, i.e. a state satisfying $\rho_{A \tilde{A} B \tilde{B}}^{T_{B \tilde{B}}} \succeq 0$, to a biparites PPT state~\cite{rains1999bound,rains2001semidefinite}. In~\cite{rains2001semidefinite} it was shown that $\mathcal{E}_{A'B' \leftarrow AB}$ is PPT-preserving if and only if $\left[ J(\mathcal{E}_{A'B' \leftarrow AB}) \right]^{T_{B B'}} \succeq 0$. We refer to~\cite{leung2015power} for a detailed discussion of other properties of quantum channels possible to be expressed with constraints in SDPs.
	
	In summary, the PSD condition is essential in the Choi-Jamiołkowski isomorphism as it ensures the validity of the represented quantum channels. It provides a mathematical framework to analyze and manipulate quantum processes, allowing for the exploration of various properties and applications in quantum information theory. The discussed methods are, in particular, used to express bounds on various types of channel capacities as SDPs~\cite{wang2016semidefinite,sutter2017approximate}.

	\subsection{Sum of squares decomposition of polynomials}
	\label{sec:SoS}
	
	The \textit{sum of squares}\index{sum of squares} (SoS) technique is a powerful tool used in SDP to represent nonnegative polynomials as sums of squares of other polynomials. Recall that in SDP, the goal is to optimize a linear objective function subject to LMI constraints. However, many optimization problems involve nonnegative polynomials, and checking the nonnegativity of a polynomial can be challenging. The SoS technique provides a way to approximate these nonnegative polynomials using sums of squares, which can be readily handled in SDP. This technique is useful in many areas of mathematics, including optimization, control theory, and signal processing. One application of the SoS decomposition is in optimization problems. In particular, it can be used to determine whether a polynomial is non-negative over a given domain. This is important in optimization because many optimization problems involve minimizing or maximizing a polynomial subject to certain constraints. By using the SoS decomposition, one can determine whether the polynomial is non-negative over the feasible region, which can help in finding the optimal solution. The method found various applications in multiple areas of science and  is particularly useful for polynomial optimization, robust control, and polynomial system analysis, as it allows for tractable representation and computation of nonnegative polynomials in SDP frameworks~\cite{prajna2002introducing,parrilo2003semidefinite,papachristodoulou2005tutorial,jarvis2005control,parrilo2008approximation,lofberg2009pre,chernyshenko2014polynomial}.
	
	The main idea behind the SoS technique is to express a nonnegative polynomial as a SoSs of lower-degree polynomials. This is achieved by introducing additional variables and using semidefinite constraints. Specifically, the nonnegative polynomial is decomposed into a SoSs of polynomials, where each polynomial is multiplied by a semidefinite matrix. The positivity of the original polynomial is then guaranteed by the positivity of the semidefinite matrices. Let's consider a homogeneous polynomial $h(x)$, where $x \in \mathbb{R}^n$, and all terms of $h(x)$ have a degree of $2m$. We say that $h(x)$ is a SoS polynomial if and only if, for some $k$ there exists a set $\{ g_{i} \}_{i \in [k]}$, where each $g_{i}$ is a polynomial of degree $m$ and $h(x) = \sum_{i \in [k]} g_{i}(x)^{2}$. Obviously, any SoS polynomial is always positive. Moreover, the interesting property of the SoS polynomials is that any real non-negative polynomial can be approximated arbitrarily closely by a sequence of SoS polynomials, known as the SoS hierarchy~\cite{lasserre2007sum}.
	For any polynomial $g_i(x)$, we can express it as the inner product of a vector $v_i$ and the basis of monomials $x^{(m)}$ of degree $m$. Mathematically, this can be written as $g_i(x) = v_i^T \cdot x^{(m)}$. The basis $x^{(m)}$ consists of monomials of degree $m$, and the dimension of this basis is given by $d \equiv {{n+m-1} \choose {m}}$, thus $v_i \in \mathbb{R}^d$.
	If SoS exist, then a set $\mathcal{S} \equiv \{v_i\}_{i=1}^{k}$ also exist and:
	\begin{equation}
		\label{eq:SoS}
		\begin{aligned}
			h(x) &= \sum_{i \in [k]} g_{i}(x)^{2} = \sum_{i \in [k]} \left(v_i^T \cdot x^{(m)}\right)^2 = \sum_{i \in [k]} \left(x^{(m) T} \cdot v_i\right) \left(v_i^T \cdot x^{(m)}\right) \\
			&= \sum_{i \in [k]} x^{(m) T} \left(v_i v_i^T\right) x^{(m)} = x^{(m) T} \left(\sum_{i \in [k]} v_i v_i^T\right) x^{(m)} \equiv x^{(m) T} M x^{(m)}.
		\end{aligned}
	\end{equation}
	Hence, even if the set $\mathcal{S}$ is not explicitly known, it is evident that $M \equiv \sum_{i \in [k]} v_i v_i^T \succeq 0$, $M \in \mathbb{S}^{d \times d}$. This observation implies that verifying whether a polynomial $h(x)$ is SoS is equivalent to determining the existence of a PSD matrix $M \succeq 0$ that satisfies the relation $h(x) = x^{(m) T} M x^{(m)}$.
	
	Now, let us consider a symmetric matrix $H \in \mathbb{S}^d$, not necessarily PSD, such that $h(x) = x^{(m)T} H x^{(m)}$. Constructing such a matrix is straightforward, as it suffices to assign the relevant coefficients from $h(x)$ onto the diagonal elements of $H$. Let $\{N_i\}_{i \in [D]}$, for some $D$, $\bigforall_{i} N_i \in \mathbb{S}^d$, be a basis of the space of all symmetric $d$ by $d$ matrices satisfying the equation $x^{(m)T} N_i x^{(m)} = 0$. The dimension of this space depends on $m$, i.e. the degree of the polynomial $h(x)$, as well as $n$, the number of variables.
	The objective now is to verify the feasibility of the so-called Gram representation\index{Gram representation} of the SoS polynomial. This representation is expressed as $H + \sum_{i} y_i N_{i} \succeq 0$, where $y_i$ are coefficients.
	Finally, the feasibility of the Gram representation of the SoS polynomial is examined by checking if the matrix $H + \sum_{i} y_i N_{i}$ is PSD. This assessment allows for the determination of whether the given polynomial can be represented as a SoS.
	
	The non-commutative analog of SoS, called \textit{sum of Hermitian squares} or \textit{non-commutative sum of squares}~\cite{mccullough2005noncommutative} was introduced in~\cite{helton2002positive}. We say that a Hermitian polynomial $p(X)$ in non-commutative variables $\mathbf{X} = (X_i)_i$ is a non-commutative SoS (or, simply, SoS) when there exist polynomials $(r_j(X))_j$ such that $p(X) = \sum_j r_j^\dagger r_j$. For operators used as non-commuting variables, being SoS means that the polynomial of the operators is PSD, meaning, in particular, that its expectation value is non-negative for all quantum states. The non-commutative polynomial is \textit{weighted SoS} (WSoS) generated by a collection of Hermitian polynomials in non-commutative variables $\mathcal{P}$ when it is of the form~\cite{doherty2008quantum}:
	\begin{equation}
		\label{eq:WSoS}
		\underbrace{\sum_j r_j^\dagger r_j}_{\text{SoS}} + \underbrace{\sum_k \sum_l s_{k,l}^\dagger p_k s_{k,l}}_{\text{weighted term}}
	\end{equation}
	for $p_i \in \mathcal{P}$, and some polynomials $(r_j(X))_j$ and $(s_{k,l}(X))_{k,l}$.
	An algorithm for finding a sum of Hermitian squares decompositions for Hermitian polynomials in non-commuting variables based on SDP was given in~\cite{klep2010semidefinite}.
	The SoS decompositions were used to provide the so-called self-testing of quantum states and measurements in~\cite{yang2013robust,bamps2015sum}.
	
	As a direct example of the application of SoS in quantum information, we will analyze the so-called quantum moment problem. In this problem, we ask whether, for a given probability distribution, there exist a quantum state and measurements that produce such distribution (see sec.~\ref{sec:probDistr}). A complementary problem is to decide whether a given instance of the quantum moment problem is unsatisfiable. A crucial tool for this purpose is one of the versions of the Positivstellensatz. Positivstellensatze are theorems in real algebraic geometry that provides a way to determine whether a polynomial is positive on a semi-algebraic set, or if it can be written as an SoS; see~\cite[Chap.~4]{bochnak2013real} of an introduction for the commutative variable case.
	
	Suppose that the quantum setup of interest involves the measurement operators $\mathbf{X} = (X_i)_i$. We briefly show that many of the desired properties of measurements can be expressed as a requirement that certain polynomials of these operators vanish. One of the requirements is that the measurements over different subsystems commute. Indeed, the condition $[X_i, X_j] = 0$ is equivalant to statement the Hermitian polynomial $i [X_i, X_j]$ is equal zero. If for certain $\mathcal{M}$ the operators $\{X_i\}_{i \in \mathcal{M}}$ for a projective measurement, then the normalization condition is expressed as the requirement that the Hermitian polynomial $\openone - \sum_{i \in \mathcal{M}} X_i$ is equal zero; similarly the idempotency condition of $X_i$ is expressed as $X_i^2 - X_i$ is equal zero. Another simple constraint possible to be expressed with vanishing Hermitian polynomials is the requirement that $X$ has eigenvalues in $\{+1,-1\}$; this is equivalent to the requirement that $\openone -X^2$ is equal zero. Let us define $\mathcal{P}$ of all Hermitian polynomials of one of these forms, as well as their negatives, imposing the constraints which are desired in a given scenario.
	
	Consider an expression $G \equiv \sum_{\mathbf{a}, \mathbf{x}} \alpha_{\mathbf{a}|\mathbf{x}} p_{\mathbf{a}|\mathbf{x}}[\mathbf{X}]$. We are interested in finding the Tsirelson bound~\cite{cirel1980quantum} of such an expression. If for any quantum state $\ket{\Phi}$ and measurement operators $\mathbf{X}$ it holds $\bra{\Phi} G[\mathbf{X}] \ket{\Phi} \leq q$ for some $q \in \mathbb{R}$, then trivially $q \cdot \openone - G \succeq 0$. In~\cite[Theorem~4.3]{doherty2008quantum} the following form of Positivstellensatz was given: If $\left[ \bigforall_{\mathbf{X}} \left( \bigforall_{p \in \mathcal{P}} p[\mathbf{X}]=0 \implies \bigforall_{\ket{\Phi}} \bra{\Phi} (q \cdot \openone - G) \ket{\Phi} > 0 \right) \right]$, then $v \cdot \openone - G$ is WSoS. In other words, if the Hermitian polynomial $q \cdot \openone - G$ in non-commuting variables $\mathbf{X}$ is a PD operator (expressed as $\bigforall_{\ket{\Phi}} \bra{\Phi} (q \cdot \openone - G) \ket{\Phi} > 0$) under the assumption that $\mathbf{X}$ are quantum measurements (expressed as $\bigforall_{p \in \mathcal{P}} p[\mathbf{X}]=0$), then it can be written in the WSoS form. Thus, this Positivstellensatz says that if there exist no quantum state and measurements attaining the value $q$, then $q \cdot \openone - G$ is WSoS.
	
	Now, the question is, how to derive the value of $q$ using SDP. We do not have information regarding the degree of polynomials $(r_j(X))_j$ and $(s_{k,l}(X))_{k,l}$ occurring in~\eqref{eq:WSoS}. The result of~\cite[sec.~5]{doherty2008quantum} is a hierarchy of relaxations allowing to get a sequence $q_n$ such that $\lim_{n \to \infty} q_n = q$, with $q_n \geq q$. For the level $n$ of the relaxation, we require that $r_j$ and $s_{k,l}$ are of degree at most $n$ and $n-1$, respectively.
	
	Consider the first level, $n=1$ in the CHSH scenario, with $G = A_1 B_1 + A_1 B_2 + A_2 B_1 - A_2 B_2$, where $E^a_x$ and $F^b_y$ are commuting projective measurement operators of Alice and Bob, respectively, see sec.~\ref{sec:probDistr}, and $A_x \equiv E^0_x - E^1_x$ and $B_y \equiv F^0_y - F^1_y$. The basis of polynomials of degree $1$ in variables occurring in $G$ is e.g. $x^{(1)} = [A_1; A_2; B_1; B_2]$. The SoS part of WSoS~\eqref{eq:WSoS} is thus $\sum_j r_j^\dagger r_j = x^{(1)T} M x^{(1)}$, cf.~\eqref{eq:SoS}, for some $M \succeq 0$. The only constraint expressed in $\mathcal{P}$ are those imposing that $A_x$ and $B_y$ have eigenvalues in $\{+1,-1\}$; let us denote the constraint polynomials as $p^{(A)}_x \in \mathcal{P}$ and $p^{(B)}_y \in \mathcal{P}$, respectively. Since $s_{k,l}$ are at this level all of degree $n-1=0$, they are real numbers; let us denote them as $\gamma^{(A)}_x \in \mathbb{R}$ and $\gamma^{(B)}_y \in \mathbb{R}$, respectively. The WSoS decomposition~\eqref{eq:WSoS} is now of the form:
	\begin{equation}
		\label{eq:WSoS_CHSH}
		q_1 \cdot \openone - (A_1 B_1 + A_1 B_2 + A_2 B_1 - A_2 B_2) = x^{(1)T} M x^{(1)} + \sum_{x \in [2]} \gamma^{(A)}_x A_x +  + \sum_{y \in [2]} \gamma^{(B)}_y B_y.
	\end{equation}
	To find the value $q_1$ we can solve the problem of minimizing $q_1$ subject to~\eqref{eq:WSoS_CHSH} and $M \succeq 0$; this clearly is an SDP. The result is
	\begin{equation}
		2 \sqrt{2} \cdot \openone - (A_1 B_1 + A_1 B_2 + A_2 B_1 - A_2 B_2) = \frac{1}{2 \sqrt{2}} (r_1^\dagger r_1 + r_2^\dagger r_2) + \frac{1}{\sqrt{2}} (A_1 + A_2 + B_1 + B_2),
	\end{equation}
	where $r_1 \equiv A_1 + A_2 - \sqrt{2} B_1$ and $r_2 \equiv A_1 - A_2 - \sqrt{2} B_2$. The Tsirelson bound is thus $2 \sqrt{2}$.
	
	% TODO: opisać SoS z Appendix https://arxiv.org/pdf/1009.1567.pdf
	
	% https://yalmip.github.io/tutorial/sumofsquaresprogramming/
	
% TODO: napisać tę sekcję, zwłaszcza na podstawie artykułów Watrous'a
%	\subsection{Norms}
%	
%	% tu przegląd norm macierzy przedstawialnych jako SDP, np. https://arxiv.org/pdf/1207.5726.pdf
%	
%	Trace norm optimization:
%	\begin{lstlisting}[language=MATLAB]
%		A = rand(3,4) + 1i * rand(3,4)
%		Y = sdpvar(4,4,'hermitian','complex')
%		Z = sdpvar(3,3,'hermitian','complex')
%		M = [Y A'; A Z]
%		optimize([M >= 0], 0.5*(trace(Y)+trace(Z)))
%	\end{lstlisting}

	\subsection{Lov{\'a}sz theta and contextuality}
	\label{sec:Lovasz}

% TODO: pracować nad tymi 3 uwagami:
%	najpierw wprowadzenie z Lov{\'a}sz, On the Shannon capcity of a graph%, (do Tw. 5 włącznie)
%	potem, że uogólnienie wg. \cite{GLS86} s.4
%	potem wprowadzić theta body \cite{GLS86}, s.8
	
	The concept of zero-error capacity plays a significant role in information theory as it pertains to the flawless transmission of information through a communication channel. This notion is of utmost importance as it guarantees the reliability and precision of data transfer, which holds immense significance in diverse domains, including telecommunications, computer networking, and cryptography.
	The notion of a zero-error capacity of a channel represented by a graph was introduced by Shannon in 1956 in the paper \textit{The zero error capacity of a noisy channel}~\cite{Shannon56}, as defined below.
	The calculation of this entity, unfortunately, poses significant challenges. Lov{\'a}sz addressed this problem in~\cite{Lovasz79} by formulating an SDP relaxation known as the Lov{\'a}sz $\theta$ function\index{Lov{\'a}sz theta}, or Lov{\'a}sz number\index{Lov{\'a}sz number}. The introduction of this function had a profound influence not only in classical and quantum information theories~\cite{citeLovasz1,citeLovasz2,citeLovaszQuantum} but also in related fields such as graph theory~\cite{Goemans98,KMS98}. Its impact transcends disciplinary boundaries, highlighting its importance and wide-ranging implications.
	
	Consider an alphabet consisting of $n$ letters that need to be communicated through an erroneous channel. We can represent this communication scenario using a graph $G$, where each vertex corresponds to a letter from the alphabet. The edges of the graph indicate the possible confusion between letters, based on the communication model being considered. It is evident that the count of one-letter messages that are guaranteed not to be confused is equivalent to the size of the largest independent set in the graph, denoted as $\alpha(G)$.
	
	In the context of zero-error communication in the asymptotic limit, where multiple uses of the communication channel and non-trivial coding schemes are allowed, it often becomes possible to transmit a higher average number of letters per channel use. To illustrate this concept, let us consider the number of $k$-letter messages that can be transmitted without confusion, denoted as $\alpha(G^{k})$. It is observed that $\alpha(G^{k}) \geq \alpha(G)^{k}$, indicating that the size of the largest independent set, $\alpha(G)$, raised to the power of $k$ provides a lower bound on the number of distinct messages that can be encoded without the risk of confusion. For instance, if a single-letter message allows for $l = \alpha(G)$ different messages without confusion, then with $k$ letters, we can encode at least $l^k$ distinct messages without the risk of confusion. As an example, consider the cycle graph $C_5$ with 5 vertices, where $\alpha(C_5) = 2$ and $\alpha(C_5^2) = 5$.
	
	The \textit{Shannon capacity}\index{Shannon capacity} of a graph $G$ is a measure defined as follows. It is denoted by $\Theta(G)$ and is given by the supremum over all values of $k$ of the expression $\alpha(G^{k})^{\frac{1}{k}}$, i.e. over all possible lengths of messages encoding a chunk of information:
	\begin{equation}
		\Theta(G) = \sup_{k} \alpha(G^{k})^{\frac{1}{k}}.
	\end{equation}
	Here, $\alpha(G^{k})$ represents the size of the largest independent set in the graph $G^{k}$, which is obtained by taking the strong product of $G$ with itself $k$ times. The Shannon capacity provides an important characterization of the graph's ability to transmit information without errors, and it is widely used in the field of information theory.
	
	To provide the formulation of SDP relaxation of the Shannon capacity, and thus also of the independence number of a graph, introduce the concept of the \textit{strong product}\index{strong product} of two graphs. Let us consider a pair of graphs, $G$ and $H$, and define their strong product as $G \boxtimes H$. The vertex set of $G \boxtimes H$, denoted as $V(G \boxtimes H)$, is the Cartesian product of the vertex sets of $G$ and $H$, i.e., $V(G \boxtimes H) = V(G) \times V(H)$. In this construction, a vertex $(x_1, y_1)$ in $G \boxtimes H$ is adjacent to another vertex $(x_2, y_2)$ if and only if one of the following conditions holds:
	\begin{itemize}
		\item vertex $x_1$ is adjacent to vertex $x_2$ in $G$, and vertex $y_1$ is adjacent to vertex $y_2$ in $H$;
		\item vertex $x_1$ is equal to vertex $x_2$, and vertex $y_1$ is adjacent to vertex $y_2$ in $H$; or 
		\item vertex $x_1$ is adjacent to vertex $x_2$ in $G$, and vertex $y_1$ is equal to vertex $y_2$.
	\end{itemize}
	This construction is known as the strong product of graphs. To extend this notion, we define $G^1 = G$, and for $k+1$, we have $G^{k+1} = G^k \boxtimes G$.
	
	An \textit{orthonormal representation}\index{orthonormal representation} (OR) of $G$ in $\mathbb{R}^d$ for some $d$ is a set of vectors $\{\ket{u_i}\}_{i \in V(G)} \subset \mathbb{R}^d$ satisfying $\bkc{u_i}{u_j} = 0$ for all pairs of non-adjacent vertices $i,j \in V(G)$. We denote by $\mathcal{OR}(G)$ the set of all OR of $G$ in any dimension. The \textit{value}\index{orthonormal representation!value} of the OR is
	\begin{equation}
		\label{eq:ORV}
		\min_{\substack{\ket{\Psi} \in \mathbb{R}^d, \\ \Ab{\ket{\Psi} = 1}}} \max_{i \in V(G)} \frac{1}{\Ab{\bkc{\Psi}{u_i}}^2}.
	\end{equation}
	Any $\Psi$ for that the minimum in~\eqref{eq:ORV} is attained, is called a \textit{handle}\index{orthonormal representation!handle} of OR.
	If $\{\ket{u_i}\}_{i \in V(G)}$ and $\{\ket{v_j}\}_{j \in V(H)}$ are OR of graphs $G$ and $H$, then $\{\ket{u_i} \otimes \ket{v_j}\}_{i \in V(G), j \in V(H)}$ is an OR of the graph $G \boxtimes H$~\cite[p.2]{Lovasz79}. If the vectors are in $\mathbb{C}^d$ instead of $\mathbb{R}^d$, then the term \textit{orthogonal embedding}\index{orthogonal embedding} is used instead of OR. The \textit{orthogonal rank}\index{orthogonal rank} of $G$, denoted $\xi(G)$ is the smallest positive integer $d$ such that there exists an orthogonal embedding~\cite{cameron2006quantum,briet2016orthogonal}.
	Lov{\'a}sz's function, denoted as $\theta(G)$, is defined as the minimum of~\eqref{eq:ORV} over $\mathcal{OR}(G)$.
	
	% $\theta$ function
	Lov{\'a}sz's $\theta(G)$ plays an important role in the study of the Shannon capacity. It possesses the property that the Shannon capacity $\Theta(G)$, is bounded from above by $\theta(G)$, $\Theta(G) \leq \theta(G)$. To provide an SDP formulation of $\theta(G)$, let's consider a set $\mathcal{A}$ consisting of all symmetric matrices $\{A_i\}_i$ that satisfy the following conditions: For any two nodes $i$ and $j$ in the graph $G$, if $i=j$ or if $i$ and $j$ are not adjacent in $G$, then the entry $A_{ij}$ is set to $1$. The remaining entries of these matrices are left unconstrained. The value of $\theta(G)$ is equal to the minimum of the largest eigenvalue among all matrices in the set $\mathcal{A}$~\cite[Theorem~3]{Lovasz79}. This relaxation technique provides an effective approach to approximate the Shannon capacity of a graph and is possible to be expressed as an SDP. Indeed, the constraints defining the set $\mathcal{A}$ are linear, and the SDP is
	\begin{align}
		\begin{split}
			\text{minimize } &\null \lambda \\
			\text{subject to } &\null X \in \mathcal{A} \\
			&\null \lambda \cdot \openone - X \succeq 0.
		\end{split}
	\end{align}
	Alternatively, it can be shown~\cite[Theorem~5]{Lovasz79} that 
	\begin{equation}
		\label{eq:Thm5Lovasz79}
		\theta(G) = \max_{\{\ket{u_i}\}_i \in \mathcal{OR}(\bar{G})} \max_{\substack{\ket{\Psi} \in \mathbb{R}^d, \\ \Ab{\ket{\Psi}} = 1}} \sum_{i \in V(G)} \Ab{\bkc{\Psi}{u_i}}^2,
	\end{equation}
	where $\bar{G}$ is the complementary graph of $G$.
	
	In~\cite{grotschel1981ellipsoid} Gr{\"o}tschel, Lov{\'a}sz and Schrijver introduced the weighted version of $\theta$ function, intending to derive the maximum weight independent sets in perfect graphs. Let $\mathbf{w} = (w_i)_{i \in V(G)}$ be weights of nodes in a weighted graph $G$. The generalization of the $\theta$ function~\eqref{eq:ORV} to weighted graphs is defined as~\cite[p.~4]{GLS86}:
	\begin{equation}
		\label{eq:wORV}
		\theta(G, \mathbf{w}) \equiv \min_{\{\ket{u_i}\}_i \in \mathcal{OR}(G)} \min_{\substack{\ket{\Psi} \in \mathbb{R}^d, \\ \Ab{\ket{\Psi}} = 1}} \max_{i \in V(G)} \frac{w_i}{\Ab{\bkc{\Psi}{u_i}}^2}.
	\end{equation}
	A direct generalization of~\eqref{eq:Thm5Lovasz79} is~\cite[Theorem~2.3]{GLS86}:
	\begin{equation}
		\label{eq:Thm23GLS86}
		\theta(G, \mathbf{w}) = \max_{\{\ket{u_i}\}_i \in \mathcal{OR}(\bar{G})} \max_{\substack{\ket{\Psi} \in \mathbb{R}^d, \\ \Ab{\ket{\Psi}} = 1}} \sum_{i \in V(G)} w_i \Ab{\bkc{\Psi}{u_i}}^2.
	\end{equation}
	The value of~\eqref{eq:Thm23GLS86} can also be expressed as SDP~\cite{grotschel1981ellipsoid}. % TODO: jak to policzyć jako SDP?
	
	The fundamental role of the Lov{\'a}sz $\theta$ for quantum correlations and contextuality was recognized in~\cite{cabello2010non} and developed in~\cite{cabello2014graph}. The results state that the maximal value of correlations allowed by quantum mechanics is given by the Lov{\'a}sz number of the so-called \textit{exclusivity graph}\index{exclusivity graph}. Here we briefly sketch the results, and we refer to~\cite{acin2015combinatorial} for a discussion and further advancements.
	
	The exclusivity graph\index{exclusivity graph} of a multi-partite correlation experiment represents the possible events with vertices and the exclusion of pairs of events by edges. We say that the events $e_1$ and $e_2$ are exclusive if and only if there exist two jointly measurable observables (tests) $\mu_i$ and $\mu_j$ that distinguish between them. The experiments with space-like separated tests are Bell inequalities~\cite{bell1964einstein,clauser1969proposed} as discussed in sec.~\ref{sec:probDistr}. More general scenarios are non-contextual inequalities, which distinguish between theories in which outcomes are predefined from contextual theories, including quantum mechanics~\cite{gleason1975measures,kochen1990problem,spekkens2005contextuality}. For instance, in the CHSH Bell experiment~\cite{clauser1969proposed} there are four tests, each providing binary results, \textit{viz.} two measurements performed by Alice, and two measurements performed by Bob.
	
\begin{figure}
		\centering
		\resizebox{0.65\textwidth}{!}{
			\begin{tikzpicture}
				
				% settings (x=0,y=0)
				\node[circle, fill] (v0000) at (0,8) {};
				\node[above left] at (v0000.west) {00|00};
				
				\node[circle, draw] (v0100) at (0,6) {};
				\node[below left] at (v0100.west) {01|00};
				
				\node[circle, draw] (v1000) at (2,8) {};
				\node[above right] at (v1000.east) {10|00};
				
				\node[circle, fill] (v1100) at (2,6) {};
				\node[below right] at (v1100.east) {11|00};

				% settings (x=1,y=0)
				\node[circle, draw] (v0110) at (6,8) {};
				\node[above left] at (v0110.west) {01|10};
				
				\node[circle, fill] (v0010) at (6,6) {};
				\node[below left] at (v0010.west) {00|10};
				
				\node[circle, fill] (v1110) at (8,8) {};
				\node[above right] at (v1110.east) {11|10};
				
				\node[circle, draw] (v1010) at (8,6) {};
				\node[below right] at (v1010.east) {10|10};

				% settings (x=1,y=1)
				\node[circle, fill] (v1011) at (6,2) {};
				\node[above left] at (v1011.west) {10|11};
				
				\node[circle, draw] (v1111) at (6,0) {};
				\node[below left] at (v1111.west) {11|11};
				
				\node[circle, draw] (v0011) at (8,2) {};
				\node[above right] at (v0011.east) {00|11};
				
				\node[circle, fill] (v0111) at (8,0) {};
				\node[below right] at (v0111.east) {01|11};

				% settings (x=0,y=1)
				\node[circle, fill] (v1101) at (0,2) {};
				\node[above left] at (v1101.west) {11|01};
				
				\node[circle, draw] (v1001) at (0,0) {};
				\node[below left] at (v1001.west) {10|01};
				
				\node[circle, draw] (v0101) at (2,2) {};
				\node[above right] at (v0101.east) {01|01};
				
				\node[circle, fill] (v0001) at (2,0) {};
				\node[below right] at (v0001.east) {00|01};

				% hyperedges
				\draw (0, 8) -- (8, 8) node[label={[xshift=50]right:{$y=0, b=x$}}] {}; % y=0, b=x
				\draw (0, 6) -- (8, 6) node[label={[xshift=50]right:{$y=0, b=\neg x$}}] {}; % y=0, b=\neg x
				\draw (0, 2) -- (8, 2) node[label={[xshift=50]right:{$y=1, b=\neg x$}}] {}; % y=1, b=\neg x
				\draw (0, 0) -- (8, 0) node[label={[xshift=50]right:{$y=1, b=x$}}] {}; % y=1, b=x
				
				\draw (0, 8) -- (0, 0) node[align=center, text width=2, label={[below, yshift=-30]:{$\begin{aligned}x&=0, \\ a&=y\end{aligned}$}}] {}; % x=0, a=y
				\draw (2, 8) -- (2, 0) node[align=center, label={[below, yshift=-30]:{$\begin{aligned}x&=0, \\ a&=\neg y\end{aligned}$}}] {}; % x=0, a=\neg y
				\draw (6, 8) -- (6, 0) node[align=center, label={[below, yshift=-30]:{$\begin{aligned}x&=1, \\ a&=y\end{aligned}$}}] {}; % x=1, a=y
				\draw (8, 8) -- (8, 0) node[align=center, label={[below, yshift=-30]:{$\begin{aligned}x&=1, \\ a&=\neg y\end{aligned}$}}] {}; % x=1, a=\neg y
				
				% same settings
				\draw (1,7) circle ({sqrt(2)});
				\draw (7,7) circle ({sqrt(2)});
				\draw (7,1) circle ({sqrt(2)});
				\draw (1,1) circle ({sqrt(2)});
				
			\end{tikzpicture}
			}
		\caption{Exclusivity graph for two-partite Bell scenario with two settings and two outcomes for each of the parties. The events are labeled as $ab|xy$, with the settings of Alice and Bob denoted as $x$ and $y$, and their outcomes as $a$ and $b$. Sets of pairwise exclusive events are those lying on the same line or in the same circle. The exclusivity graph for the positive non-contextual game expression~\eqref{eq:PGEchsh} is the induced subgraph containing only the black nodes.}
		\label{fig:CHSH_exclusivityGraph}
	\end{figure}
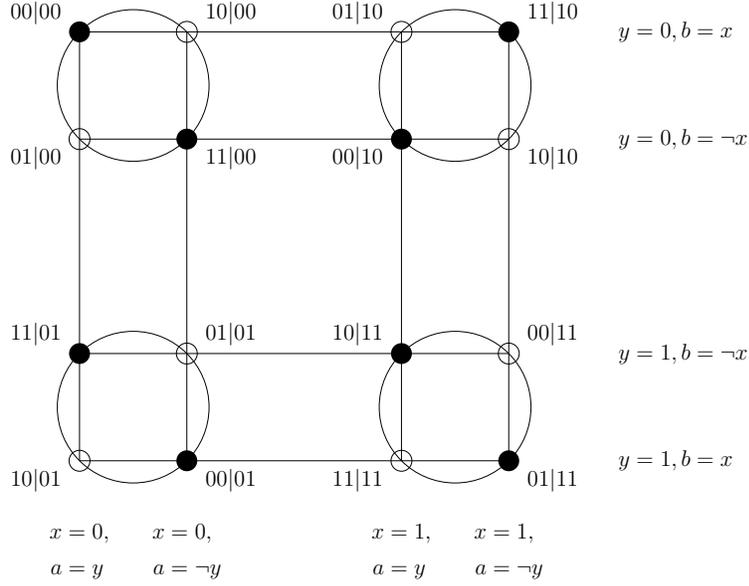

	Consider a positive linear combination of events, or positive non-contextual game expression\index{positive non-contextual game expression}, of the form $\sum_i w_i P(e_i)$, with all $w_i > 0$. The CHSH Bell inequality~\cite{clauser1969proposed} can be rewritten in this form as
	\begin{equation}
		\label{eq:PGEchsh}
		\sum_{a \in \{0,1\}} P(a,a|0,0) + P(a,a|0,1) + P(a,a|1,0) + P(a,\neg a|1,1)\leq 3.
	\end{equation}
	The exclusivity graph of the positive non-contextual game expression is the induced subgraph of the exclusivity graph of the experiment, see Fig.~\ref{fig:CHSH_exclusivityGraph}.
	In~\cite{cabello2014graph} it was shown that from~\eqref{eq:Thm23GLS86} it follows that the attainable upper bound on the positive non-contextual game expression in quantum mechanics is exactly $\theta(G, \mathbf{w})$, where $G$ is the exclusivity graph of the positive non-contextual game expression~\cite{sadiq2013bell}. % TODO: Result 2 z cabello2014graph i komentarze po nim, że to nie odnosi się do Bell
	
	% TODO: rozwinąć to, podać bezpośrednie wyniki
	The mentioned above orthogonal rank\index{orthogonal rank} $\xi(G)$ has a direct relation to the one-round quantum communication complexity of calculation of a function $f(x,y)$, which is equal to $\lceil \log_2(\xi(G)) \rceil$, with the promise that the joint input $(x,y) \in \mathcal{D} \subset V(G) \times \mathcal{Y}$. The vertices $i$ and $j$ in $G$ are connected if and only if $\bigexists_{y \in \mathcal{Y}} (i,y) \in \mathcal{D} \wedge (j,y) \in \mathcal{D}$~\cite[Theorem~8.5.2]{de2001quantum}. % 1608.06113p2
	As discussed in~\cite{Navascues2015a}, the so-called \textit{almost quantum}\index{almost quantum} or $\mathfrak{Q}_{1+AB}$ SDP relaxation of the set of quantum probabilities, see sec.~\ref{sec:NPA}, is also closely related to the Lov{\'a}sz $\theta$.

	\subsection{Correlation matrices, moment matrices, and optimization over non-commuting variables}
	\label{sec:NPA}
	
	% TODO
	% https://en.wikipedia.org/wiki/Moment_matrix
	% https://encyclopediaofmath.org/wiki/Moment_matrix
	% MONIQUE LAUREN, SUMS OF SQUARES, MOMENT MATRICES ANDOPTIMIZATION OVER POLYNOMIALS - https://homepages.cwi.nl/~monique/files/moment-ima-update-new.pdf
	% w [10.1103/PhysRevLett.111.030501] jako referencję do "matrix of moments for continuous variable systems" podają 28-32 a także 20-23 do moment matrces ogólnie
	
	Consider a sequence of real-valued random variables $\mathcal{S} = (x_1, \cdots x_n)$. The covariance matrix\index{covariance matrix} of $\mathcal{S}$ is defined as the matrix whose entries are given by relevant covariations of pairs of variables, $\Cov[\mathcal{S}] = \left[ \cov[x_i, x_j] \right]_{x_i, x_j \in \mathcal{S}} \in \mathbb{R}^{\mathcal{S} \times \mathcal{S}}$. The rows and columns in this case could be indexed with the numbers of the variables, and thus belong to $[n]$ and then it would be $\Cov[\mathcal{S}] \in \mathbb{R}^{n \times n}$. Instead, we take a more generic approach and index the rows and columns with the variables themselves. Such matrices, indexed by labels or other expressions are called \textit{moment matrices}\index{moment matrix}. Now, consider a vector of constant coefficients $v = (v_1, \cdots v_n) \in \mathbb{R}^{\mathcal{S} \times 1}$. We have
	\begin{equation}
		v^T \Cov[\mathcal{S}] v = \sum_{i,j \in [n]} v_i \cdot \cov[x_i, x_j] \cdot v_j = \cov \left[ \sum_{i \in [n]} x_i, \sum_{j \in [n]} x_j \right] = \var \left[ \sum_{i \in [n]} x_i \right] \geq 0,
	\end{equation}
	and thus the covariance matrix is PSD. The correlation matrix $\corr(\mathcal{S})$ is the matrix whose entries are given by relevant correlation of pairs of variables, i.e. $\Corr[\mathcal{S}] = \left[ \corr[x_i, x_j] \right]_{x_i \in \mathcal{S}, x_j \in \mathcal{S}} \in \mathbb{R}^{\mathcal{S} \times \mathcal{S}}$. Equivalently, the correlation matrix is equal to the covariance matrix of all variables rescaled to have variance $1$, $\Corr[\mathcal{S}] = \Cov[\bar{\mathcal{S}}]$, for $\bar{\mathcal{S}} = (x_1 / \standardDeviation{x_1}, \cdots x_n / \standardDeviation{x_n})$. In consequence, all diagonal values of the correlation matrix are $1$ and the matrix is also PSD. If variables are linearly independent, then it is positive definite. A sample numerical calculation with a correlation matrix is given in~\ref{app:corrYalmip}. The concept of moment matrices is used in the methods discussed further in this section.

%	\section{Moment matrices in quantum information}
%	\label{sec:momentMatricesInQI}

% TODO: napisać te sekcje
%	to chyba jest istotne: [https://link.springer.com/article/10.1007/s00220-019-03382-y]	
%	\subsection{Fidelity}
%	% poza innymi, to też się wiąże z SDP i ma 300 cytowań: https://arxiv.org/pdf/quant-ph/0008047.pdf
%	
%	\subsection{Quantum channels}
%	% o tym wystarczy wzmiankę, nie trzeba explicite: https://arxiv.org/pdf/2001.02668.pdf
%	
%	\subsection{Quantum proofs}
%	% długa książka jest tu: https://arxiv.org/pdf/1610.01664.pdf ale SDP dotyczy rozdział 4 i część 6

	\subsubsection{The Navascués-Pironio-Acín hierarchy}~\\

	Now, we provide a brief introduction to the optimization over non-commuting variables, with a concentration on the so-called Navascués-Pironio-Acín (NPA) method. The NPA method is based on SDP, as presented in the paper \textit{Bounding the Set of Quantum Correlations} by Navascués \textit{et al.} (2007)~\cite{NPA07} and~\cite{NPA08}. Moment matrices are a basic element of the technique. The concept was inspired by the seminal work by Lasserre~\cite{lasserre2001global}.
	
	The problem at hand is to find a way to characterize, at least approximately, the class of all quantum probability distributions $\mathfrak{Q}$ without resorting to the formalism of quantum theory. To this end, the NPA method introduces a hierarchy of SDP problems $\left\{\mathfrak{Q}_k\right\}_{k=1}^{\infty}$. Each level of the hierarchy corresponds to a specific SDP problem, where higher levels yield more accurate solutions. In other words, as we increase the level $k$, the set $\mathfrak{Q}_{k+1}$ becomes a subset of $\mathfrak{Q}_k$, $\mathfrak{Q}_{k+1} \subset \mathfrak{Q}_k$, providing a progressively better approximation of $\mathfrak{Q}$. However, as the level increases, the SDPs become more complex and computationally demanding.
	It is important to note that the hierarchy of SDP problems converges to the quantum set $\mathfrak{Q}$, meaning that the intersection of all sets in the hierarchy is equal to $\mathfrak{Q}$, \textit{viz.} $\cap_{k=1}^{\infty} \mathfrak{Q}_k = \mathfrak{Q}$. By considering all levels of the hierarchy, we can accurately capture the entire quantum set $\mathfrak{Q}$ without explicit involvement of the formalism of Hilbert spaces, and when we restrict considerations to a particular level, then we are able to effectively approximate the optimization over all quantum probability distributions $\mathfrak{Q}$ using SDP.
	
	A sequence of operators, which is formed by concatenating projective measurement operators, plays a crucial role in the context of quantum systems. Consider an illustrative example of such a sequence, denoted as $E^1_2 E^3_2 F^2_1 E^1_1$, consisting of four operators. The operators associated with Alice, denoted as $E^a_x$, commute with the operators corresponding to Bob, denoted as $F^b_y$. This allows us to rearrange the sequence without altering the original action of the operators. Thus, we can rewrite the sequence as $E^1_2 E^3_2 E^1_1 F^2_1$ by interchanging the operators while respecting the commutation relationship between Alice's and Bob's measurements. This reordering is made possible by the commutativity property exhibited by the operators belonging to Alice's and Bob's measurements.
	
	In a sequence of operators, we can exploit the orthogonality property $E^a_x E^{a^{\prime}}_x = 0$ and $F^b_y F^{b^{\prime}}_y = 0$ for $a \neq a^{\prime}$ and $b \neq b^{\prime}$. Applying this property and utilizing the commutation property, we can rearrange operators within the sequence. For instance, let us consider the expression $E^2_1 F^3_3 E^1_1$. By utilizing the commutation property, we can rearrange the operators as $E^2_1 E^1_1 F^3_3$, which equals zero due to the orthogonality between $E^2_1$ and $E^1_1$. Additionally, it is worth noting that since $E^a_x$ and $F^b_y$ are projectors, we have the property $(E^a_x)^k = E^a_x$ for any $k \geq 1$, and the same holds for $F^b_y$. This property further aids in simplifying the expressions involving repeated application of the projectors. To characterize the length of a sequence of operators\index{sequence of operators!length}, we define it as the minimum number of projectors required to represent the sequence. In this context, we consider the identity operator $\openone$ as the \textit{null sequence}\index{sequence of operators!null}, denoting no projectors, and its length is defined to be zero. This notion of length provides a measure of the complexity or number of steps involved in a sequence of operators.
	
	Let us consider an $n$-element set $\mathcal{S}$ consisting of sequences of operators. For instance, we can define $\mathcal{S}_{1+AB} = \left\{ \openone, E^a_x, F^b_y, E^a_x F^b_y \right\}_{a,b,x,y}$. Using the NPA method, we can construct a hierarchy of relaxations by choosing different sets of sequences. Specifically, we define a set $\mathcal{S}_k$ to be the set of all sequences of operators $\{E^a_x, F^b_y\}$ with the length of at most $k$. It is worth noting that $\mathcal{S}_{1+AB} = \mathcal{S}_1 \cup \{ E^a_x F^b_y \}$, where $\mathcal{S}_{1+AB}$ corresponds to the specific example mentioned above. Hierarchy of level $\mathfrak{Q}_{2}$ means that the set $\mathcal{S}$ consists of all sequences of measurement operators of length $2$, whereas in level $\mathfrak{Q}_{1+AB}$, $\mathcal{S}$ is a set of all sequences of length $1$ and sequences with one operator of Alice and one of Bob. $\mathfrak{Q}_{1+AB}$ revealed to be so efficient that it is called an \textit{almost quantum}\index{almost quantum} set of correlations~\cite{Navascues2015a}.

	The key idea of the NPA method can be summarized as follows. Consider a joint probability distribution $\{ P(a,b|x,y) \}$ and suppose that it is quantum. This means that there exists a specific realization involving a quantum state $\ket{\psi}$ and projective measurements $\{ E^a_x, F^b_y \}$ such that, for all settings $x$ and $y$, and outcomes $a$ and $b$, the following relation~\eqref{eq:PabxyPsiEF} holds, and expresses the probability of obtaining outcomes $a$ and $b$ when measurements $x$ and $y$ are performed on the quantum state $\ket{\psi}$. In the context of the NPA method, the notion of moment matrices\index{moment matrix} is used as follows. For any operators $O_i$ and $O_j$ belonging to the set $\mathcal{S}$, we define the element of the moment matrix as:
	\begin{equation}
		\label{eq:GammaIJ}
		\Gamma_{O_i, O_j} \equiv \bra{\psi} O_i^{\dagger} O_j \ket{\psi}.
	\end{equation}
	This equation establishes a connection between certain elements of the moment matrix and the joint probability distribution. Specifically, we have $\Gamma_{E^a_x, F^b_y} = P(a,b|x,y)$, which demonstrates that the elements of the moment matrix correspond to the probabilities of obtaining outcomes $a$ and $b$ for measurements $x$ and $y$. Additionally, we have the element $\Gamma_{\openone, \openone} = 1$, which represents the identity operator, indicating that its contribution to the moment matrix is unity, and $\ket{\psi}$ is normalized. This definition results in an $n \times n$ moment matrix, where the rows and columns are indexed by the elements of the set $\mathcal{S}$. Hence, the moment matrix serves as a representation of the moments associated with the considered joint probability distribution.
	
	We can observe that the elements of the moment matrix $\Gamma$ are subject to the following linear constraints.
	For any indices $i$, $j$, $k$, and $l$, the equality $\Gamma_{O_i,O_j} = \Gamma_{O_k,O_l}$ holds whenever the corresponding operators satisfy $O_i^{\dagger} O_j = O_k^{\dagger} O_l$, i.e. $O_i^{\dagger} O_j = O_k^{\dagger} O_l \implies \Gamma_{O_i,O_j} = \Gamma_{O_k,O_l}$. This constraint ensures that the inner products of identical operator sequences yield equal moments.
	Similarly, if $O_i^{\dagger} O_j$ results in the zero operator, then it implies that $\Gamma_{O_i,O_j} = 0$. This condition ensures that the moments associated with operator sequences resulting in the null operator\index{null operator} are also zero.
	These linear constraints provide necessary relations between the elements of the moment matrix, allowing us to impose consistency and capture important properties of the joint probability distribution.
	Positive semi-definiteness of $\Gamma$ is a direct consequence of~\eqref{eq:GammaIJ}. Indeed, let $v \in \mathbb{C}^n$. For $V = \sum_j v_j O_j$ we have
	\begin{equation}
		v^\dagger \Gamma v = \sum_{i,j} v_i^{*} \Gamma_{O_i,O_j} v_j = \sum_{i,j} v_i^{*} \bra{\psi} O_i^{\dagger} O_j \ket{\psi} v_j = \bra{\psi} V^{\dagger} V \ket{\psi} = \Ab{V \ket{\psi}}^2 \geq 0,
	\end{equation}
	and thus $\Gamma \succeq 0$.
	% TODO: odkomentować, gdy będzie gotowe: We provide an explicit example of the formulation of the NPA problem as an SDP in~\ref{App:NPA}.
	To summarize, we observe the following:
	\begin{itemize}
		\item The sets $\mathcal{S}_1 \subset \mathcal{S}_2 \cdots \subset \mathcal{S}_{\infty}$ form an increasing sequence, where each set contains longer sequences of measurement operators.
		\item The hierarchy levels $\mathfrak{Q}_1 \supset \mathfrak{Q}_2 \cdots \supset \mathfrak{Q}$ form a decreasing sequence, indicating a refinement of the approximation to the quantum set.
		\item The quantum set $\mathfrak{Q}$ is equal to the intersection of all levels $Q_k$, i.e. $\mathfrak{Q} = \bigcap_{k=1}^{\infty} Q_k$.
	\end{itemize}
	The final equality, which pertains to convergence to the quantum set, has been proven in~\cite{NPA08}. The sizes of the sets $\mathcal{S}_k$ and, consequently, the sizes of the $\Gamma$ matrices, grow exponentially, specifically as $O \left( (\Ab{A} \cdot \Ab{X} + \Ab{B} \cdot \Ab{Y})^k \right)$. In practice, sets beyond $Q_3$ are rarely utilized. The set $Q{1+AB}$ is generally sufficient for most purposes and is often referred to as the \textit{almost quantum set}\index{almost quantum}. A comparison of the primal and dual approaches for imposing the aforementioned operator constraints is presented in Tab.~\ref{tab:NPAsize}. A comprehensive discussion on this topic can be found in sec.~2.3.1 of~\cite{myThesis}. It is worth noting that when expressing the constraint of NPA optimizations, the parameter $m$ representing the size of the canonical SDP form in~\eqref{SDP-primal} and~\eqref{SDP-dual} is significantly smaller if we opt for the latter approach.
	
	% How fast does it grow?
	\begin{table}
		\begin{tabular}{|l|r|r|r||r|}
			\hline
			hierarchy level & \multicolumn{1}{l|}{n} & \multicolumn{1}{l|}{m-dual} & \multicolumn{1}{l|}{average density (dual)} & \multicolumn{1}{l|}{m-primal} \\ \hline
			$\mathfrak{Q}_2$ & 13 & 31 & 0.183 & 137 \\ \hline
			$\mathfrak{Q}_3$ & 25 & 61 & 0.098 & 563 \\ \hline
			$\mathfrak{Q}_4$ & 41 & 101 & 0.060 & 1579 \\ \hline
			$\mathfrak{Q}_5$ & 61 & 151 & 0.041 & 3569 \\ \hline
			$\mathfrak{Q}_6$ & 85 & 211 & 0.029 & 7013 \\ \hline
			$\mathfrak{Q}_7$ & 113 & 281 & 0.022 & 12487 \\ \hline
			$\mathfrak{Q}_8$ & 145 & 361 & 0.017 & 20663 \\ \hline
			$\mathfrak{Q}_9$ & 181 & 451 & 0.014 & 32309 \\ \hline
			$\mathfrak{Q}_{10}$ & 221 & 551 & 0.011 & 48289 \\ \hline
			$\mathfrak{Q}_{11}$ & 265 & 661 & 0.009 & 69563 \\ \hline
			$\mathfrak{Q}_{12}$ & 313 & 781 & 0.008 & 97187 \\ \hline
			$\mathfrak{Q}_{13}$ & 365 & 911 & 0.007 & 132313 \\ \hline
			$\mathfrak{Q}_{14}$ & 421 & 1051 & 0.006 & 176189 \\ \hline
			$\mathfrak{Q}_{15}$ & 481 & 1201 & 0.005 & 230159 \\ \hline
		\end{tabular}
		\caption{\label{tab:NPAsize} Comparison of sizes of SDP formulations of the levels of the NPA in a scenario with two parties, each with two binary measurements, when the constraints are expressed in terms of primal or dual canonical SDP forms.}
	\end{table}

% TODO: zrobić te subsekcje lub inaczej umieścić to w tekście
%	\subsubsection{Applications}~\\
%	
%	NPA found applications in device independent cryptography [], in particular in randomness certification [Pironio2010]. We give an example of randomness certification in~\ref{App:Randomess}.
%	
%	
%	\subsubsection{XOR games}~\\
%	
%	% czy coś tu będzie?
	
	There are multiple variants and extensions of NPA. In~\cite{pironio2010convergent} the inventors of NPA showed how to apply their techniques to the general problem of polynomial optimization over non-commuting variables.
	In~\cite{johnston2016extended} it was shown how to use NPA to analyze the so-called extended non-local games. These games involve three parties, \textit{viz.} Alice, Bob, and a referee. Initially, Alice and Bob share a tripartite quantum state with the referee. In these games, the conditions for Alice and Bob to win may depend not only on their answers to randomly selected questions but also on the outcomes of measurements performed by the referee on its portion of the shared quantum state.
	In a recent work~\cite{pozas2019bounding} a method for analysis of classical and quantum correlations in networks with causally independent parties was introduced, providing a way to use NPA in complex quantum networks. Another work analyzing generalizations of NPA for characterization of the quantum network correlations, together with convergence results was given in~\cite{renou2022two}, see~\cite{tavakoli2022bell} for an overview.
	
	The almost quantum\index{almost quantum} correlations are applied and discussed in~\cite{sainz2015postquantum,hoban2018channel}. In~\cite[Appendix A]{sainz2015postquantum} the NPA method has been modified to express a relaxation of the set of quantum assemblages~\cite{cavalcanti2015detection}, i.e. sets of unnormalized states towards which a multipartite state can be steered to~\cite{wiseman2007steering}. In this method, the $\Gamma$ matrix' entries are not numbers but matrices themselves allowing for introduction in particular the so-called \textit{almost quantum assemblages}, see~\cite{sainz2018formalism} for a physical definition and discussion for its applications.
	A hierarchy for analysis of quantum steering was also given in~\cite{kogias2015hierarchy}. The proposed method enables the derivation of steering witnesses for arbitrary families of quantum states. A framework for the analytical derivation of non-linear steering criteria was also presented.
	
	The work~\cite{moroder2013device} introduces the so-called Moroder's hierarchy\index{Moroder's hierarchy}, where in addition to the NPA constraints, more restrictions of a certain form regarding entanglement can be imposed on the quantum state. From Moroder's hierarchy a further variant, allowing to impose of constraints on Bob's measuring devices was given in~\cite{pusey2013negativity}.
	
	A prominent application of NPA is in device-independent quantum cryptography, in particular in quantum randomness certification. The initial methods used a single parameter, the Bell inequality violation, as the certificate for this task~\cite{pironio2010random,pironio2013security,fehr2013security}. In~\cite{nieto2014using,bancal2014more} it was shown how to modify NPA so that the full experimental statistics can be imposed as SDP constraints with the method called \textit{more randomness from the same data} or the Nieto-Silleras hierarchy\index{Nieto-Silleras hierarchy}. What is more, the method allow to use the dual optimization task to derive a new Bell expression suited to provide the most randomness from the particular experimental realization.
	
	An alternative approach to optimization over non-commutative polynomials is given in~\cite{burgdorf2013tracial}. In that work, the problem of minimization of a trace of a given polynomial function in non-commuting variables using SDP is considered. Next, in~\cite{klep2016constrained} a method for constrained trace and eigenvalue optimization of noncommutative polynomials was introduced. The results were used in~\cite{tavakoli2022informationally} to characterize the classical and quantum correlations that arise in prepare-and-measure experiments when communication is informationally restricted.
		
% TODO: może coś z tego rozwinąć do pełnej subsekcji?
%	\subsubsection{More randomness from the same data}
%	\label{sec:moreRandomness}
%	\cite{nieto2014using,bancal2014more}
%	
%	Task: We have experimental data, and want to find a Bell expression that will have a sens of a guessing probability and will be as low as possible for our data.
%	Idea: Take dual formulation, and optimize Bell coefficients for constraints given by your experiment.
%	
%	\subsubsection{Moroder's hierarchy}
%	\label{sec:Moroder}
%	\cite{moroder2013device}
%	
%	\subsubsection{Steering}
%	\label{sec:steering}
%	\cite{kogias2015hierarchy}
	
	\subsubsection{Optimization of von~Neumann entropy}~\\
	
	In particular, the NPA method can be used to calculate the lower bound of the von Neumann conditional entropy given any kind of knowledge (classical or quantum) that an eavesdropper may possess if is subject to the laws of quantum mechanics. The method uses the SDP representations of the logarithm function discussed in sec.~\ref{sec:SDPreps} together with the Gauss-Radau quadrature rule for the lower bound. Let $w_i$ and $t_i$ be the nodes and weights defined by this quadrature. 
	Specifically, this method can be employed to compute a lower bound on the von Neumann conditional entropy under the presence of an eavesdropper, considering both classical and quantum knowledge, while adhering to the principles of quantum mechanics. Let $A$, $B$, and $E$ denote the Hilbert spaces corresponding to Alice's, Bob's, and the eavesdropper's devices, respectively. The quadrature rule provides a set of nodes, $\{t_i\}_i$, and weights, $\{w_i\}_i$, that are utilized in the computation. The lower bound formula for the settings selection $x^{*}$ and $y^{*}$ is given as~\cite{Brown2022}:
	\begin{equation}
		\label{eq:opt}
		\sum_{i} c_i \left( \sum_{a,b=0,1} \inf_{\substack{Z_{a,b} \in B(Q_E), \\ \text{cond}(P)}} \left( 1 + \phi[E^a_{x^{*}}, F^b_{y^{*}}, Z_{a,b}, t_i] \right) \right),
	\end{equation}
	where $\phi[E^a_{x^{*}}, F^b_{y^{*}}, Z_{a,b}, t_i]$ is defined as
	\begin{equation}
		\Tr \left[ \rho_{ABE} \left( E^a_{x^{*}} \otimes F^b_{y^{*}} \otimes \left( Z_{a,b} + Z_{a,b}^{\dagger} + (1 - t_i) Z_{a,b} Z_{a,b}^{\dagger} \right) +  t_i \left( \openone_{AB} \otimes Z_{a,b} Z_{a,b}^{\dagger} \right) \right) \right].
	\end{equation}
	The expression $\text{cond}(P)$ means that the probability distribution $P(a,b|x,y)$ satisfies a set of linear constraints defined by the protocol, and $c_i$ are coefficients calculated from Gauss-Radau quadrature as $c_i \equiv w_i / (t_i \log(2))$. The $i$ index in the sum~\eqref{eq:opt} assumes values that index nodes in quadrature, skipping the last one.
	
	\subsubsection{Self-testing with SWAP method}~\\
	\label{sec:self-testing}
	
	The phenomenon of self-testing is characterized by the ability to assess both states and measurements of certain quantum devices in a black-box setting, relying solely on observed statistics without the need for prior device calibration. However, prior to the work~\cite{yang2014robust} the existing examples of self-testing are limited in their applicability, as they only provide meaningful assessments for devices that closely resemble the ideal case. In~\cite{yang2014robust} these limitations were overcome by adopting a novel approach to self-testing\index{self-testing}, utilizing an SDP hierarchy for the characterization of quantum correlations. This approach allows for a more comprehensive and robust assessment of quantum devices, enabling meaningful evaluations even in scenarios where the devices deviate from the ideal case.
	
	We illustrate the method with a two partite Bell scenario. Suppose we have gatherd experimental description of a device $P(a,b|x,y)$, and that we expect that these statistics should have been obtained using a particular quantum state $\ket{\bar{\psi}}_{AB}$ and projective measurements $\{ \bar{E}^a_x, \bar{F}^b_y \}$. We would like to have a quantitative way of estimating, how close is the actual state that was prepared in the laboratory, considered as a black box described only by the statistics, to the theoretical one $\ket{\bar{\psi}}_{AB}$. The work~\cite{yang2014robust} proposed a method to perform such self-testing where the content of the black box is hypothetized to be swapped with a trusted system in a thought experiment; the method itself is called SWAP\index{self-testing!SWAP method}. Suppose that it is possible to formulate four linear functions $\mathcal{F}_{E,\sigma_x}$, $\mathcal{F}_{E,\sigma_z}$, $\mathcal{F}_{F,\sigma_x}$, and $\mathcal{F}_{F,\sigma_z}$, such that $\mathcal{F}_{E,\sigma_x}[\{ \bar{E}^a_x \}] = \sigma_x$, $\mathcal{F}_{E,\sigma_z}[\{ \bar{E}^a_x \}] = \sigma_z$, $\mathcal{F}_{F,\sigma_x}[\{ \bar{F}^b_y \}] = \sigma_x$, and $\mathcal{F}_{F,\sigma_z}[\{ \bar{F}^b_y \}] = \sigma_z$. Recall the SWAP operator given in~\eqref{eq:SWAP}. For $d_S = 2$ it takes the form
	\begin{equation}
		\label{eq:SWAP_2}
		\begin{bmatrix} 1 & 0 & 0 & 0 \\ 0 & 0 & 1 & 0 \\ 0 & 1 & 0 & 0 \\0 & 0 & 0 & 1 \end{bmatrix} = U_{S_1 S_2}[\sigma_x] \cdot V_{S_1 S_2}[\sigma_z] \cdot U_{S_1 S_2}[\sigma_x],
	\end{equation}
	where
	\begin{subequations}
		\begin{equation}
			U_{S_1 S_2}[X] \equiv \openone_{2} \otimes \proj{0} + X \otimes \proj{1}, \text{ and}
		\end{equation}
		\begin{equation}
			V_{S_1 S_2}[X] \equiv \frac{1}{2} \left( (\openone_{2} + X) \otimes \openone_{2} + (\openone_{2} - X) \otimes \sigma_x \right).
		\end{equation}
	\end{subequations}
	The black box is possibly using some other quantum state $\ket{\psi}_{AB}$ and measurements $\{ E^a_x, F^b_y \}$ to implement $P(a,b|x,y)$, but if the considered scenario posseses the self-testing property, then one can expect that the state and measurements will be not very far from the theoretical ones $\ket{\bar{\psi}}$ and $\{ \bar{E}^a_x, \bar{F}^b_y \}$. Now, let us perform a thought experiment with hypothetized swapping of the black-box state $\ket{\psi}_{AB}$ with a trusted ancillary states $\ket{\bar{\phi}_1}_{A'}$ and $\ket{\bar{\phi}_2}_{B'}$ using the black-box measurements $\{ E^a_x, F^b_y \}$ together with trusted operations on the ancillas, where we denote by $A$ and $B$ their black-box subsystems, and by $A'$ and $B'$ the subsystems of their trusted ancillas. Let us consider the ancillas to be qubits, so that we can apply~\eqref{eq:SWAP_2} to each parties, Alice and Bob. Since the operations $\{ E^a_x\}$ on the black-box subsystem of Alice are expected to approximate to some extend $\{ \bar{E}^a_x\}$, we may expect that $\mathcal{F}_{E,\sigma_x}[\{ E^a_x \}] \approx \sigma_x$ and $\mathcal{F}_{E,\sigma_z}[\{ E^a_x \}] \approx \sigma_z$; and similarly for Bob, \textit{viz.} $\mathcal{F}_{F,\sigma_x}[\{ F^b_y \}] \approx \sigma_x$ and $\mathcal{F}_{F,\sigma_z}[\{ F^b_y \}] \approx \sigma_z$. Thus for:
	\begin{subequations}
		\label{eqs:SWAP_for_sigmas}
		\begin{equation}
			\mathcal{S}_{A A'} \equiv U_{A A'}\left[\mathcal{F}_{E,\sigma_x}[\{ E^a_x \}]\right] \cdot V_{A A'}\left[ \mathcal{F}_{E,\sigma_z}[\{ E^a_x \}] \right] \cdot U_{A A'}\left[\mathcal{F}_{E,\sigma_x}[\{ E^a_x \}]\right],
		\end{equation}
		\begin{equation}
			\mathcal{S}_{B B'} \equiv U_{B B'}\left[\mathcal{F}_{F,\sigma_x}[\{ F^b_y \}]\right] \cdot V_{B B'}\left[ \mathcal{F}_{F,\sigma_z}[\{ F^b_y \}] \right] \cdot U_{B B'}\left[\mathcal{F}_{F,\sigma_x}[\{ F^b_y \}]\right],
		\end{equation}
	\end{subequations}
	we have $SWAP(A, A') \approx \mathcal{S}_{A A'}$ and $SWAP(B, B') \approx \mathcal{S}_{B B'}$. Define $\mathcal{S}_{ABA'B'} \equiv \mathcal{S}_{A A'} \otimes \mathcal{S}_{B B'}$ (with proper ordering of the subsystems), and consider the hypothetical state:
	\begin{equation}
		\label{eq:SWAP_rho}
		\rho_{SWAP} \equiv \Tr_{AB} \left[ \mathcal{S}_{ABA'B'} \cdot \left( \proj{\psi}_{AB} \otimes \proj{\bar{\phi}_1}_{A'} \otimes \proj{\bar{\phi}_2}_{B'} \right) \cdot \mathcal{S}_{ABA'B'}^\dagger \right].
	\end{equation}
	Knowing the explicit form of the trusted ancillas $\ket{\bar{\phi}_1}_{A'}$ and $\ket{\bar{\phi}_2}_{B'}$ and applying algebraic calculations with~\eqref{eqs:SWAP_for_sigmas} on $\ket{\psi}_{AB}$ we can derive that the formula for $\rho_{SWAP}$ depending on the elements possible to be retrieved from the NPA moment matrix, see sec.~\ref{sec:NPA}. This way e.g. the fidelity of the state or other linear functions of its element can be optimized with the same technique as NPA, as proposed in the seminal paper~\cite{yang2014robust}, or NV (see sec.~\ref{sec:NV}) as shown in~\cite{tavakoli2018self}.
	
	The SWAP method has found multiple applications, in particular in analysis of experimental data.
	The work~\cite{coladangelo2017all} used the SWAP method to show that for every bipartite entangled quantum state in arbitrary dimension, there exist a set of joint probability distributions $\{P(a,b|x,y)\}$ (see sec.~\ref{sec:probDistr}) allowing for self-testing of the state.
	In~\cite{wang2018multidimensional} the method was applied to self-test arbitrary qutrit states of the form $(2 + \gamma^2)^{-\frac{1}{2}} (\ket{00} + \gamma \ket{11} + \ket{22})$ and used for analysis of a large scale quantum optical circuitry.

	\subsection{Non-commuting variables with dimension constraints}
	\label{sec:dimConstraint}
	
	In this section, we will explore two distinct approaches that leverage moment matrices for optimizing over states and operators of a fixed dimension. The first approach, discussed in Section~\ref{sec:MLP}, builds directly upon the NPA method and incorporates the dimension constraint as additional linear constraints. On the other hand, the second technique, presented in Section~\ref{sec:NV}, shares a similar structure with NPA but adopts a randomized approach to construct the basis of the space of SDP variables.
	
	\subsubsection{Dimension constraints imposed on Navascués-Pironio-Acin hierarchy}~\\
	\label{sec:MLP}
	
	We present our method, which was introduced and developed in our previous works~\cite{li2013relationship,mironowicz2014properties}, and is reffered as Mironowicz-Li-Pawłowski (MLP) hierarchy. This method enables the analysis of Semi-Device-Independent~\cite{pawlowski2011semi} scenarios using the powerful techniques of SDP by reducing the problem to a Device-Independent~\cite{mayers1998quantum} framework modeled in the NPA hierarchy. 
	
	To introduce the SDP relaxation, we first consider a device $\textbf{D0}$, consisting of two distinct black boxes assigned to Alice and Bob, respectively. We know the dimension of the messages exchanged between the two parts. The box of Alices generates and emits quantum states from some $\{\rho_{x}\}_{x \in \bar{X}}$. Bob is provided with a separate device that includes settings corresponding to measurement choices involving measurements denoted as $\left\{ \{M^b_y\}_{b \in \bar{B}} \right\}_{y \in \bar{Y}}$. We denote the conditional probability of obtaining an outcome $b$ when settings $x$ and $y$ are selected as $P_{\textbf{D0}}(b|x,y)$. Suppose we are provided with a dimension witness $W$ in the form presented in~\eqref{eq:DW} earlier. Assume this dimension witness yields an average value of $W_0$ in experiments conducted on the $\textbf{D0}$ device.
	
	Although we do not know the specification of the device $\textbf{D0}$, we can consider an alternative device denoted as $\textbf{D1}$, as follows. The device $\textbf{D1}$ comprises two components, each equipped with buttons labeled the same way as in $\textbf{D0}$. In $\textbf{D1}$ we assume that both parts share a singlet state of dimension $d$. Alice's component performs a projective measurement with outcomes $0$ (indicating successful projection) or $1$ (otherwise), depending on the chosen input $x$. This measurement projects Alice's part of the singlet onto the state $\rho_{x}$, which corresponds to the relevant state in the device $\textbf{D0}$. If the projection succeeds, which occurs with a probability of $\frac{1}{d}$, the device returns $a = 0$ and transforms Alice's side into the state $\rho_{x}$. Otherwise, it returns $a = 1$. Since the shared state is a singlet, this measurement prepares the same $d$-dimensional state on Bob's side. Bob subsequently performs the same measurements $\{M^b_y\}_{b}$ as the device $\textbf{D0}$ would perform, and he returns the outcome $b$. Let us denote the probability that Alice obtains outcome $a$ with setting $x$, while Bob obtains outcome $b$ with setting $y$, as $P_{\textbf{D1}}(a,b|x,y)$. The case where $a = 1$ giving the probabilities $P_{\textbf{D1}}(1,b|x,y)$ are not utilized in our relaxation. Trivially, we have:
	\begin{equation}
		P_{\textbf{D0}}(b|x,y) = d \cdot P_{\textbf{D1}}(0,b|x,y).
	\end{equation}
	
	Let's now turn our attention to a third device, denoted as $\textbf{D2}$, which shares the same interface as $\textbf{D1}$. In contrast to the previous devices, the internal workings of $\textbf{D2}$ are unrestricted, meaning that we make no assumptions about the measurements performed. Both Alice's and Bob's parts are now allowed to be in an arbitrary state $\rho$ of any dimension. For this device, we give an additional constraint
	\begin{equation}
		\label{eq:MLPdimConstr}
		\forall_{x} P_{\textbf{D2}}(0|x) = \frac{1}{d},
	\end{equation}
	where $P_{\textbf{D2}}(a|x)$ is the probability of getting the outcome $a$ by Alice with the setting $x$ with the device $\textbf{D2}$. It is important to highlight that the description of this device falls under the category of Device-Independent\index{device-independent} scenarios. Consequently, we can utilize the NPA method to mathematically represent its behavior, specifically the probability distributions exhibited by device $\textbf{D2}$.
	In the case of the third device, we can represent the probability of obtaining outcomes $a$ and $b$ for a given combination of settings $x$ and $y$ for Alice and Bob, respectively, as $P_{\textbf{D2}}(a,b|x,y)$.
	
	It is evident that all the conditional probability distributions achievable by the device $\textbf{D1}$, and equivalently by device $\textbf{D0}$, can also be obtained by device $\textbf{D2}$. Furthermore, as device $\textbf{D2}$ is a relaxed variant of the original device $\textbf{D0}$. It is important to note that one of the key characteristics of the set of joint probability distributions is its efficient approximation using the NPA hierarchy. An advantageous aspect of this method is that it provides a bound for any dimension of the communicated system, with the linear bound being the only parameter that needs adjustment. We conclude that using the relation $P(b|x,y) = d \cdot P(0,b|x,y)$, we can impose a dimension constraint to an existing implementation of the NPA as in~\eqref{eq:MLPdimConstr}, i.e. $P(0|x) = 1/d$.

	\subsubsection{Navascués-Vertesi hierarchy}~\\
	\label{sec:NV}
	
	Navascués and Vertesi (2014) proposed~\cite{MiguelVertesi,Navascues2015} a hierarchy of SDPs aimed to upper bound quantum correlation in scenarios with dimension constraint, where similarly as in NPA, the improved accuracy is obtained when increasing the hierarchy level. The Navascués-Vertesi hierarchy (NV) is particularly useful in dimension bounded scenarios, where the number of dimensions of the quantum systems involved is limited. In such scenarios, the full characterization of quantum correlations becomes computationally challenging due to the exponential growth of the dimension. NV provides a systematic and tractable approach to approximate and quantify quantum correlations in these scenarios.
	
	Consider a sequence of operators representing states and measurements, denoted as $\mathcal{S} = (O_1, \dots, O_n)$, where each operator corresponds to a specific quantum state or measurement or their polynomial function, for some $N$. For instance, we can have a sequence that includes identity operator $\openone$, pure states $\rho_{00}$, $\rho_{01}$, $\rho_{10}$, $\rho_{11}$, and measurement projectors $P^1_0$, $P^1_1$, $P^2_0$, $P^2_1$. It is important to note that all states in the sequence are pure and all measurements are projectors with fixed rank.
	In NV method, a crucial component is the moment matrix denoted as $M = [M]_{\mathcal{S},\mathcal{S}}$. This matrix is indexed by the sequence $\mathcal{S}$ and is defined similarly to the moment matrix used in the NPA. The entries of the moment matrix are given by the inner products of the operators in the sequence, specifically $M_{O_i, O_j} = \Tr(O_i^{\dagger} O_j)$.
	It is worth noting that when the sequence $\mathcal{S}$ contains both states and measurements, the moment matrix will have entries that correspond to the probabilities $P(b|x,y)$ of obtaining outcome $b$ when performing measurement $M^y_b$ on the state $\rho_x$. This allows the moment matrix to capture the statistical information of the correlations between states and measurements, providing a framework for characterizing and quantifying the quantum correlations present in the system.

	The implementation of NV method involves a randomized approach to construct a set of moment matrices. The first step is to randomize the moment matrices, which will define the optimization problem in the SDP framework. The goal is to optimize a linear combination $G$ of the entries $P(b|x,y)$, $G = \sum_{b,x,y} \beta_{b,x,y} P(b|x,y)$ for some fixed $\{ \beta_{b,x,y} \} \subset \mathbb{R}$, which may, for instance, correspond to the average success probability of a quantum random access code. To provide more detailed steps, the implementation begins with an initialization phase where a basis of moment matrices is created:
	\begin{enumerate}
		\item The set $\mathcal{M}$ is initialized as an empty set to store the moment matrices, $\mathcal{M} = \emptyset$.
		\item Operators in the sequence, specific to the given hierarchy level, are randomized to form the sequence $\mathcal{S}$.
		\item The moment matrix $\tilde{\Gamma}$ is constructed by evaluating the inner products of the operators in the randomized sequence $\mathcal{S}$, $\tilde{\Gamma} \equiv \left[ \Tr(O_i^{\dagger} O_j) \right]_{\mathcal{S},\mathcal{S}}$.
		\item The matrix $\tilde{\Gamma}^{\perp}$ is obtained by projecting $\tilde{\Gamma}$ onto the subspace orthogonal to the span of the moment matrices in $\mathcal{M}$.
		\item If $\tilde{\Gamma}^{\perp}$ is a zero matrix, the process of extending the basis of moment matrices $\mathcal{M}$ is terminated.
		\item Otherwise, the orthogonalized moment matrix $\tilde{\Gamma}^{\perp}$ is added to the set $\mathcal{M}$, $\mathcal{M} = \mathcal{M} \cup \{\tilde{\Gamma}^{\perp}\}$. The process returns to step 2 to generate the next randomized sequence.
	\end{enumerate}
	By iteratively adding orthogonalized moment matrices to $\mathcal{M}$, the NV implementation constructs a basis of moment matrices that captures the possible correlations in the considered system. These moment matrices form the foundation for the subsequent SDP optimization, where the objective is to find the optimal values for the entries $P(b|x,y)$ in order to maximize the value of $G$. The optimization has the form:
	\begin{align}
		\begin{split}
			\text{maximize } &\null \Tr \left[ \hat{\mathcal{B}} \Gamma \right] \\
			\text{subject to } &\null \Gamma \in \text{span}\left( \mathcal{M} \right), \\
			&\null \left(\Gamma\right)_{\openone, \openone} = 1, \\
			&\null \Gamma \succeq 0,
		\end{split}
	\end{align}
	where we call $\hat{\mathcal{B}}$ the \textit{game matrix} of the expression $G$, and construct it to select from $\Gamma$ the relevant values $\Tr[\rho_x M^y_b]$ with coefficients defined by probability functional we are considering.
	% TODO: odkomentować, gdy sekcja {App:MUBs} będzie gotowa
	%An example of problem solved with NV is given in the Appendix~\ref{App:MUBs}.
	The hierarchy can get a significant boost of performance when symmetries of $G$ are exploited~\cite{aguilar2018connections,tavakoli2019enabling}.

	\subsection{The see-saw iterative non-linear optimization}
	\label{sec:seesaw}
	
	The see-saw method is an iterative optimization technique used in SDP to find approximate solutions for certain optimization problems~\cite{seesaw1,seesaw2}. It is particularly effective for problems involving quantum states and measurements. The method involves alternating optimization over states and measurements, refining the solutions at each iteration until convergence is achieved.	
	In the see-saw method, the optimization problem is first formulated as an SDP, where the objective function and constraints are expressed in terms of quantum states and measurements. The method starts with an initial guess for either the states or the measurements, which is usually a simple randomization. In each iteration, the method optimizes over the states while keeping the measurements fixed, and then optimizes over the measurements while keeping the states fixed. This alternating optimization continues until a convergence criterion is met, such as a small change in the objective function or constraints. The see-saw method leverages the interplay between states and measurements in quantum systems. By iteratively optimizing over states and measurements, the method explores different combinations that lead to improved solutions. When a set of states is given, the expression~\eqref{eq:DW} can be optimized using a similar approach as in Quantum State Discrimination\index{Quantum State Discrimination} (QSD)~\cite{helstrom1969quantum,ivanovic1987differentiate,bae2015quantum} by employing SDPs, see~\ref{App:QSD}. Similarly, if measurements are provided, SDP can be utilized with states as variables to optimize the expression. In cases where neither states nor measurements are known, SDP can be employed to simultaneously find optimal solutions for both. The see-saw method operates based on the following outline:
	\begin{enumerate}
		\item Initially, a set of states is chosen randomly as an initial guess.
		\item Using SDP, the method optimizes over measurements while keeping the states fixed. The objective is to find measurements that maximize the target expression.
		\item Next, the method optimizes over states while keeping the measurements fixed. It uses SDP to find states that maximize the target expression.
		\item The process iterates by returning to step 2 if certain stopping criteria are not satisfied. The stopping criteria could be based on the convergence of the objective function or other specified conditions.
	\end{enumerate}
	It is important to note that the see-saw method does not guarantee finding the global optimum of the optimization problem. Instead, it provides an approximate solution that can be improved iteratively. To enhance the chances of finding better solutions, the method suggests restarting the process multiple times with different initial states. By repeating the see-saw method with various initial states, the hope is to explore different regions of the solution space and potentially find better solutions. Although the method may not guarantee the global optimum, it offers a practical approach for approximating the optimal solution to the optimization problem at hand.
	% TODO: odkomentować, gdy będzie gotowe: An example of a see-saw implementation in Matlab is given in~\ref{App:Seesaw}.
	
	% TODO: odkomentować, gdy sekcja {App:MUBs} będzie gotowa
	%A more complicated example of problem solved with NV is given in the Appendix~\ref{App:MUBs}.
	
	% TODO: \paragraph{Channel inside}
	% Choi, jak z Debą
	% Example of see-saw with channel: DQRACs - see directory KCIK\talks\Siegen2019\code_examples\seesaw\DQRACs

	% TODO: dokończyć np. według \cite{bamps2015sum}
	% A different approach for self-testing is based on SoS decomposition described in sec.~\ref{sec:SoS}. The method was developed in~\cite{yang2013robust,bamps2015sum}. Consider a Bell operator $I = \sum_{a,b,x,y} E^a_x \cdot F^b_y$ with commuting projective measurements $\{E^a_x\}$ and $\{F^b_y\}$.
	
	% TODO: opisać SoS dla self-testing
	%	\subsubsection{SoS}~\\
	%	
	%	https://arxiv.org/pdf/1904.10042.pdf box 4.1 (s. 14) - i przejrzeć artykuł
	%	może też np. https://arxiv.org/pdf/1504.06960.pdf

	\section{Conclusions}
	
	In conclusion, this paper has delved into the realm of SDP within the context of quantum information, offering a comprehensive exploration of their mathematical foundations and practical applications. By elucidating the concepts of convex optimization, duality, and SDP formulations, the study has equipped researchers and practitioners with powerful tools to address optimization challenges in quantum systems. The insights gained from the research of SDP have proven invaluable in advancing the field of quantum information, enabling the characterization and manipulation of quantum correlations, optimization of quantum states, and design of efficient quantum algorithms and protocols. The practical implementation of SDP discussed in the paper, has empowered researchers to effectively formulate and solve optimization problems in quantum systems, fostering the development of more efficient quantum communication protocols, self-testing methods, and a deeper understanding of quantum entanglement. Overall, this study has elaborated on the intersection of optimization and quantum information, paving the way for further advancements and discoveries in this exciting field.

	\section*{List of Abbreviations}
	\addcontentsline{toc}{section}{List of Abbreviations}
	
	\begin{abbrv}
		\item[DPS] Doherty-Parillo-Spedalieri
		\item[IPM] interior point method
		\item[LP] linear programming
		\item[LMI] linear matrix inequalities
		\item[MLP] Mironowicz-Li-Pawłowski
		\item[NPA] Navascués-Pironio-Acín
		\item[NV] Navascués-Vertesi
		\item[PD] positive definite
		\item[PPT] positive partial transpose
		\item[PSD] positive semi-definite
		\item[QSD] quantum state discrimination
		\item[SDP] semi-definite programming
		\item[SoS] sum of squares
	\end{abbrv}

	\section*{Acknowledgements}
	
	The work on this review started in 2019. The current support from the Knut and Alice Wallenberg Foundation through the Wallenberg Centre for Quantum Technology (WACQT), the Swedish research council (VR), and NCBiR QUANTERA/2/2020 (www.quantera.eu) an ERA-Net co-fund in Quantum Technologies under the project eDICT is acknowledged. In the period 2019-2022 the work was partially conducted at the Department of Algorithms and System Modeling at Gdańsk University of Technology, and was also supported by the Foundation for Polish Science (IRAP project, ICTQT, contract no. 2018/MAB/5, co-financed by EU within Smart Growth Operational Programme).
	
	Recently we learned about an independent work pertaining to SDP in quantum information~\cite{skrzypczyk2023semidefinite}. The content of the book and the current work is to a large extent complimentary. The current work attempts to sketch a more general framework to provide the context of SDP, current bibliographical and frontiers notes, endows a detailed treatment of moment matrices with non-commuting variables, SDP representations, SoS decompositions and Lov{'a}sz theta topics, and covers the discussion of implementations of solvers and is intended primarily for active researchers in quantum information.\newline
	
	\bibliographystyle{siam}

\begin{thebibliography}{100}

\bibitem{mosek21}
{\em {MOSEK Modeling Cookbook}}, MOSEK ApS, 3.2.3~ed., 2021.

\bibitem{MATLAB2022}
{\em {MATLAB 9.13.0.2126072 (R2022b) Update 3}}, The MathWorks Inc., Natick,
  Massachusetts, 2022.

\bibitem{acin2015combinatorial}
{\sc A.~Ac{\'\i}n, T.~Fritz, A.~Leverrier, and A.~B. Sainz}, {\em A
  combinatorial approach to nonlocality and contextuality}, Communications in
  Mathematical Physics, 334 (2015), pp.~533--628.

\bibitem{agarwal2022brief}
{\sc H.~Agarwal and I.~Garg}, {\em {A Brief Review of Operator Monotone and
  Operator Convex Functions}}, in Journal of Physics: Conference Series,
  vol.~2267, IOP Publishing, 2022, p.~012087.

\bibitem{agresti2021experimental}
{\sc I.~Agresti, B.~Polacchi, D.~Poderini, E.~Polino, A.~Suprano,
  I.~{\v{S}}upi{\'c}, J.~Bowles, G.~Carvacho, D.~Cavalcanti, and F.~Sciarrino},
  {\em Experimental robust self-testing of the state generated by a quantum
  network}, PRX Quantum, 2 (2021), p.~020346.

\bibitem{aguilar2018connections}
{\sc E.~A. Aguilar, J.~J. Borka{\l}a, P.~Mironowicz, and M.~Paw{\l}owski}, {\em
  Connections between mutually unbiased bases and quantum random access codes},
  Physical Review Letters, 121 (2018), p.~050501.

\bibitem{A91}
{\sc F.~Alizadeh}, {\em Combinatorial optimization with interior point methods
  and semi-definite matrices}, Ph. D. thesis, University of Minnesota,  (1991).

\bibitem{A95}
\leavevmode\vrule height 2pt depth -1.6pt width 23pt, {\em Interior point
  methods in semidefinite programming with applications to combinatorial
  optimization}, SIAM journal on Optimization, 5 (1995), pp.~13--51.

\bibitem{AHO98}
{\sc F.~Alizadeh, J.-P.~A. Haeberly, and M.~L. Overton}, {\em Primal-dual
  interior-point methods for semidefinite programming: convergence rates,
  stability and numerical results}, SIAM Journal on Optimization, 8 (1998),
  pp.~746--768.

\bibitem{ando1978topics}
{\sc T.~Ando}, {\em Topics on operator inequalities}, Ryukyu University.
  Lecture Note Series,  (1978).

\bibitem{anjos2011handbook}
{\sc M.~F. Anjos and J.~B. Lasserre}, {\em Handbook on semidefinite, conic and
  polynomial optimization}, vol.~166, Springer Science \& Business Media, 2011.

\bibitem{aspect1982experimental2}
{\sc A.~Aspect, J.~Dalibard, and G.~Roger}, {\em Experimental test of {B}ell's
  inequalities using time-varying analyzers}, Physical Review Letters, 49
  (1982), p.~1804.

\bibitem{aspect1981experimental}
{\sc A.~Aspect, P.~Grangier, and G.~Roger}, {\em Experimental tests of
  realistic local theories via {B}ell's theorem}, Physical Review Letters, 47
  (1981), p.~460.

\bibitem{aspect1982experimental}
\leavevmode\vrule height 2pt depth -1.6pt width 23pt, {\em Experimental
  realization of {E}instein-{P}odolsky-{R}osen-{B}ohm {G}edankenexperiment: a
  new violation of {B}ell's inequalities}, Physical Review Letters, 49 (1982),
  p.~91.

\bibitem{baccari2017efficient}
{\sc F.~Baccari, D.~Cavalcanti, P.~Wittek, and A.~Ac{\'\i}n}, {\em Efficient
  device-independent entanglement detection for multipartite systems}, Physical
  Review X, 7 (2017), p.~021042.

\bibitem{bae2015quantum}
{\sc J.~Bae and L.-C. Kwek}, {\em Quantum state discrimination and its
  applications}, Journal of Physics A: Mathematical and Theoretical, 48 (2015),
  p.~083001.

\bibitem{bamps2015sum}
{\sc C.~Bamps and S.~Pironio}, {\em Sum-of-squares decompositions for a family
  of {Clauser-Horne-Shimony-Holt}-like inequalities and their application to
  self-testing}, Physical Review A, 91 (2015), p.~052111.

\bibitem{bancal2014more}
{\sc J.-D. Bancal, L.~Sheridan, and V.~Scarani}, {\em More randomness from the
  same data}, New Journal of Physics, 16 (2014), p.~033011.

\bibitem{beavis1990optimisation}
{\sc B.~Beavis and I.~Dobbs}, {\em Optimisation and stability theory for
  economic analysis}, Cambridge University Press, 1990.

\bibitem{beck2014introduction}
{\sc A.~Beck}, {\em Introduction to nonlinear optimization: Theory, algorithms,
  and applications with {MATLAB}}, SIAM, 2014.

\bibitem{bell1964einstein}
{\sc J.~S. Bell}, {\em On the {E}instein {P}odolsky {R}osen paradox}, Physics,
  1 (1964), p.~195.

\bibitem{bell2004speakable}
{\sc J.~S. Bell}, {\em Speakable and unspeakable in quantum mechanics:
  Collected papers on quantum philosophy}, Cambridge University Press, 2004.

\bibitem{ben2011lectures}
{\sc A.~Ben-Tal and A.~Nemirovski}, {\em Lectures on modern convex optimization
  (2012)}, SIAM, Philadelphia, PA. Google Scholar Google Scholar Digital
  Library Digital Library,  (2011).

\bibitem{bengtsson2017geometry}
{\sc I.~Bengtsson and K.~{\.Z}yczkowski}, {\em Geometry of quantum states: An
  introduction to quantum entanglement}, Cambridge University Press, 2017.

\bibitem{DSDP}
{\sc S.~J. Benson and Y.~Ye}, {\em {DSDP5} user guide-software for semidefinite
  programming}, Tech. Rep. ANL/MCS-P1289-0905, Argonne National Lab.(ANL),
  Argonne, IL (United States), 2006.

\bibitem{bernards2020generalizing}
{\sc F.~Bernards and O.~G{\"u}hne}, {\em Generalizing optimal {B}ell
  inequalities}, Physical Review Letters, 125 (2020), p.~200401.

\bibitem{bhatia2009positive}
{\sc R.~Bhatia}, {\em Positive definite matrices}, in Positive Definite
  Matrices, Princeton University Press, 2009.

\bibitem{bochnak2013real}
{\sc J.~Bochnak, M.~Coste, and M.-F. Roy}, {\em Real algebraic geometry},
  vol.~36, Springer Berlin, Heidelberg, 2013.

\bibitem{CSDP}
{\sc B.~Borchers}, {\em {CSDP}, a {C} library for semidefinite programming},
  Optimization methods and Software, 11 (1999), pp.~613--623.

\bibitem{BL06}
{\sc J.~M. Borwein and A.~S. Lewis}, {\em Convex {A}nalysis and {N}onlinear
  {O}ptimization}, CMS Books in Mathematics, Springer New York, 2006.

\bibitem{BZ05}
{\sc J.~M. Borwein and Q.~J. Zhu}, {\em Techniques of {V}ariational
  {A}nalysis}, CMS Books in Mathematics, Springer New York, 2005.

\bibitem{borwein2006variational}
\leavevmode\vrule height 2pt depth -1.6pt width 23pt, {\em Variational methods
  in convex analysis}, Journal of Global Optimization, 35 (2006), pp.~197--213.

\bibitem{boyd1994linear}
{\sc S.~Boyd, L.~El~Ghaoui, E.~Feron, and V.~Balakrishnan}, {\em Linear matrix
  inequalities in system and control theory}, SIAM, 1994.

\bibitem{Boyd04}
{\sc S.~P. Boyd and L.~Vandenberghe}, {\em Convex optimization}, Cambridge
  University Press, 2004.

\bibitem{brandao2004robust}
{\sc F.~G. S.~L. Brandao and R.~O. Vianna}, {\em Robust semidefinite
  programming approach to the separability problem}, Physical Review A, 70
  (2004), p.~062309.

\bibitem{briet2016orthogonal}
{\sc J.~Bri{\"e}t and J.~Zuiddam}, {\em On the orthogonal rank of {C}ayley
  graphs and impossibility of quantum round elimination}, Quantum Information
  and Computation, 17 (2016), p.~0106.

\bibitem{brown2019constructions}
{\sc P.~Brown}, {\em On constructions of quantum-secure device-independent
  randomness expansion protocols}, PhD thesis, University of York, 2019.

\bibitem{brown2021computing}
{\sc P.~Brown, H.~Fawzi, and O.~Fawzi}, {\em Computing conditional entropies
  for quantum correlations}, Nature Communications, 12 (2021), pp.~1--12.

\bibitem{brown2021device}
\leavevmode\vrule height 2pt depth -1.6pt width 23pt, {\em Device-independent
  lower bounds on the conditional von {N}eumann entropy}, preprint
  arXiv:2106.13692,  (2021).

\bibitem{Brown2022}
{\sc P.~J. Brown}, {\em Example scripts for computing rates of
  device-independent protocols}.
\newblock \url{https://github.com/peterjbrown519/DI-rates}, 2022.

\bibitem{brown2019framework}
{\sc P.~J. Brown, S.~Ragy, and R.~Colbeck}, {\em A framework for quantum-secure
  device-independent randomness expansion}, IEEE Transactions on Information
  Theory, 66 (2019), pp.~2964--2987.

\bibitem{brunner2008testing}
{\sc N.~Brunner, S.~Pironio, A.~Acin, N.~Gisin, A.~A. M{\'e}thot, and
  V.~Scarani}, {\em Testing the dimension of {H}ilbert spaces}, Physical Review
  Letters, 100 (2008), p.~210503.

\bibitem{bu2017maximum}
{\sc K.~Bu, U.~Singh, S.-M. Fei, A.~K. Pati, and J.~Wu}, {\em Maximum relative
  entropy of coherence: an operational coherence measure}, Physical Review
  Letters, 119 (2017), p.~150405.

\bibitem{burgdorf2013tracial}
{\sc S.~Burgdorf, K.~Cafuta, I.~Klep, and J.~Povh}, {\em The tracial moment
  problem and trace-optimization of polynomials}, Mathematical programming, 137
  (2013), pp.~557--578.

\bibitem{cabello2021converting}
{\sc A.~Cabello}, {\em Converting contextuality into nonlocality}, Physical
  Review Letters, 127 (2021), p.~070401.

\bibitem{cabello2010non}
{\sc A.~Cabello, S.~Severini, and A.~Winter}, {\em {(Non-)} contextuality of
  physical theories as an axiom}, preprint arXiv:1010.2163,  (2010).

\bibitem{cabello2014graph}
{\sc A.~Cabello, S.~Severini, and A.~Winter}, {\em Graph-theoretic approach to
  quantum correlations}, Physical Review Letters, 112 (2014), p.~040401.

\bibitem{cameron2006quantum}
{\sc P.~J. Cameron, A.~Montanaro, M.~W. Newman, S.~Severini, and A.~Winter},
  {\em On the quantum chromatic number of a graph}, Electronic Journal of
  Combinatorics, 14 (2007).

\bibitem{carlen2010trace}
{\sc E.~Carlen}, {\em Trace inequalities and quantum entropy: an introductory
  course}, Entropy and the quantum, 529 (2010), pp.~73--140.

\bibitem{cavalcanti2016quantum}
{\sc D.~Cavalcanti and P.~Skrzypczyk}, {\em Quantum steering: a review with
  focus on semidefinite programming}, Reports on Progress in Physics, 80
  (2016), p.~024001.

\bibitem{cavalcanti2015detection}
{\sc D.~Cavalcanti, P.~Skrzypczyk, G.~H. Aguilar, R.~V. Nery, P.~H.~S. Ribeiro,
  and S.~P. Walborn}, {\em Detection of entanglement in asymmetric quantum
  networks and multipartite quantum steering}, Nature Communications, 6 (2015),
  p.~7941.

\bibitem{chaturvedi2021characterising}
{\sc A.~Chaturvedi, M.~Farkas, and V.~J. Wright}, {\em Characterising and
  bounding the set of quantum behaviours in contextuality scenarios}, Quantum,
  5 (2021), p.~484.

\bibitem{chaturvedi2020quantum}
{\sc A.~Chaturvedi and D.~Saha}, {\em Quantum prescriptions are more
  ontologically distinct than they are operationally distinguishable}, Quantum,
  4 (2020), p.~345.

\bibitem{chaturvedi2022extending}
{\sc A.~Chaturvedi, G.~Viola, and M.~Paw{\l}owski}, {\em Extending
  loophole-free nonlocal correlations to arbitrarily large distances}, preprint
  arXiv:2211.14231,  (2022).

\bibitem{cheng2006implementation}
{\sc J.~T.~W. Cheng and S.~Zhang}, {\em On implementation of a self-dual
  embedding method for convex programming}, Optimization Methods and Software,
  21 (2006), pp.~75--103.

\bibitem{chernyshenko2014polynomial}
{\sc S.~I. Chernyshenko, P.~Goulart, D.~Huang, and A.~Papachristodoulou}, {\em
  Polynomial sum of squares in fluid dynamics: a review with a look ahead},
  Philosophical Transactions of the Royal Society A: Mathematical, Physical and
  Engineering Sciences, 372 (2014), p.~20130350.

\bibitem{chesi2010lmi}
{\sc G.~Chesi}, {\em {LMI} techniques for optimization over polynomials in
  control: a survey}, IEEE transactions on Automatic Control, 55 (2010),
  pp.~2500--2510.

\bibitem{choi1975completely}
{\sc M.-D. Choi}, {\em Completely positive linear maps on complex matrices},
  Linear algebra and its applications, 10 (1975), pp.~285--290.

\bibitem{citeLovasz2}
{\sc F.~R. Chung}, {\em {Spectral Graph Theory}}, CBMS Regional Conference
  Series in Mathematics, University of Pennsylvania,  (1997).

\bibitem{cirel1980quantum}
{\sc B.~S. Cirel'son}, {\em Quantum generalizations of {B}ell's inequality},
  Letters in Mathematical Physics, 4 (1980), pp.~93--100.

\bibitem{clauser1969proposed}
{\sc J.~F. Clauser, M.~A. Horne, A.~Shimony, and R.~A. Holt}, {\em Proposed
  experiment to test local hidden-variable theories}, Physical review letters,
  23 (1969), p.~880.

\bibitem{coladangelo2017all}
{\sc A.~Coladangelo, K.~T. Goh, and V.~Scarani}, {\em All pure bipartite
  entangled states can be self-tested}, Nature Communications, 8 (2017),
  p.~15485.

\bibitem{citeLovasz1}
{\sc T.~M. Cover and J.~A. Thomas}, {\em Elements of information theory}, John
  Wiley \& Sons, 2012.

\bibitem{cubitt2011zero}
{\sc T.~S. Cubitt, D.~Leung, W.~Matthews, and A.~Winter}, {\em Zero-error
  channel capacity and simulation assisted by non-local correlations}, IEEE
  Transactions on Information Theory, 57 (2011), pp.~5509--5523.

\bibitem{Dantzig90}
{\sc G.~B. Dantzig}, {\em Origins of the simplex method}, in A history of
  scientific computing, S.~G. Nash, ed., Association for Computing Machinery,
  New York, NY, United States, 1990, pp.~141--151.

\bibitem{dattorro2010convex}
{\sc J.~Dattorro}, {\em Convex optimization \& {E}uclidean distance geometry},
  Lulu. com, 2010.

\bibitem{de2001quantum}
{\sc R.~De~Wolf}, {\em Quantum computing and communication complexity},
  University of Amsterdam, 2001.

\bibitem{doherty2008quantum}
{\sc A.~C. Doherty, Y.-C. Liang, B.~Toner, and S.~Wehner}, {\em The quantum
  moment problem and bounds on entangled multi-prover games}, in 2008 23rd
  Annual IEEE Conference on Computational Complexity, IEEE, 2008, pp.~199--210.

\bibitem{DPS02}
{\sc A.~C. Doherty, P.~A. Parrilo, and F.~M. Spedalieri}, {\em Distinguishing
  separable and entangled states}, Physical Review Letters, 88 (2002),
  p.~187904.

\bibitem{DPS04}
\leavevmode\vrule height 2pt depth -1.6pt width 23pt, {\em Complete family of
  separability criteria}, Physical Review A, 69 (2004), p.~022308.

\bibitem{citeLovaszQuantum}
{\sc R.~Duan, S.~Severini, and A.~Winter}, {\em Zero-error communication via
  quantum channels, noncommutative graphs, and a quantum {L}ov{\'a}sz number},
  IEEE Transactions on Information Theory, 59 (2012), pp.~1164--1174.

\bibitem{octave}
{\sc J.~W. Eaton, D.~Bateman, S.~Hauberg, and R.~Wehbring}, {\em {GNU Octave}
  version 6.1.0 manual: a high-level interactive language for numerical
  computations}, 2020.

\bibitem{ebadian2011perspectives}
{\sc A.~Ebadian, I.~Nikoufar, and M.~Eshaghi~Gordji}, {\em Perspectives of
  matrix convex functions}, Proceedings of the National Academy of Sciences,
  108 (2011), pp.~7313--7314.

\bibitem{effros2014non}
{\sc E.~Effros and F.~Hansen}, {\em Non-commutative perspectives}, Annals of
  Functional Analysis, 5 (2014), pp.~74--79.

\bibitem{effros2009matrix}
{\sc E.~G. Effros}, {\em A matrix convexity approach to some celebrated quantum
  inequalities}, Proceedings of the National Academy of Sciences, 106 (2009),
  pp.~1006--1008.

\bibitem{eldar2003semidefinite}
{\sc Y.~C. Eldar}, {\em A semidefinite programming approach to optimal
  unambiguous discrimination of quantum states}, IEEE Transactions on
  information theory, 49 (2003), pp.~446--456.

\bibitem{fannes1988symmetric}
{\sc M.~Fannes, J.~T. Lewis, and A.~Verbeure}, {\em Symmetric states of
  composite systems}, Letters in mathematical physics, 15 (1988), pp.~255--260.

\bibitem{fawzi2018efficient}
{\sc H.~Fawzi and O.~Fawzi}, {\em Efficient optimization of the quantum
  relative entropy}, Journal of Physics A: Mathematical and Theoretical, 51
  (2018), p.~154003.

\bibitem{fawzi2017lieb}
{\sc H.~Fawzi and J.~Saunderson}, {\em Lieb's concavity theorem, matrix
  geometric means, and semidefinite optimization}, Linear Algebra and its
  Applications, 513 (2017), pp.~240--263.

\bibitem{fawzi2019semidefinite}
{\sc H.~Fawzi, J.~Saunderson, and P.~A. Parrilo}, {\em Semidefinite
  approximations of the matrix logarithm}, Foundations of Computational
  Mathematics, 19 (2019), pp.~259--296.

\bibitem{fehr2013security}
{\sc S.~Fehr, R.~Gelles, and C.~Schaffner}, {\em Security and composability of
  randomness expansion from {B}ell inequalities}, Physical Review A, 87 (2013),
  p.~012335.

\bibitem{fenchel_1949}
{\sc W.~Fenchel}, {\em On conjugate convex functions}, Canadian Journal of
  Mathematics, 1 (1949), p.~73–77.

\bibitem{FiaccoMcCormick90}
{\sc A.~V. Fiacco and G.~P. McCormick}, {\em Nonlinear programming: sequential
  unconstrained minimization techniques}, vol.~4, SIAM, 1990.

\bibitem{frerot2022unveiling}
{\sc I.~Fr{\'e}rot, F.~Baccari, and A.~Ac{\'\i}n}, {\em Unveiling quantum
  entanglement in many-body systems from partial information}, PRX Quantum, 3
  (2022), p.~010342.

\bibitem{FM00}
{\sc R.~M. Freund and S.~Mizuno}, {\em Interior point methods: current status
  and future directions}, in High performance optimization, Springer, 2000,
  pp.~441--466.

\bibitem{fujii1992operator}
{\sc J.~I. Fujii}, {\em Operator means and the relative operator entropy}, in
  Operator Theory and Complex Analysis: Workshop on Operator Theory and Complex
  Analysis Sapporo (Japan) June 1991, Springer, 1992, pp.~161--172.

\bibitem{fujii1989relative}
{\sc J.~I. Fujii and E.~Kamei}, {\em Relative operator entropy in
  noncommutative information theory}, Math. Japon., 34 (1989), pp.~341--348.

\bibitem{fujii2018relative}
{\sc J.~I. Fujii and Y.~Seo}, {\em The relative operator entropy and the
  {K}archer mean}, Linear Algebra and its Applications, 542 (2018), pp.~4--34.

\bibitem{fujii2022relative}
\leavevmode\vrule height 2pt depth -1.6pt width 23pt, {\em Relative operator
  entropy}, in Operator and Norm Inequalities and Related Topics, Springer,
  2022, pp.~69--95.

\bibitem{Fujisawa97}
{\sc K.~Fujisawa, M.~Kojima, and K.~Nakata}, {\em Exploiting sparsity in
  primal-dual interior-point methods for semidefinite programming},
  Mathematical Programming, 79 (1997), pp.~235--253.

\bibitem{SDPA}
{\sc K.~Fujisawa, M.~Kojima, K.~Nakata, and M.~Yamashita}, {\em {SDPA
  (SemiDefinite Programming Algorithm) user’s manual—version 6.2}},
  Research Report B-308, Department of Mathematical and Computing SciencesTokyo
  Institute of Technology, 2-12-1 Oh-Okayama, Meguro-ku, Tokyo 152-0033, Japan,
  1995.

\bibitem{fujisawa2002sdpa}
\leavevmode\vrule height 2pt depth -1.6pt width 23pt, {\em {SDPA} (semidefinite
  programming algorithm) user’s manual—version 6.2. 0}, Department of
  Mathematical and Com-puting Sciences, Tokyo Institute of Technology. Research
  Reports on Mathematical and Computing Sciences Series B: Operations Research,
   (2002).

\bibitem{gallego2010device}
{\sc R.~Gallego, N.~Brunner, C.~Hadley, and A.~Ac{\'\i}n}, {\em
  Device-independent tests of classical and quantum dimensions}, Physical
  Review Letters, 105 (2010), p.~230501.

\bibitem{gallier2020schur}
{\sc J.~Gallier}, {\em The {S}chur complement and symmetric positive
  semidefinite (and definite) matrices}.
\newblock 2019.

\bibitem{gartner2012approximation}
{\sc B.~G{\"a}rtner and J.~Matousek}, {\em Approximation algorithms and
  semidefinite programming}, Springer Science \& Business Media, 2012.

\bibitem{gilbert1991positive}
{\sc G.~T. Gilbert}, {\em Positive definite matrices and {S}ylvester's
  criterion}, The American Mathematical Monthly, 98 (1991), pp.~44--46.

\bibitem{GMSTW86}
{\sc P.~E. Gill, W.~Murray, M.~A. Saunders, J.~A. Tomlin, and M.~H. Wright},
  {\em On projected {N}ewton barrier methods for linear programming and an
  equivalence to {K}armarkar’s projective method}, Mathematical Programming,
  36 (1986), pp.~183--209.

\bibitem{gleason1975measures}
{\sc A.~M. Gleason}, {\em Measures on the closed subspaces of a {H}ilbert
  space}, The Logico-Algebraic Approach to Quantum Mechanics: Volume I:
  Historical Evolution,  (1975), pp.~123--133.

\bibitem{goemans2001approximation}
{\sc M.~X. Goemans and D.~Williamson}, {\em Approximation algorithms for
  {MAX-3-CUT} and other problems via complex semidefinite programming}, in
  Proceedings of the thirty-third annual ACM symposium on Theory of computing,
  2001, pp.~443--452.

\bibitem{goldfarb1998interior}
{\sc D.~Goldfarb and K.~Scheinberg}, {\em Interior point trajectories in
  semidefinite programming}, SIAM Journal on Optimization, 8 (1998),
  pp.~871--886.

\bibitem{Gondzio12}
{\sc J.~Gondzio}, {\em Interior point methods 25 years later}, European Journal
  of Operational Research, 218 (2012), pp.~587--601.

\bibitem{GT92}
{\sc C.~C. Gonzaga and M.~J. Todd}, {\em {An O(nL)-Iteration Large-Step
  Primal-Dual Affine Algorithm for Linear Programming}}, SIAM Journal on
  Optimization, 2 (1992), pp.~349--359.

\bibitem{grant2014cvx}
{\sc M.~Grant and S.~Boyd}, {\em Cvx: Matlab software for disciplined convex
  programming, version 2.1}, 2014.

\bibitem{grant2011cvx}
{\sc M.~Grant, S.~Boyd, and Y.~Ye}, {\em Cvx: Matlab software for disciplined
  convex programming}, 2011.

\bibitem{grotschel1981ellipsoid}
{\sc M.~Gr{\"o}tschel, L.~Lov{\'a}sz, and A.~Schrijver}, {\em The ellipsoid
  method and its consequences in combinatorial optimization}, Combinatorica, 1
  (1981), pp.~169--197.

\bibitem{GLS86}
\leavevmode\vrule height 2pt depth -1.6pt width 23pt, {\em Relaxations of
  vertex packing}, Journal of Combinatorial Theory, Series B, 40 (1986),
  pp.~330--343.

\bibitem{guimaraes2015tutorial}
{\sc D.~A. Guimaraes, G.~H.~F. Floriano, and L.~S. Chaves}, {\em A tutorial on
  the {CVX} system for modeling and solving convex optimization problems}, IEEE
  Latin America Transactions, 13 (2015), pp.~1228--1257.

\bibitem{gupta2023quantum}
{\sc S.~Gupta, D.~Saha, Z.-P. Xu, A.~Cabello, and A.~S. Majumdar}, {\em Quantum
  contextuality provides communication complexity advantage}, Physical Review
  Letters, 130 (2023), p.~080802.

\bibitem{hansen2003jensen}
{\sc F.~Hansen and G.~K. Pedersen}, {\em Jensen's operator inequality},
  Bulletin of the London Mathematical Society, 35 (2003), pp.~553--564.

\bibitem{hansson2014sampling}
{\sc A.~Hansson and L.~Vandenberghe}, {\em Sampling method for semidefinite
  programmes with non-negative {P}opov function constraints}, International
  Journal of Control, 87 (2014), pp.~330--345.

\bibitem{BC10}
{\sc P.~L.~C. Heinz H.~Bauschke}, {\em Convex {A}nalysis and {M}onotone
  {O}perator {T}heory in {H}ilbert {S}paces}, CMS Books in Mathematics,
  Springer Cham, 2010.

\bibitem{B00}
{\sc C.~Helmberg}, {\em Semidefinite programming for combinatorial
  optimization}, PhD thesis, Konrad-Zuse-Zentrum fur Informationstechnik
  Berlin, Konrad-Zuse-Zentrum für Informationstechnik, 2000.

\bibitem{HRVW96}
{\sc C.~Helmberg, F.~Rendl, R.~J. Vanderbei, and H.~Wolkowicz}, {\em An
  interior-point method for semidefinite programming}, SIAM Journal on
  optimization, 6 (1996), pp.~342--361.

\bibitem{helstrom1969quantum}
{\sc C.~W. Helstrom}, {\em Quantum detection and estimation theory}, Journal of
  Statistical Physics, 1 (1969), pp.~231--252.

\bibitem{helton2002positive}
{\sc J.~W. Helton}, {\em {"Positive"} noncommutative polynomials are sums of
  squares}, Annals of Mathematics,  (2002), pp.~675--694.

\bibitem{helton2009sufficient}
{\sc J.~W. Helton and J.~Nie}, {\em Sufficient and necessary conditions for
  semidefinite representability of convex hulls and sets}, SIAM Journal on
  Optimization, 20 (2009), pp.~759--791.

\bibitem{helton2010semidefinite}
\leavevmode\vrule height 2pt depth -1.6pt width 23pt, {\em Semidefinite
  representation of convex sets}, Mathematical Programming, 122 (2010),
  pp.~21--64.

\bibitem{helton2007linear}
{\sc J.~W. Helton and V.~Vinnikov}, {\em Linear matrix inequality
  representation of sets}, Communications on Pure and Applied Mathematics: A
  Journal Issued by the Courant Institute of Mathematical Sciences, 60 (2007),
  pp.~654--674.

\bibitem{hoban2018channel}
{\sc M.~J. Hoban and A.~B. Sainz}, {\em A channel-based framework for steering,
  non-locality and beyond}, New Journal of Physics, 20 (2018), p.~053048.

\bibitem{matrixAnalysis}
{\sc R.~A. Horn and C.~R. Johnson}, {\em Matrix analysis}, Cambridge University
  Press, 2012.

\bibitem{horodecki1996separability}
{\sc M.~Horodecki, P.~Horodecki, and R.~Horodecki}, {\em Separability of mixed
  quantum states: necessary and sufficient conditions}, Physics Letters A, 223
  (1996), pp.~1--8.

\bibitem{howard2014contextuality}
{\sc M.~Howard, J.~Wallman, V.~Veitch, and J.~Emerson}, {\em Contextuality
  supplies the ‘magic’for quantum computation}, Nature, 510 (2014),
  pp.~351--355.

\bibitem{ivanovic1987differentiate}
{\sc I.~D. Ivanovic}, {\em How to differentiate between non-orthogonal states},
  Physics Letters A, 123 (1987), pp.~257--259.

\bibitem{jameson1972convex}
{\sc G.~J.~O. Jameson}, {\em Convex series}, Mathematical Proceedings of the
  Cambridge Philosophical Society, 72 (1972), pp.~37--47.

\bibitem{jamiolkowski1972linear}
{\sc A.~Jamio{\l}kowski}, {\em Linear transformations which preserve trace and
  positive semidefiniteness of operators}, Reports on Mathematical Physics, 3
  (1972), pp.~275--278.

\bibitem{jarvis2005control}
{\sc Z.~Jarvis-Wloszek, R.~Feeley, W.~Tan, K.~Sun, and A.~Packard}, {\em
  Control applications of sum of squares programming}, Positive polynomials in
  control,  (2005), pp.~3--22.

\bibitem{jbilou2004some}
{\sc K.~Jbilou, A.~Messaoudi, and K.~Taba{\^a}}, {\em Some {S}chur complement
  identities and applications to matrix extrapolation methods}, Linear algebra
  and its applications, 392 (2004), pp.~195--210.

\bibitem{jeyakumar2016convergent}
{\sc V.~Jeyakumar, J.~B. Lasserre, G.~Li, and T.~S. Pham}, {\em Convergent
  semidefinite programming relaxations for global bilevel polynomial
  optimization problems}, SIAM Journal on Optimization, 26 (2016),
  pp.~753--780.

\bibitem{jevzek2002finding}
{\sc M.~Je{\v{z}}ek, J.~{\v{R}}eh{\'a}{\v{c}}ek, and J.~Fiur{\'a}{\v{s}}ek},
  {\em Finding optimal strategies for minimum-error quantum-state
  discrimination}, Physical Review A, 65 (2002), p.~060301.

\bibitem{jiang2013channel}
{\sc M.~Jiang, S.~Luo, and S.~Fu}, {\em Channel-state duality}, Physical Review
  A, 87 (2013), p.~022310.

\bibitem{johnston2016extended}
{\sc N.~Johnston, R.~Mittal, V.~Russo, and J.~Watrous}, {\em Extended non-local
  games and monogamy-of-entanglement games}, Proceedings of the Royal Society
  A: Mathematical, Physical and Engineering Sciences, 472 (2016), p.~20160003.

\bibitem{KMS98}
{\sc D.~Karger, R.~Motwani, and M.~Sudan}, {\em Approximate graph coloring by
  semidefinite programming}, Journal of the ACM (JACM), 45 (1998),
  pp.~246--265.

\bibitem{Karmarkar84}
{\sc N.~Karmarkar}, {\em A new polynomial-time algorithm for linear
  programming}, in Proceedings of the sixteenth annual ACM symposium on Theory
  of computing, 1984, pp.~302--311.

\bibitem{kempe2011entangled}
{\sc J.~Kempe, H.~Kobayashi, K.~Matsumoto, B.~Toner, and T.~Vidick}, {\em
  Entangled games are hard to approximate}, SIAM Journal on Computing, 40
  (2011), pp.~848--877.

\bibitem{Khachian79}
{\sc L.~G. Khachian}, {\em A polynomial time algorithm for linear programing},
  Soviet Math. Dokl., 244 (1979), p.~1093–1096.

\bibitem{kheirfam2015adaptive}
{\sc B.~Kheirfam}, {\em An adaptive infeasible interior-point algorithm with
  full {N}esterov-{T}odd step for semidefinite optimization}, Journal of
  Mathematical Modelling and Algorithms in Operations Research, 14 (2015),
  pp.~55--66.

\bibitem{KleeMinty70}
{\sc V.~Klee and G.~J. Minty}, {\em How good is the simplex algorithm},
  Inequalities, 3 (1972), pp.~159--175.

\bibitem{Goemans98}
{\sc J.~Kleinberg and M.~X. Goemans}, {\em The {L}ov{\'a}sz theta function and
  a semidefinite programming relaxation of vertex cover}, SIAM Journal on
  Discrete Mathematics, 11 (1998), pp.~196--204.

\bibitem{klep2021sparse}
{\sc I.~Klep, V.~Magron, and J.~Povh}, {\em Sparse noncommutative polynomial
  optimization}, Mathematical Programming,  (2021), pp.~1--41.

\bibitem{klep2010semidefinite}
{\sc I.~Klep and J.~Povh}, {\em Semidefinite programming and sums of hermitian
  squares of noncommutative polynomials}, Journal of Pure and Applied Algebra,
  214 (2010), pp.~740--749.

\bibitem{klep2016constrained}
\leavevmode\vrule height 2pt depth -1.6pt width 23pt, {\em Constrained
  trace-optimization of polynomials in freely noncommuting variables}, Journal
  of Global Optimization, 64 (2016), pp.~325--348.

\bibitem{kobayashi2007conversion}
{\sc K.~Kobayashi, K.~Nakata, and M.~Kojima}, {\em A conversion of an {SDP}
  having free variables into the standard form {SDP}}, Computational
  optimization and Applications, 36 (2007), pp.~289--307.

\bibitem{kochen1990problem}
{\sc S.~Kochen and E.~P. Specker}, {\em The problem of hidden variables in
  quantum mechanics}, Ernst Specker Selecta,  (1990), pp.~235--263.

\bibitem{kogias2015hierarchy}
{\sc I.~Kogias, P.~Skrzypczyk, D.~Cavalcanti, A.~Ac{\'\i}n, and G.~Adesso},
  {\em Hierarchy of steering criteria based on moments for all bipartite
  quantum systems}, Physical Review Letters, 115 (2015), p.~210401.

\bibitem{kojima1993primal}
{\sc M.~Kojima, N.~Megiddo, and S.~Mizuno}, {\em A primal—dual
  infeasible-interior-point algorithm for linear programming}, Mathematical
  programming, 61 (1993), pp.~263--280.

\bibitem{kojima1989primal}
{\sc M.~Kojima, S.~Mizuno, and A.~Yoshise}, {\em A primal-dual interior point
  algorithm for linear programming}, Springer, 1989.

\bibitem{KSH97}
{\sc M.~Kojima, S.~Shindoh, and S.~Hara}, {\em Interior-point methods for the
  monotone semidefinite linear complementarity problem in symmetric matrices},
  SIAM Journal on Optimization, 7 (1997), pp.~86--125.

\bibitem{kraus1936konvexe}
{\sc F.~Kraus}, {\em {\"U}ber konvexe matrixfunktionen}, Mathematische
  Zeitschrift, 41 (1936), pp.~18--42.

\bibitem{kubo1980means}
{\sc F.~Kubo and T.~Ando}, {\em Means of positive linear operators},
  Mathematische Annalen, 246 (1980), pp.~205--224.

\bibitem{kueng2016comparing}
{\sc R.~Kueng, D.~M. Long, A.~C. Doherty, and S.~T. Flammia}, {\em Comparing
  experiments to the fault-tolerance threshold}, Physical Review Letters, 117
  (2016), p.~170502.

\bibitem{lasserre2001global}
{\sc J.~B. Lasserre}, {\em Global optimization with polynomials and the problem
  of moments}, SIAM Journal on optimization, 11 (2001), pp.~796--817.

\bibitem{lasserre2007sum}
\leavevmode\vrule height 2pt depth -1.6pt width 23pt, {\em A sum of squares
  approximation of nonnegative polynomials}, SIAM review, 49 (2007),
  pp.~651--669.

\bibitem{leifer2014quantum}
{\sc M.~S. Leifer}, {\em Is the quantum state real? an extended review of
  psi-ontology theorems}, Quanta, 3 (2014), p.~67.

\bibitem{leung2015power}
{\sc D.~Leung and W.~Matthews}, {\em On the power of {PPT}-preserving and
  non-signalling codes}, IEEE Transactions on Information Theory, 61 (2015),
  pp.~4486--4499.

\bibitem{lewenstein1998separability}
{\sc M.~Lewenstein and A.~Sanpera}, {\em Separability and entanglement of
  composite quantum systems}, Physical Review Letters, 80 (1998), p.~2261.

\bibitem{li2013relationship}
{\sc H.-W. Li, P.~Mironowicz, M.~Paw{\l}owski, Z.-Q. Yin, Y.-C. Wu, S.~Wang,
  W.~Chen, H.-G. Hu, G.-C. Guo, and Z.-F. Han}, {\em Relationship between
  semi-and fully-device-independent protocols}, Physical Review A, 87 (2013),
  p.~020302.

\bibitem{li2011semi}
{\sc H.-W. Li, Z.-Q. Yin, Y.-C. Wu, X.-B. Zou, S.~Wang, W.~Chen, G.-C. Guo, and
  Z.-F. Han}, {\em Semi-device-independent random-number expansion without
  entanglement}, Physical Review A, 84 (2011), p.~034301.

\bibitem{lin2022naturally}
{\sc P.-S. Lin, T.~V{\'e}rtesi, and Y.-C. Liang}, {\em Naturally restricted
  subsets of nonsignaling correlations: typicality and convergence}, Quantum, 6
  (2022), p.~765.

\bibitem{yalmip}
{\sc J.~L{\"{o}}fberg}, {\em {YALMIP : A Toolbox for Modeling and Optimization
  in MATLAB}}, in In Proceedings of the CACSD Conference, Taipei, Taiwan, 2004.

\bibitem{dualizeIt}
{\sc J.~L{\"o}fberg}, {\em Dualize it: software for automatic primal and dual
  conversions of conic programs}, Optimization Methods \& Software, 24 (2009),
  pp.~313--325.

\bibitem{lofberg2009pre}
{\sc J.~Lofberg}, {\em Pre-and post-processing sum-of-squares programs in
  practice}, IEEE transactions on automatic control, 54 (2009), pp.~1007--1011.

\bibitem{Lovasz79}
{\sc L.~Lov{\'a}sz}, {\em On the {S}hannon capacity of a graph}, IEEE
  Transactions on Information theory, 25 (1979), pp.~1--7.

\bibitem{lowner1934monotone}
{\sc K.~L{\"o}wner}, {\em {\"U}ber monotone matrixfunktionen}, Mathematische
  Zeitschrift, 38 (1934), pp.~177--216.

\bibitem{L06}
{\sc R.~Lucchetti}, {\em Convexity and {W}ell-{P}osed {P}roblems}, CMS Books in
  Mathematics, Springer New York, 2006.

\bibitem{magesan2012characterizing}
{\sc E.~Magesan, J.~M. Gambetta, and J.~Emerson}, {\em Characterizing quantum
  gates via randomized benchmarking}, Physical Review A, 85 (2012), p.~042311.

\bibitem{magnus2019matrix}
{\sc J.~R. Magnus and H.~Neudecker}, {\em Matrix differential calculus with
  applications in statistics and econometrics}, John Wiley \& Sons, 2019.

\bibitem{mayers1998quantum}
{\sc D.~Mayers and A.~Yao}, {\em Quantum cryptography with imperfect
  apparatus}, in Proceedings 39th Annual Symposium on Foundations of Computer
  Science (Cat. No. 98CB36280), IEEE, 1998, pp.~503--509.

\bibitem{mccullough2005noncommutative}
{\sc S.~McCullough and M.~Putinar}, {\em Noncommutative sums of squares},
  Pacific J. Math, 218 (2005), pp.~167--171.

\bibitem{Mehrotra92}
{\sc S.~Mehrotra}, {\em On the implementation of a primal-dual interior point
  method}, SIAM Journal on optimization, 2 (1992), pp.~575--601.

\bibitem{Mercer1909}
{\sc J.~Mercer}, {\em Functions of positive and negative type and their
  connection with the theory of integral equations}, Philosophical Transactions
  of the Royal Society A, 209 (1909), pp.~415--446.

\bibitem{merkel2013self}
{\sc S.~T. Merkel, J.~M. Gambetta, J.~A. Smolin, S.~Poletto, A.~D.
  C{\'o}rcoles, B.~R. Johnson, C.~A. Ryan, and M.~Steffen}, {\em
  Self-consistent quantum process tomography}, Physical Review A, 87 (2013),
  p.~062119.

\bibitem{meszaros1998free}
{\sc C.~M{\'e}sz{\'a}ros}, {\em On free variables in interior point methods},
  Optimization Methods and Software, 9 (1998), pp.~121--139.

\bibitem{matrixAnalysis2}
{\sc C.~D. Meyer}, {\em Matrix analysis and applied linear algebra}, vol.~71,
  SIAM, 2000.

\bibitem{Miltenberger_mittelmann-plots_-_Interactive_2021}
{\sc M.~Miltenberger}, {\em {mittelmann-plots - Interactive Visualizations of
  Mittelmann benchmarks}}, June 2021.

\bibitem{myThesis}
{\sc P.~Mironowicz}, {\em Applications of semi-definite optimization in quantum
  information protocols}, PhD thesis, Gdańsk University of Technology, 2015.

\bibitem{mironowicz2014properties}
{\sc P.~Mironowicz, H.-W. Li, and M.~Paw{\l}owski}, {\em Properties of
  dimension witnesses and their semidefinite programming relaxations}, Physical
  Review A, 90 (2014), p.~022322.

\bibitem{MittelmannBenchmark}
{\sc H.~Mittelmann}, {\em {Decision Tree for Optimization Software}}, 2023.

\bibitem{Mittelmann12}
{\sc H.~D. Mittelmann}, {\em The state-of-the-art in conic optimization
  software}, in Handbook on Semidefinite, Conic and Polynomial Optimization,
  M.~Anjos and J.~Lasserre, eds., Springer, 2012, pp.~671--686.

\bibitem{M97}
{\sc R.~D.~C. Monteiro}, {\em Primal--dual path-following algorithms for
  semidefinite programming}, SIAM Journal on Optimization, 7 (1997),
  pp.~663--678.

\bibitem{M98}
\leavevmode\vrule height 2pt depth -1.6pt width 23pt, {\em Polynomial
  convergence of primal-dual algorithms for semidefinite programming based on
  the {M}onteiro and {Z}hang family of directions}, SIAM Journal on
  Optimization, 8 (1998), pp.~797--812.

\bibitem{monteiro1989interiorI}
{\sc R.~D.~C. Monteiro and I.~Adler}, {\em {Interior path following primal-dual
  algorithms. Part I: Linear programming}}, Mathematical programming, 44
  (1989), pp.~27--41.

\bibitem{monteiro1989interiorII}
\leavevmode\vrule height 2pt depth -1.6pt width 23pt, {\em {Interior path
  following primal-dual algorithms. Part II: Convex quadratic programming}},
  Mathematical Programming, 44 (1989), pp.~43--66.

\bibitem{MZ98}
{\sc R.~D.~C. Monteiro and Y.~Zhang}, {\em A unified analysis for a class of
  long-step primal-dual path-following interior-point algorithms for
  semidefinite programming}, Mathematical Programming, 81 (1998), pp.~281--299.

\bibitem{moroder2013device}
{\sc T.~Moroder, J.-D. Bancal, Y.-C. Liang, M.~Hofmann, and O.~G{\"u}hne}, {\em
  Device-independent entanglement quantification and related applications},
  Physical Review Letters, 111 (2013), p.~030501.

\bibitem{nakata2010numerical}
{\sc M.~Nakata}, {\em A numerical evaluation of highly accurate
  multiple-precision arithmetic version of semidefinite programming solver:
  {SDPA-GMP,-QD and-DD}.}, in 2010 IEEE International Symposium on
  Computer-Aided Control System Design, IEEE, 2010, pp.~29--34.

\bibitem{navascues2014characterization}
{\sc M.~Navascu{\'e}s, G.~de~la Torre, and T.~V{\'e}rtesi}, {\em
  Characterization of quantum correlations with local dimension constraints and
  its device-independent applications}, Physical Review X, 4 (2014), p.~011011.

\bibitem{Navascues2015}
{\sc M.~Navascu\'es, A.~Feix, M.~Ara\'ujo, and T.~V\'ertesi}, {\em
  Characterizing finite-dimensional quantum behavior}, Physical Review A, 92
  (2015), p.~042117.

\bibitem{Navascues2015a}
{\sc M.~Navascu{\'e}s, Y.~Guryanova, M.~J. Hoban, and A.~Ac{\'i}n}, {\em Almost
  quantum correlations}, Nature Communications, 6 (2015), pp.~6288 EP --.
\newblock Article.

\bibitem{NPA07}
{\sc M.~Navascu{\'e}s, S.~Pironio, and A.~Ac{\'\i}n}, {\em Bounding the set of
  quantum correlations}, Physical Review Letters, 98 (2007), p.~010401.

\bibitem{NPA08}
\leavevmode\vrule height 2pt depth -1.6pt width 23pt, {\em A convergent
  hierarchy of semidefinite programs characterizing the set of quantum
  correlations}, New Journal of Physics, 10 (2008), p.~073013.

\bibitem{MiguelVertesi}
{\sc M.~Navascu\'es and T.~V\'ertesi}, {\em {Bounding the Set of Finite
  Dimensional Quantum Correlations}}, Physical Review Letters, 115 (2015),
  p.~020501.

\bibitem{N04}
{\sc A.~Nemirovski}, {\em Interior point polynomial time methods in convex
  programming}, Lecture notes, 42 (2004), pp.~3215--3224.

\bibitem{nemirovski2007advances}
\leavevmode\vrule height 2pt depth -1.6pt width 23pt, {\em Advances in convex
  optimization: conic programming}, in International Congress of
  Mathematicians, vol.~1, 2007, pp.~413--444.

\bibitem{nesterov2018lectures}
{\sc Y.~Nesterov}, {\em Lectures on convex optimization}, vol.~137 of Springer
  Optimization and Its Applications, Springer, ii~ed., 2018.

\bibitem{NN94}
{\sc Y.~Nesterov and A.~Nemirovskii}, {\em Interior-point polynomial algorithms
  in convex programming}, SIAM, 1994.

\bibitem{NN92}
{\sc Y.~Nesterov and A.~Nemirovsky}, {\em Conic formulation of a convex
  programming problem and duality}, Optimization Methods and Software, 1
  (1992), pp.~95--115.

\bibitem{NT97}
{\sc Y.~E. Nesterov and M.~J. Todd}, {\em Self-scaled barriers and
  interior-point methods for convex programming}, Mathematics of Operations
  research, 22 (1997), pp.~1--42.

\bibitem{NT98}
\leavevmode\vrule height 2pt depth -1.6pt width 23pt, {\em Primal-dual
  interior-point methods for self-scaled cones}, SIAM Journal on optimization,
  8 (1998), pp.~324--364.

\bibitem{netzer2010semidefinite}
{\sc T.~Netzer}, {\em On semidefinite representations of non-closed sets},
  Linear algebra and its applications, 432 (2010), pp.~3072--3078.

\bibitem{netzer2009note}
{\sc T.~Netzer and R.~Sinn}, {\em A note on the convex hull of finitely many
  projections of spectrahedra}, preprint arXiv:0908.3386,  (2009).

\bibitem{nieto2014using}
{\sc O.~Nieto-Silleras, S.~Pironio, and J.~Silman}, {\em Using complete
  measurement statistics for optimal device-independent randomness evaluation},
  New Journal of Physics, 16 (2014), p.~013035.

\bibitem{seesaw2}
{\sc K.~F. P\'al and T.~V\'ertesi}, {\em Maximal violation of a bipartite
  three-setting, two-outcome {B}ell inequality using infinite-dimensional
  quantum systems}, Physical Review A, 82 (2010), p.~022116.

\bibitem{papachristodoulou2005tutorial}
{\sc A.~Papachristodoulou and S.~Prajna}, {\em A tutorial on sum of squares
  techniques for systems analysis}, in Proceedings of the 2005, American
  Control Conference, 2005., IEEE, 2005, pp.~2686--2700.

\bibitem{parrilo2003semidefinite}
{\sc P.~A. Parrilo}, {\em Semidefinite programming relaxations for
  semialgebraic problems}, Mathematical programming, 96 (2003), pp.~293--320.

\bibitem{parrilo2008approximation}
{\sc P.~A. Parrilo and A.~Jadbabaie}, {\em Approximation of the joint spectral
  radius using sum of squares}, Linear Algebra and its Applications, 428
  (2008), pp.~2385--2402.

\bibitem{parrilo2003minimizing}
{\sc P.~A. Parrilo and B.~Sturmfels}, {\em Minimizing polynomial functions},
  Algorithmic and quantitative real algebraic geometry, DIMACS Series in
  Discrete Mathematics and Theoretical Computer Science, 60 (2003), pp.~83--99.

\bibitem{pawlowski2011semi}
{\sc M.~Paw{\l}owski and N.~Brunner}, {\em Semi-device-independent security of
  one-way quantum key distribution}, Physical Review A, 84 (2011), p.~010302.

\bibitem{peres1996separability}
{\sc A.~Peres}, {\em Separability criterion for density matrices}, Physical
  Review Letters, 77 (1996), p.~1413.

\bibitem{piani2016robustness}
{\sc M.~Piani, M.~Cianciaruso, T.~R. Bromley, C.~Napoli, N.~Johnston, and
  G.~Adesso}, {\em Robustness of asymmetry and coherence of quantum states},
  Physical Review A, 93 (2016), p.~042107.

\bibitem{pironio2010random}
{\sc S.~Pironio, A.~Ac{\'\i}n, S.~Massar, A.~B. de~La~Giroday, D.~N.
  Matsukevich, P.~Maunz, S.~Olmschenk, D.~Hayes, L.~Luo, and T.~A. Manning},
  {\em Random numbers certified by {B}ell’s theorem}, Nature, 464 (2010),
  pp.~1021--1024.

\bibitem{pironio2013security}
{\sc S.~Pironio and S.~Massar}, {\em Security of practical private randomness
  generation}, Physical Review A, 87 (2013), p.~012336.

\bibitem{pironio2010convergent}
{\sc S.~Pironio, M.~Navascu{\'e}s, and A.~Acin}, {\em Convergent relaxations of
  polynomial optimization problems with noncommuting variables}, SIAM Journal
  on Optimization, 20 (2010), pp.~2157--2180.

\bibitem{potra2000interior}
{\sc F.~A. Potra and S.~J. Wright}, {\em Interior-point methods}, Journal of
  computational and applied mathematics, 124 (2000), pp.~281--302.

\bibitem{pozas2019bounding}
{\sc A.~Pozas-Kerstjens, R.~Rabelo, {\L}.~Rudnicki, R.~Chaves, D.~Cavalcanti,
  M.~Navascu{\'e}s, and A.~Ac{\'\i}n}, {\em Bounding the sets of classical and
  quantum correlations in networks}, Physical Review Letters, 123 (2019),
  p.~140503.

\bibitem{prajna2002introducing}
{\sc S.~Prajna, A.~Papachristodoulou, and P.~A. Parrilo}, {\em Introducing
  {SOSTOOLS}: A general purpose sum of squares programming solver}, in
  Proceedings of the 41st IEEE Conference on Decision and Control, 2002.,
  vol.~1, IEEE, 2002, pp.~741--746.

\bibitem{pusey2013negativity}
{\sc M.~F. Pusey}, {\em Negativity and steering: {A} stronger {P}eres
  conjecture}, Physical Review A, 88 (2013), p.~032313.

\bibitem{pusz1975functional}
{\sc W.~Pusz and S.~L. Woronowicz}, {\em Functional calculus for sesquilinear
  forms and the purification map}, Reports on Mathematical Physics, 8 (1975),
  pp.~159--170.

\bibitem{pyatnitskiy1982numerical}
{\sc Y.~S. Pyatnitskiy and V.~Skorodinskiy}, {\em Numerical methods of
  {L}yapunov function construction and their application to the absolute
  stability problem}, Systems \& Control Letters, 2 (1982), pp.~130--135.

\bibitem{quarteroni2007foundations}
{\sc A.~Quarteroni, R.~Sacco, and F.~Saleri}, {\em Foundations of matrix
  analysis}, in Numerical Mathematics, Springer, 2007, pp.~1--32.

\bibitem{raggio1988quantum}
{\sc G.~A. Raggio and R.~F. Werner}, {\em Quantum statistical mechanics of
  general mean field systems}, Helvetica Physica Acta, 62 (1988), p.~980.

\bibitem{rains1999bound}
{\sc E.~M. Rains}, {\em Bound on distillable entanglement}, Physical Review A,
  60 (1999), p.~179.

\bibitem{rains2001semidefinite}
\leavevmode\vrule height 2pt depth -1.6pt width 23pt, {\em A semidefinite
  program for distillable entanglement}, IEEE Transactions on Information
  Theory, 47 (2001), pp.~2921--2933.

\bibitem{ramana1995some}
{\sc M.~Ramana and A.~J. Goldman}, {\em Some geometric results in semidefinite
  programming}, Journal of Global Optimization, 7 (1995), pp.~33--50.

\bibitem{regula2021fundamental}
{\sc B.~Regula and R.~Takagi}, {\em Fundamental limitations on distillation of
  quantum channel resources}, Nature Communications, 12 (2021), p.~4411.

\bibitem{renou2022two}
{\sc M.-O. Renou and X.~Xu}, {\em Two convergent {NPA}-like hierarchies for the
  quantum bilocal scenario}, preprint arXiv:2210.09065,  (2022).

\bibitem{RW09}
{\sc R.~T. Rockafellar and R.~J.-B.~R. Wets}, {\em Variational {A}nalysis},
  Grundlehren der mathematischen Wissenschaften, Springer Berlin, Heidelberg,
  2009.

\bibitem{QDimSumGitHub}
{\sc D.~Rosset}, {\em {QDimSum}: {S}ymmetric {SDP} relaxations for qudits
  systems}.
\newblock \url{https://github.com/denisrosset/qdimsum}.

\bibitem{sadiq2013bell}
{\sc M.~Sadiq, P.~Badzi{\k{a}}g, M.~Bourennane, and A.~Cabello}, {\em Bell
  inequalities for the simplest exclusivity graph}, Physical Review A, 87
  (2013), p.~012128.

\bibitem{sagnol2013semidefinite}
{\sc G.~Sagnol}, {\em On the semidefinite representation of real functions
  applied to symmetric matrices}, Linear Algebra and its Applications, 439
  (2013), pp.~2829--2843.

\bibitem{sainz2018formalism}
{\sc A.~B. Sainz, L.~Aolita, M.~Piani, M.~J. Hoban, and P.~Skrzypczyk}, {\em A
  formalism for steering with local quantum measurements}, New Journal of
  Physics, 20 (2018), p.~083040.

\bibitem{sainz2015postquantum}
{\sc A.~B. Sainz, N.~Brunner, D.~Cavalcanti, P.~Skrzypczyk, and
  T.~V{\'e}rtesi}, {\em Postquantum steering}, Physical Review Letters, 115
  (2015), p.~190403.

\bibitem{scheiderer2018semidefinite}
{\sc C.~Scheiderer}, {\em Semidefinite representation for convex hulls of real
  algebraic curves}, SIAM Journal on Applied Algebra and Geometry, 2 (2018),
  pp.~1--25.

\bibitem{scheiderer2018spectrahedral}
\leavevmode\vrule height 2pt depth -1.6pt width 23pt, {\em Spectrahedral
  shadows}, SIAM Journal on Applied Algebra and Geometry, 2 (2018), pp.~26--44.

\bibitem{schur1917potenzreihen}
{\sc J.~Schur}, {\em {\"U}ber {P}otenzreihen, die im {I}nnern des
  {E}inheitskreises beschr{\"a}nkt sind.}, Journal für die reine und
  angewandte Mathematik, 147 (1917), pp.~205--232.

\bibitem{seidenberg1954new}
{\sc A.~Seidenberg}, {\em A new decision method for elementary algebra}, Annals
  of Mathematics,  (1954), pp.~365--374.

\bibitem{Shannon56}
{\sc C.~Shannon}, {\em The zero error capacity of a noisy channel}, IRE
  Transactions on Information Theory, 2 (1956), pp.~8--19.

\bibitem{skrzypczyk2023semidefinite}
{\sc P.~Skrzypczyk and D.~Cavalcanti}, {\em {Semidefinite Programming in
  Quantum Information Science}},  (2023).

\bibitem{slater2013lagrange}
{\sc M.~Slater}, {\em Lagrange multipliers revisited}, in Traces and emergence
  of nonlinear programming, Springer, 2013, pp.~293--306.

\bibitem{smania2020experimental}
{\sc M.~Smania, P.~Mironowicz, M.~Nawareg, M.~Paw{\l}owski, A.~Cabello, and
  M.~Bourennane}, {\em Experimental certification of an informationally
  complete quantum measurement in a device-independent protocol}, Optica, 7
  (2020), pp.~123--128.

\bibitem{spekkens2005contextuality}
{\sc R.~W. Spekkens}, {\em Contextuality for preparations, transformations, and
  unsharp measurements}, Physical Review A, 71 (2005), p.~052108.

\bibitem{SeDuMiGitHub}
{\sc J.~F. Sturm}, {\em {SQLP/Sedumi: Sedumi: A linear/quadratic/semidefinite
  solver for MATLAB and octave}}.
\newblock \url{https://github.com/sqlp/sedumi}.

\bibitem{SeDuMi}
\leavevmode\vrule height 2pt depth -1.6pt width 23pt, {\em Using {SeDuMi} 1.02,
  {A} {MATLAB} toolbox for optimization over symmetric cones}, Optimization
  methods and software, 11 (1999), pp.~625--653.

\bibitem{Sturm02}
\leavevmode\vrule height 2pt depth -1.6pt width 23pt, {\em Implementation of
  interior point methods for mixed semidefinite and second order cone
  optimization problems}, Optimization methods and software, 17 (2002),
  pp.~1105--1154.

\bibitem{Sturm99}
{\sc J.~F. Sturm and S.~Zhang}, {\em Symmetric primal-dual path-following
  algorithms for semidefinite programming}, Applied Numerical Mathematics, 29
  (1999), pp.~301--315.

\bibitem{sun2020sdpnal+}
{\sc D.~Sun, K.-C. Toh, Y.~Yuan, and X.-Y. Zhao}, {\em {SDPNAL+: A} matlab
  software for semidefinite programming with bound constraints (version 1.0)},
  Optimization Methods and Software, 35 (2020), pp.~87--115.

\bibitem{sutter2017approximate}
{\sc D.~Sutter, V.~B. Scholz, A.~Winter, and R.~Renner}, {\em Approximate
  degradable quantum channels}, IEEE Transactions on Information Theory, 63
  (2017), pp.~7832--7844.

\bibitem{Tarski1949-TARADM-3}
{\sc A.~Tarski}, {\em {A Decision Method for Elementary Algebra and Geometry}},
  Journal of Symbolic Logic, 14 (1949), pp.~188--188.

\bibitem{tavakoli2022informationally}
{\sc A.~Tavakoli, E.~Z. Cruzeiro, E.~Woodhead, and S.~Pironio}, {\em
  Informationally restricted correlations: a general framework for classical
  and quantum systems}, Quantum, 6 (2022), p.~620.

\bibitem{tavakoli2018self}
{\sc A.~Tavakoli, J.~Kaniewski, T.~V{\'e}rtesi, D.~Rosset, and N.~Brunner},
  {\em Self-testing quantum states and measurements in the prepare-and-measure
  scenario}, Physical Review A, 98 (2018), p.~062307.

\bibitem{tavakoli2022bell}
{\sc A.~Tavakoli, A.~Pozas-Kerstjens, M.-X. Luo, and M.-O. Renou}, {\em Bell
  nonlocality in networks}, Reports on Progress in Physics, 85 (2022),
  p.~056001.

\bibitem{tavakoli2019enabling}
{\sc A.~Tavakoli, D.~Rosset, and M.-O. Renou}, {\em Enabling computation of
  correlation bounds for finite-dimensional quantum systems via
  symmetrization}, Physical Review Letters, 122 (2019), p.~070501.

\bibitem{Terlaky96}
{\sc T.~Terlaky}, {\em Interior point methods of mathematical programming},
  vol.~5, Springer Science \& Business Media, 2013.

\bibitem{tilly2022variational}
{\sc J.~Tilly, H.~Chen, S.~Cao, D.~Picozzi, K.~Setia, Y.~Li, E.~Grant,
  L.~Wossnig, I.~Rungger, G.~H. Booth, and J.~Tennyson}, {\em The variational
  quantum eigensolver: a review of methods and best practices}, Physics
  Reports, 986 (2022), pp.~1--128.

\bibitem{T99}
{\sc M.~J. Todd}, {\em A study of search directions in primal-dual
  interior-point methods for semidefinite programming}, Optimization methods
  and software, 11 (1999), pp.~1--46.

\bibitem{todd2001semidefinite}
{\sc M.~J. Todd}, {\em Semidefinite optimization}, Acta Numerica, 10 (2001),
  pp.~515--560.

\bibitem{TTT98}
{\sc M.~J. Todd, K.-C. Toh, and R.~H. T{\"u}t{\"u}nc{\"u}}, {\em {On the
  Nesterov--Todd Direction in Semidefinite Programming}}, SIAM Journal on
  Optimization, 8 (1998), pp.~769--796.

\bibitem{Toh02}
{\sc K.-C. Toh}, {\em A note on the calculation of step-lengths in
  interior-point methods for semidefinite programming}, Computational
  Optimization and Applications, 21 (2002), pp.~301--310.

\bibitem{TTT99}
{\sc K.-C. Toh, M.~J. Todd, and R.~H. T{\"u}t{\"u}nc{\"u}}, {\em {SDPT3—a
  MATLAB software package for semidefinite programming, version 1.3}},
  Optimization methods and software, 11 (1999), pp.~545--581.

\bibitem{TTT12}
\leavevmode\vrule height 2pt depth -1.6pt width 23pt, {\em {On the
  implementation and usage of SDPT3--a Matlab software package for
  semidefinite-quadratic-linear programming, version 4.0}}, in Handbook on
  semidefinite, conic and polynomial optimization, Springer, 2012,
  pp.~715--754.

\bibitem{tomamichel2015quantum}
{\sc M.~Tomamichel}, {\em Quantum information processing with finite resources:
  mathematical foundations}, vol.~5, Springer, 2015.

\bibitem{SDPT3b}
{\sc R.~H. T{\"u}t{\"u}nc{\"u}, K.~C. Toh, and M.~J. Todd}, {\em Solving
  semidefinite-quadratic-linear programs using {SDPT3}}, Mathematical
  Programming, 95 (2003), pp.~189--217.

\bibitem{v1928theorie}
{\sc J.~v.~Neumann}, {\em Zur theorie der gesellschaftsspiele}, Mathematische
  annalen, 100 (1928), pp.~295--320.

\bibitem{Boyd95}
{\sc L.~Vandenberghe and S.~Boyd}, {\em A primal—dual potential reduction
  method for problems involving matrix inequalities}, Mathematical Programming,
  69 (1995), pp.~205--236.

\bibitem{sdp96}
\leavevmode\vrule height 2pt depth -1.6pt width 23pt, {\em {Semidefinite
  Programming}}, SIAM Review, 38 (1996), pp.~49--95.

\bibitem{V13}
{\sc V.~V. Vazirani}, {\em Approximation algorithms}, Springer Science \&
  Business Media, 2013.

\bibitem{vinnikov2012lmi}
{\sc V.~Vinnikov}, {\em {LMI} representations of convex semialgebraic sets and
  determinantal representations of algebraic hypersurfaces: past, present, and
  future}, Mathematical Methods in Systems, Optimization, and Control:
  Festschrift in Honor of J. William Helton,  (2012), pp.~325--349.

\bibitem{waki2006sums}
{\sc H.~Waki, S.~Kim, M.~Kojima, and M.~Muramatsu}, {\em Sums of squares and
  semidefinite program relaxations for polynomial optimization problems with
  structured sparsity}, SIAM Journal on Optimization, 17 (2006), pp.~218--242.

\bibitem{wang2009new}
{\sc G.~Q. Wang and Y.~Q. Bai}, {\em A new primal-dual path-following
  interior-point algorithm for semidefinite optimization}, Journal of
  Mathematical Analysis and Applications, 353 (2009), pp.~339--349.

\bibitem{wang2021exploiting}
{\sc J.~Wang and V.~Magron}, {\em Exploiting term sparsity in noncommutative
  polynomial optimization}, Computational Optimization and Applications, 80
  (2021), pp.~483--521.

\bibitem{wang2018multidimensional}
{\sc J.~Wang, S.~Paesani, Y.~Ding, R.~Santagati, P.~Skrzypczyk, A.~Salavrakos,
  J.~Tura, R.~Augusiak, L.~Man{\v{c}}inska, D.~Bacco, et~al.}, {\em
  Multidimensional quantum entanglement with large-scale integrated optics},
  Science, 360 (2018), pp.~285--291.

\bibitem{wang2016semidefinite}
{\sc X.~Wang and R.~Duan}, {\em A semidefinite programming upper bound of
  quantum capacity}, in 2016 IEEE International Symposium on Information Theory
  (ISIT), IEEE, 2016, pp.~1690--1694.

\bibitem{Watrous11}
{\sc J.~Watrous}, {\em {CS 867/QIC 890 Semidefinite Programming in Quantum
  Information}}.
\newblock Lecture notes, 2011.

\bibitem{watrous2018theory}
{\sc J.~Watrous}, {\em The theory of quantum information}, Cambridge University
  Press, 2018.

\bibitem{werner1989quantum}
{\sc R.~F. Werner}, {\em Quantum states with {Einstein-Podolsky-Rosen}
  correlations admitting a hidden-variable model}, Physical Review A, 40
  (1989), p.~4277.

\bibitem{seesaw1}
{\sc R.~F. Werner and M.~M. Wolf}, {\em Bell inequalities and entanglement},
  Quantum Info. Comput., 1 (2001), pp.~1--25.

\bibitem{wilde_2017}
{\sc M.~M. Wilde}, {\em {Quantum Information Theory}}, Cambridge University
  Press, 2~ed., 2017.

\bibitem{wiseman2007steering}
{\sc H.~M. Wiseman, S.~J. Jones, and A.~C. Doherty}, {\em Steering,
  entanglement, nonlocality, and the {Einstein-Podolsky-Rosen} paradox},
  Physical Review Letters, 98 (2007), p.~140402.

\bibitem{wittek2015algorithm}
{\sc P.~Wittek}, {\em Algorithm 950: Ncpol2sdpa—sparse semidefinite
  programming relaxations for polynomial optimization problems of noncommuting
  variables}, ACM Transactions on Mathematical Software (TOMS), 41 (2015),
  pp.~1--12.

\bibitem{Ncpol2sdpaGitHub}
{\sc P.~Wittek and P.~J. Brown}, {\em Updating ncpol2sdpa after {Peter
  Wittek}}.
\newblock \url{https://github.com/peterjbrown519/ncpol2sdpa}.

\bibitem{Wright05}
{\sc M.~Wright}, {\em The interior-point revolution in optimization: history,
  recent developments, and lasting consequences}, Bulletin of the American
  mathematical society, 42 (2005), pp.~39--56.

\bibitem{yamashita2012latest}
{\sc M.~Yamashita, K.~Fujisawa, M.~Fukuda, K.~Kobayashi, K.~Nakata, and
  M.~Nakata}, {\em Latest developments in the {SDPA} family for solving
  large-scale {SDP}s}, Handbook on semidefinite, conic and polynomial
  optimization,  (2012), pp.~687--713.

\bibitem{yamashita2003implementation}
{\sc M.~Yamashita, K.~Fujisawa, and M.~Kojima}, {\em Implementation and
  evaluation of {SDPA} 6.0 (semidefinite programming algorithm 6.0)},
  Optimization Methods and Software, 18 (2003), pp.~491--505.

\bibitem{yamashita2010high}
{\sc M.~Yamashita, K.~Fujisawa, K.~Nakata, M.~Nakata, M.~Fukuda, K.~Kobayashi,
  and K.~Goto}, {\em A high-performance software package for semidefinite
  programs: {SDPA} 7}, Tokyo, Japan,  (2010).

\bibitem{yang2015sdpnal}
{\sc L.~Yang, D.~Sun, and K.-C. Toh}, {\em {SDPNAL++: A majorized semismooth
  Newton-CG augmented Lagrangian method for semidefinite programming with
  nonnegative constraints}}, Mathematical Programming Computation, 7 (2015),
  pp.~331--366.

\bibitem{yang2013robust}
{\sc T.~H. Yang and M.~Navascu{\'e}s}, {\em Robust self-testing of unknown
  quantum systems into any entangled two-qubit states}, Physical Review A, 87
  (2013), p.~050102.

\bibitem{yang2014robust}
{\sc T.~H. Yang, T.~V{\'e}rtesi, J.-D. Bancal, V.~Scarani, and
  M.~Navascu{\'e}s}, {\em Robust and versatile black-box certification of
  quantum devices}, Physical Review Letters, 113 (2014), p.~040401.

\bibitem{yu2021complete}
{\sc X.-D. Yu, T.~Simnacher, N.~Wyderka, H.~C. Nguyen, and O.~G{\"u}hne}, {\em
  A complete hierarchy for the pure state marginal problem in quantum
  mechanics}, Nature Communications, 12 (2021), pp.~1--7.

\bibitem{zalinescu2002convex}
{\sc C.~Zalinescu}, {\em Convex analysis in general vector spaces}, World
  Scientific, 2002.

\bibitem{Schur06}
{\sc F.~Zhang}, {\em The {S}chur complement and its applications}, vol.~4,
  Springer Science \& Business Media, 2006.

\bibitem{Z98}
{\sc Y.~Zhang}, {\em {On Extending Some Primal--Dual Interior-Point Algorithms
  From Linear Programming to Semidefinite Programming}}, SIAM Journal on
  Optimization, 8 (1998), pp.~365--386.

\bibitem{zhang1992superlinear}
{\sc Y.~Zhang, R.~A. Tapia, and J.~E. Dennis, Jr}, {\em On the superlinear and
  quadratic convergence of primal-dual interior point linear programming
  algorithms}, SIAM Journal on Optimization, 2 (1992), pp.~304--324.

\bibitem{zhao2010newton}
{\sc X.-Y. Zhao, D.~Sun, and K.-C. Toh}, {\em A {Newton-CG} augmented
  {L}agrangian method for semidefinite programming}, SIAM Journal on
  Optimization, 20 (2010), pp.~1737--1765.

\end{thebibliography}

	\appendix
	
	\section{Proof of the Decoupling Lemma}
	\label{sec:decouplingProof}
	
	We will now provide a proof of the Decoupling Lemma used in sec.~\ref{sec:FR}. We follow the line of~\cite{BZ05,borwein2006variational}. The lemma is used as constraint qualification condition, i.e. it provides the strong duality sufficient criteria of the Fenchel-Rockafellar scheme. To this end we first introduce the concept of convex series.
	
	\paragraph{Convex series.}
	
	The notion of convex series was introduced in~\cite{jameson1972convex}; see~\cite[p.113nn]{BZ05} for a detailed discussion.
	\textit{Convex series of $C$}\index{convex!series of a set} are series of the form $\sum_{i = 1}^{+\infty} \lambda_i x_i$ with $\bigforall_{i} x_i \in C$, $\bigforall_{i} \lambda_i \geq 0$ and $\sum_{i = 1}^{+\infty} \lambda_i = 1$.
	$C$ is defined to be \textit{convex series closed}\index{convex!series closed set} if the sum of every convergent convex series of $C$ is contained in $C$.
	One can check that if $C$ is convex series closed, then~\cite[p.116]{BZ05}:
	\begin{equation}
		\label{eq:CSintEQcore}
		\interior{C} = \core{C}.
	\end{equation}
	$C$ is defined to be \textit{convex series compact}\index{convex!series compact set} if every convex series of $C$ converges to an element of $C$.
	It can be shown that $C$ is convex series compact if and only if, it is convex series closed and bounded. Thus, $\ball_X$ is convex series compact.
	
	\paragraph{Preliminary comment.}
	$F$ defined as~\eqref{eq:FxyFR} is also $\lsc$, since $A$ is continuous. Without loss of generality, we also assume that $f(0) = g(0) = 0$, since if it is not the case, we can trivially transform the primal problem~\eqref{eq:FR-primal} shifting it by a constant value. From this it follows $F(0,0) = 0$ and
	\begin{equation}
		\label{eq:Fl_leq _lF}
		\bigforall{\lambda \in [0,1], x \in X, y \in Y} F(\lambda x, \lambda y) \leq \lambda F(x,y).
	\end{equation}
	
	\paragraph{Step 1: Define the convex set $S$.}
	Let us define~\cite[p.127]{BZ05}
	\begin{equation}
		\label{eq:S_in_DecouplingLemma}
		S \equiv \bigcup_{x \in \ball_X} \left\{ y \in Y : F(x,y) \leq 1 \right\}.
	\end{equation}
	Let $y_0, y_1 \in S$. Then there exist $x_0, x_1 \in \ball_X$ such that $F(x_i, y_i) \leq 1$ for $i = 0,1$. For any $\lambda \in [0,1]$ from the convexity of $F$ we have $F(\lambda x_0 + (1-\lambda) x_1, \lambda y_0 + (1-\lambda) y_1) \leq \lambda F(x_0, y_0) + (1-\lambda) F(x_1, y_1) \leq 1$ and thus $\lambda y_0 + (1-\lambda) y_1 \in S$ implying convexity of $S$.
	
	\paragraph{Step 2: Show that $0 \in \core{S}$.}
	From the definition~\eqref{eq:coreOfSet} and the assumption given in~\eqref{eq:decouplingLemmaCond1} we get
	%\begin{equation}
	$\bigforall_{y \in Y} \bigexists_{T_y > 0} (T_y y) \in \dom{\theta} = \dom{g} - A \dom{f}$.
	%\end{equation}
	Consider arbitrary $y \in Y$ and take any $x_y \in \dom{f}$ such that $T_y y + A x_y \in \dom{g}$. Then $k_y \equiv F(x_y, T_y y) < +\infty$.
	We want to ensure that $\bigexists_{\alpha_y > 0} \alpha_y y \in S$ so that $S$ is absorbing.
	Indeed, let $\alpha_y = \min(1/{k_y}, 1/\Norm{x_y})$. This implies that $F(\alpha_y x_y, \alpha_y y) \leq 1$ (by~\eqref{eq:Fl_leq _lF}) and that $\alpha_y x_y \in \ball_X$, and thus $\alpha_y y \in S$. Since $y$ represents arbitrary direction in $Y$, so $0 \in \core{S}$ (and, what is more, $S$ is an absorbing set).
	
	\paragraph{Step 3: Show that $\core{S} = \interior{S}$.}
	We will use~\eqref{eq:CSintEQcore}. To this end we check that $S$ is convex series closed. Indeed, consider any convergent convex series of $S$, $\sum_{i = 1}^{+\infty} \lambda_i y_i$ with $\bigforall_{i} y_i \in S$, summing to some $y$.	
	It sufficies to show that $y \in S$.
	From~\eqref{eq:S_in_DecouplingLemma} it follows that
	%\begin{equation}
	$\bigforall_{i} \bigexists_{x_i \in \ball_X} F(x_i, y_i) \leq 1$.
	%\end{equation}
	The series $\sum_{i = 1}^{+\infty} \lambda_i x_i$ converges to some $x$ since $\ball_X$ is convex series compact.
	We have
	%\begin{equation}
	$F(x, y) = \sum_{i = 1}^{+\infty} F(\lambda_i  x_i, \lambda_i  y_i) \leq \sum_{i = 1}^{+\infty} \lambda_i F(x_i, y_i) \leq \sum_{i = 1}^{+\infty} \lambda_i = 1$,
	%\end{equation}
	where for the first inequality we used the assumption that $F$ is $\lsc$ and~\eqref{eq:Fl_leq _lF}. Thus, $y \in S$, and any convergent convex series of $S$ has sum $y$ contained in $S$ meaning that $S$ is convex series closed.
	Using~\eqref{eq:CSintEQcore} we conclude that for $S$ we have $\core{S} = \interior{S}$.
	
	\paragraph{Step 4: Show that $\theta$ is continuous in the neighborhood of $0$.}
	It can be shown~\cite[p.112]{BZ05} that for a Banach space $Z$ a convex function $f: Z \rightarrow \mathbb{R} \cup \{+\infty\}$, locally bounded above at $z \in \interior(\dom{f})$ is also locally Lipschitz at $z$. Thus it is also \textit{a fortiori} continuous at $z$.
	From~\eqref{eq:valueFunctionFR} and~\eqref{eq:S_in_DecouplingLemma} it follows that $\bigforall_{y \in S} \theta(y) \leq 1$, so $\theta$ is continuous at $0 \in \interior{S}$.
	
	\section{Code samples in Matlab}
	
	% TODO: skonfigurować formatowanie i kolorowanie kodu
	
	% TODO: re-execute
	\subsection*{Illustration of simple problem formulations using YALMIP}
	\label{sec:matlab:1stYALMIP}
	
	We start code discussions with a trivial example of SDP finding the largest eigenvalue of a matrix. The purpose of this example is to provide a short overview of the syntax characteristic to usage of the YALMIP modelling toolbox~\cite{yalmip}. To install it one should follow the instructions from the repository and also install one of the supported solvers, e.g. SDPT3~\cite{TTT99} or SeDuMi~\cite{SeDuMi,SeDuMiGitHub}. %TODO adres repozytorium YALMIP i SDPT3
	
	We start with randomizing a Hermitian $3$ by $3$ matrix \Minline{X} using the ordinary Matlab syntax. We see in Listing~\ref{matlab:1stYALMIP-randEigX} that in this instance the eigenvalues were $-0.086472$, $0.68428$, and $3.227$.
\begin{lstlisting}[caption={Random matrix for the YALMIP example.},label={matlab:1stYALMIP-randEigX},language=MATLAB,float,linerange={1-9}]
>> X = rand(3) + 1i * rand(3); X = X + X';
>> eig(X)

ans =

    -0.086472
      0.68428
        3.227
\end{lstlisting}
	
	Next, we define a $3$ by $3$ Hermitian variable in Listing~\ref{matlab:1stYALMIP-sdpvar}. The function \Minline{sdpvar} is used for this purpose. The first two parameters are the number of rows $n_r$ and columns $n_c$ of the matrix. The third parameter specifies the structure of the variable. One of the possibilities include \Minline{'full'} when all entries of the matrix are parametrized independently, meaning $n_r n_c$ parameters for real, and $2 n_r n_c$ parameters for complex matrices. Another possibility for the parametrization when $n_r = n_c = n$ is \Minline{'symmetric'} meaning that the element at $i$-th row and $j$-th column is exactly equal to the one at $j$-row and $i$-th column, using $n (n+1) / 2$ parameters for real, and $n^2$ parameters for complex matrices. A third possiblity is \Minline{'hermitian'} meaning that the element at $i$-th row and $j$-th column is equal to the complex conjugate of the element at $j$-row and $i$-th column. Other possibile structures are \Minline{'diagonal'} for diagonal matrices, \Minline{'toeplitz'} for symmetric Toeplitz matrices, \Minline{'hankel'} for unsymmetric Hankel matrices, \Minline{'rhankel'} for symmetric Hankel matrices, and \Minline{'skew'} for skew-symmetric matrices. When a real square matrix variable is to be created, an abbreviated form \Minline{sdpvar(n)} can be used to create $n$ by $n$ real symmetric matrix.
\begin{lstlisting}[caption={Creation of a $3$ by $3$ Hermitian variable in YALMIP.},label={matlab:1stYALMIP-sdpvar},language=MATLAB,float,linerange={10-14}]
>> S = sdpvar(3,3,'hermitian','complex')

Linear matrix variable 3x3 (hermitian, complex, 9 variables)
Coeffiecient range: 1 to 1
\end{lstlisting}
	
	We can notice that the coefficient range is $\{1\}$, meaning that all coefficients in the variable \Minline{S} are equal $1$. It is possible to use the parametrized variables of the type \Minline{sdpvar} with various coefficent range. Usually, if the coefficients spread by several orders of magnitude, meaning that the program is mixing large and small coefficients, this lead a solver to get into numerical problems. Similar problems may happen, if the coefficients are very large or very small. The variables that occur with very small coefficients usually do not influence significantly the value of the solution, and can be removed using the \Minline{clean} function from the YALMIP, as shown in Listing~\ref{matlab:1stYALMIP-coeffs}. % https://yalmip.github.io/inside/debuggingnumerics/
	\lstinputlisting[caption={\Minline{sdpvar} with large range of coefficients, and removing small coefficients with \Minline{clean} function.},label={matlab:1stYALMIP-coeffs},language=MATLAB,float,linerange={1-8}]{short\_codes.m}
	
	Now, we will execute the optimization with the command \Minline{optimize(F = [S >= 0; trace(S) == 1], target = -trace(X * S))}. The first argument of this function specifies the constraints of the optimization, and the second is the target of the optimization. We assigned the constraints to the variable \Minline{F}. Let us investigate this variable as shown in Listing~\ref{matlab:1stYALMIP-F}. We have specified the positive semi-definiteness constraint \Minline{S >= 0} on the complex $3$ by $3$ matrix \Minline{S}, i.e. \Minline{S >= 0}. This is described as \Minline{Matrix inequality (complex) 3x3}. The second imposed constraint of trace normalization, \Minline{trace(S) == 1}, is described as \Minline{Equality constraint 1x1}.
	\lstinputlisting[caption={Investigation of a sample constraint in SDP: positive semi-definiteness of \Minline{S}, i.e. \Minline{S >= 0}, and trace normalization, i.e. \Minline{trace(S) == 1}.},label={matlab:1stYALMIP-F},language=MATLAB,float,linerange={9-17}]{short\_codes.m}
	
	An optional third argument to \Minline{optimize} can specify additional optimization parameter, like selection of the solver, specification of how much information in during the execution of the optimization should be printed to the screen, the maximal number of iterations (if the solver allows for it). For instance, to specify that from all available solvers, YALMIP should use SDPT3, do not print any information, and limit the number of iterations to $20$ one can provide a setting \Minline{sdpsettings('solver', 'sdpt3', 'verbose', 0, 'sdpt3.maxit', 20)}. Other useful option \Minline{'showprogress'} allowing to show the progress of YALMIP, which is useful for debugging purpose for very large problems. Recall that with a modelling tool, before the optimization the problem is being converted to the form suitable for the solver, usually the canonical form discussed sec.~\ref{sec:CanonicalSDP}, or SDPA form discussed in sec.~\ref{sec:BoydSDP}. This formulation in some cases may take more time than the actual solver time. The option \Minline{'removeequalities'} specifies how constraints of the equalities should be preprocessed before passing to the solver, as discussed in sec.~\ref{sec:slackAndEqual}. The option \Minline{'dualize'} tells YALMIP to fit the problem formulation to the primal  instead of the dual form. % TODO: odkomentować, gdy będzie gotowe: , as discussed further in~\ref{App:PrimalDualExample}.
	If multiple optimizations are to executed we recommend to store the settings in a separate variable and provide it is as the third optional argument to \Minline{optimize}, especially for smaller problems. The reason for this is the fact that if this argument in not provided explicitely, then YALMIP will re-create its content for each execution of \Minline{optimize}, which requires additional computational time. Turning on the option \Minline{'showprogress'} shows that the stages with the default settings (with YALMIP version 20210331) are as shown in Listing~\ref{matlab:1stYALMIP-stages}.
	\lstinputlisting[caption={Stages of YALMIP processing in the discussed example.},label={matlab:1stYALMIP-stages},language=MATLAB,float,linerange={18-23}]{short\_codes.m}
	
	At this stage, the control is passed to the specified solver. Recall from the discussion in sec.~\ref{sec:softwareAndIPM} that YALMIP provides to the solver the problem framed in the dual form~\eqref{SDP-dual}. The solver usually prints the values describing size parameters of the problem passed to solver, as shown in Listing~\ref{matlab:1stYALMIP-dim} for the case of the solver SDPT3. The Listing~\ref{matlab:1stYALMIP-dim} shows that the dimension of SDP variable is $6$. This stems from the fact, that in the stage \Minline{Converting to real constraints} YALMIP has reformulated the $n$ by $n$ complex variable to $2n$ by $2n$ real variable, as discussed in sec.~\ref{sec:complexSDP}; in the considered example $n = 3$. This reformulation as a real variables problem, requires stating a requirement that the dual SDP variable $Z$ has the form~\eqref{eq:HermitianToSymmetric}. It is easy to see that this requires $\frac{n \cdot (n + 1)}{2}$ matrices $A_i$ to express that the two blocks containing $B^R$, and $\frac{n \cdot (n - 1)}{2}$ matrices $A_i$ to express the relation $B^I = -B^I$ in the off-diagonal block. This gives in total $n^2$ matrices $A_i$, what yields $9$ in this case, displayed as \Minline{num. of constraints =  9}. Each of the problem's constraints framed in the dual form corresponds to an additional free variable in the primal problem, as discussed in sec.~\ref{sec:slackAndEqual}. The fact that there is only one equality \Minline{trace(S) == 1} in the case, is expressed as \Minline{dim. of free   var  =  1} in Listing~\ref{matlab:1stYALMIP-dim}.
\begin{lstlisting}[caption={Sample information about size of the problem to be passed to the solver as printed by SDPT3.},label={matlab:1stYALMIP-dim},language=MATLAB,float,linerange={17-19}]
num. of constraints =  9
dim. of sdp    var  =  6,   num. of sdp  blk  =  1
dim. of free   var  =  1
\end{lstlisting}
	
	We will briefly analyse the two of the settings, \textit{viz.} \Minline{'removeequalities'} and \Minline{'dualize'}. Their default values are both $0$, and this is the case analysed in the previous paragraph. If we set \Minline{'removeequalities'} to $1$, then the discussed stages will result in output given in Listing~\ref{matlab:1stYALMIP-removeequalities}. We notice that before calling the solver, two additional stages of YALMIP processing are taken, \textit{viz.} \Minline{Solving equalities} and \Minline{Converting problem to new basis}, both responsible in reduction of the number of variables and  reformulation of the dual form, as discussed in sec.~\ref{sec:slackAndEqual}. Comparing with Listing~\ref{matlab:1stYALMIP-dim} we see that the effect is the reduction of the number of matrices $A_i$ from $m = 9$ to $8 = m - n_F$ and, at the same time, removal of the $n_F = 1$ free variables. On the test platform the problem size reduction resulted in drop of the solver time from $0.48$s to $0.24$s, at the cost of additional YALMIP processing time increased to $0.14$s from $0.10$s.
	\lstinputlisting[caption={Size of the sample problem modelled with YALMIP with \Minline{'removeequalities'} set to $1$ framed in the canonical dual form.},label={matlab:1stYALMIP-removeequalities},language=MATLAB,float,linerange={25-35}]{short\_codes.m}
	If we set \Minline{'dualize'} to $1$, then the problem will be framed in the primal form~\eqref{SDP-primal} instead. Again, the complex SDP of size $n$ will be expressed as a real SDP of size $2 n$, as explained above and in sec.~\ref{sec:complexSDP}. From Listing~\ref{matlab:1stYALMIP-dualize} we see that indeed the size of the SDP variable is $6$, and there is only one constraint $\{A_i\}$ to express the requirement \Minline{trace(S) == 1}.
	\lstinputlisting[caption={Size of the sample problem modelled with YALMIP with \Minline{'dualize'} set to $1$ in order to frame the optimization problem in the canonical primal form.},label={matlab:1stYALMIP-dualize},language=MATLAB,float,linerange={39-42}]{short\_codes.m}
	
	The SDPT3 solver will provide also information about the chosen algorithm, as shown in Listing~\ref{matlab:1stYALMIP-algorithmData}. In this case the algorithm uses the HKM search direction, see~\eqref{eq:HKM_direction}. Other input parameters of SDPT3 are \Minline{gam} and \Minline{expon}, and they are used to calculate the value of the step-length $\alpha_P$ and $\text{expon\_used}$ in~\eqref{eq:PredCorr_SDPT3} for the predictor-corrector\index{predictor-corrector} mechanism, as discussed in sec.~\ref{sec:PredictorCorrector}. 
\begin{lstlisting}[caption={Sample information about size of the parameters of the algorithm used by SDPT3.},label={matlab:1stYALMIP-algorithmData},language=MATLAB,float,linerange={20-25}]
*** convert ublk to linear blk
***************************************************************************
SDPT3: homogeneous self-dual path-following algorithms
***************************************************************************
 version  predcorr  gam  expon
   HKM      1      0.000   1
\end{lstlisting}
	
	The core part of solving of an SDP is an iterative procedure of gradual improvement of the solution $(X^{(i)},y^{(i)},Z^{(i)})$, with a sample progress shown in Listing~\ref{matlab:1stYALMIP-iterate}. The first column is the iteration number, here the solver finished after $14$-th iteration. The second and the third are primal and dual step-lengths, see~\eqref{eq:stepPSDcond}, taken in each iteration. When the step-lengths are close to $1$, it means that the search direction allowed for a large change in the values of $(X^{(i)},y^{(i)},Z^{(i)})$, what usually indicates a significant improvement of the solution in the iteration. The fourth and fifth columns are residuals norms of the primal\index{primal residual norm} and dual\index{dual residual norm} solutions, as discussed in sec.~\ref{sec:IPM}. The residual norms are expected to be close to $0$ when the solution is feasible, i.e. satisfies all the imposed constraints. The sixth columns is proportional to the gap~\eqref{eq:gapTrZX}, and also should be close to $0$ when the iterates approach the solution. Due to numerical inaccuracy, one usually considers values of the gap and the residual norms close to $10^{-7}$ as satisfactory. The $7$-th columns \Minline{mean(obj)} is the mean value of the primal $\Tr(C X^{(i)})$ and dual $b^T \cdot y^{(i)}$ solutions of the current iterate. The $8$-th column provides the time passed til the current iteration (in the example it passed less than one second). The following three columns, \Minline{kap}, \Minline{tau}, \Minline{theta} provide information about specific parameters used in determination of the step-length\index{step-length}, and the last column provides information about the Cholesky factorization taken in calculation of the search direction; these information are beyond the scope of this work.
\begin{lstlisting}[caption={Iterations of interior-point method in SDPT3.},label={matlab:1stYALMIP-iterate},language=MATLAB,float,linerange={26-43},basicstyle=\tiny]
it pstep dstep pinfeas dinfeas  gap     mean(obj)    cputime    kap   tau    theta
--------------------------------------------------------------------------------------------
 0|0.000|0.000|1.1e+01|1.2e+01|6.1e+02| 4.323372e-02| 0:0:00|6.1e+02|1.0e+00|1.0e+00| chol 1  1
 1|1.000|1.000|9.5e+00|1.0e+01|7.3e+02| 9.283979e-01| 0:0:00|4.2e+02|1.0e+00|8.1e-01| chol 1  1
 2|1.000|1.000|1.5e+00|1.7e+00|7.5e+01| 5.926998e+00| 0:0:00|8.7e+01|1.1e+00|1.5e-01| chol 1  1
 3|0.881|0.881|2.5e-01|2.8e-01|9.0e+00| 3.744834e+00| 0:0:00|8.7e-01|1.4e+00|3.2e-02| chol 1  1
 4|0.885|0.885|4.2e-02|4.7e-02|1.1e+00| 2.992843e+00| 0:0:00|1.3e+00|1.6e+00|6.2e-03| chol 1  1
 5|1.000|1.000|5.0e-03|7.5e-03|1.3e-01| 3.231258e+00| 0:0:00|2.7e-01|1.7e+00|7.7e-04| chol 1  1
 6|0.862|0.862|1.2e-03|4.6e-03|2.6e-02| 3.233302e+00| 0:0:00|8.1e-02|1.8e+00|1.9e-04| chol 1  1
 7|0.994|0.994|1.4e-04|1.6e-03|4.6e-04| 3.230829e+00| 0:0:00|1.4e-02|1.8e+00|2.2e-05| chol 1  1
 8|1.000|1.000|1.6e-05|6.1e-04|2.8e-05| 3.228448e+00| 0:0:00|1.7e-03|1.8e+00|2.5e-06| chol 1  1
 9|1.000|1.000|1.8e-06|2.4e-04|3.7e-06| 3.227543e+00| 0:0:00|1.9e-04|1.8e+00|2.9e-07| chol 1  1
10|1.000|1.000|2.1e-07|9.6e-05|5.0e-07| 3.227187e+00| 0:0:00|2.2e-05|1.8e+00|3.3e-08| chol 1  1
11|1.000|1.000|2.5e-08|1.9e-05|6.1e-08| 3.226998e+00| 0:0:00|2.6e-06|1.8e+00|3.8e-09| chol 1  1
12|1.000|1.000|5.2e-09|3.9e-06|1.6e-07| 3.226960e+00| 0:0:00|3.0e-07|1.8e+00|8.2e-10| chol 1  1
13|1.000|1.000|1.1e-09|7.7e-07|3.7e-08| 3.226953e+00| 0:0:00|6.5e-08|1.8e+00|1.8e-10| chol 1  1
14|1.000|1.000|2.5e-10|7.7e-08|8.1e-09| 3.226951e+00| 0:0:00|1.4e-08|1.8e+00|4.0e-11|
  Stop: max(relative gap,infeasibilities) < 1.00e-07
\end{lstlisting}
	
	The SDPT3 solver provides a brief summary of its execution, as shown in Listing~\ref{matlab:1stYALMIP-SDPT3summary}. The data include the number of iterations, the value of the primal solution $\Tr(C X^{(i)})$, and the value of the dual solution $b^T \cdot y^{(i)}$, the value of the , gap~\eqref{eq:gapTrZX}, and the relative gap (i.e. the gap divided by $1$ plus the mean value of the primal and dual solutions), the parameters measuring infeasibility (which should be close to $0$), the norms of the solution in the last iteration $(X^{(i)},y^{(i)},Z^{(i)})$, the norms of the matrices defining the problem $\{A_i\}_i$, $b$, and $C$, the total CPU time, and the CPU time per iteration, the termination code. The succesful termination code is $0$. The values $-1$,$-5$, and $-9$ indicate lack of progress, when the improvements are too slow; $1$ and $2$ indicate dual or primal infeasibility of the solution, $-6$ indicates that the maximal number of iteration has been reached before the desired quality of the solution was obtained; there is also a couple of value indicating various numerical problems. The last information contains the so-called DIMACS statistics for standarized benchmarking purposes~\cite{MittelmannBenchmark}.
\begin{lstlisting}[caption={Summary of SDPT3 execution.},label={matlab:1stYALMIP-SDPT3summary},language=MATLAB,float,linerange={45-58}]
 number of iterations   = 14
 primal objective value =  3.22695066e+00
 dual   objective value =  3.22695103e+00
 gap := trace(XZ)       = 8.06e-09
 relative gap           = 1.91e-09
 actual relative gap    = -4.99e-08
 rel. primal infeas     = 2.45e-10
 rel. dual   infeas     = 7.73e-08
 norm(X), norm(y), norm(Z) = 4.4e+00, 8.3e-01, 1.4e+00
 norm(A), norm(b), norm(C) = 2.5e+01, 4.1e+00, 1.4e+00
 Total CPU time (secs)  = 0.40
 CPU time per iteration = 0.03
 termination code       =  0
 DIMACS: 2.5e-10  0.0e+00  7.7e-08  0.0e+00  -5.0e-08  1.1e-09
\end{lstlisting}
	
	\subsection*{Correlation matrix}
	\label{app:corrYalmip}
	
	Now, we provide an illustration of the concept of SDP optimization over correlation matrices, as discussed in sec.~\ref{sec:NPA}. Let us consider a set $\mathcal{S} = \{x_1, x_2, x_3\}$. Suppose we have obtained information, possibly from experimental data, that $\corr(x_1, x_2) = 0.7 \pm 0.03$ and $\corr(x_1, x_3) = 0.8 \pm 0.01$. The question of interest is: What is the range of possible values for $\corr(x_2, x_3)$? To answer this, we construct a real $3$ by $3$ correlation matrix with diagonal elements equal to $1$ and impose constraints on its entries based on the specified ranges. We then perform maximization and minimization of the entry corresponding to $\corr(x_2, x_3)$ to determine its possible range. The results, as shown in Listing~\ref{matlab:corr-result}, indicate that $\corr(x_2, x_3)$ lies within the interval $[0.074153, 0.99573]$.
	
	\lstinputlisting[caption={Correlation matrix example in YALMIP.},label={matlab:corr},language=MATLAB,float,linerange={47-50}]{short\_codes.m}
	
	\lstinputlisting[caption={Results of the correlation matrix example calculations.},label={matlab:corr-result},language=MATLAB,float,linerange={64-75}]{short\_codes.m}

	\subsection*{Quantum State Discrimination}
	\label{App:QSD}
	\index{Quantum State Discrimination}
	
%	How to discriminate $N$ non-orthogonal quantum states $\rho_1, \rho_2, \cdots, \rho_N$?
%	Use a measurement $M_1, M_2, \cdots, M_N$ and try to maximize average success:
%	\begin{equation}
%		\sum_i^N \Tr \left( \rho_i M_i \right).
%	\end{equation}
%	One can simply consider cases where weights of different states are specified.

	In the task of discriminating $N$ non-orthogonal quantum states, the goal is to employ a measurement strategy using operators $M_1, M_2, \cdots, M_N$ in order to maximize the average success of the discrimination. This average success is quantified by the expression $\frac{1}{N} \sum_{i \in [N]} \Tr \left( \rho_i M_i \right)$, where $\rho_i$ represents the quantum state and $M_i$ corresponds to the measurement operator for the $i$-th state. By optimizing this expression, one can effectively differentiate between the given non-orthogonal states.
	
		\begin{lstlisting}[caption={Quantum State Discrimination in YALMIP},label={matlab:QSD},language=MATLAB]
		>> state1 = RandomState(3);
		>> state2 = RandomState(3);
		>> state3 = RandomState(3);
		>> meas1 = sdpvar(3,3,'hermitian','complex');
		>> meas2 = sdpvar(3,3,'hermitian','complex');
		>> meas3 = sdpvar(3,3,'hermitian','complex');
		>> optimize([meas1 >= 0; meas2 >= 0; meas3 >= 0; meas1+meas2+meas3 == eye(3)], -trace(state1*meas1 + state2*meas2 + state3*meas3)/3)
	\end{lstlisting}
	
	The Matlab code provided in Listing~\ref{matlab:QSD} demonstrates quantum state discrimination using the YALMIP optimization toolbox. The code begins by generating three random quantum states of dimension 3: \Minline{state1}, \Minline{state2}, and \Minline{state3}. These states represent the quantum systems to be discriminated. Next, the code declares three measurement variables, \Minline{meas1}, \Minline{meas2}, and \Minline{meas3}, using the \Minline{sdpvar} function from YALMIP. These variables are Hermitian matrices of size $3$ by $3$, representing the measurement operators corresponding to each state. The optimization problem is formulated using the \Minline{optimize} function, which takes an objective function and a set of constraints as inputs. The objective function aims to maximize the average success rate of discrimination, given by the expression \Minline{-trace(state1*meas1 + state2*meas2 + state3*meas3)/3}. This expression calculates the average trace of the product of the state and measurement operators. The division by $3$ accounts for the number of states being discriminated. The constraints include ensuring that each measurement operator is positive semidefinite \Minline{(meas1 >= 0, meas2 >= 0, meas3 >= 0)} and that the sum of all measurement operators equals the identity matrix \Minline{(meas1 + meas2 + meas3 == eye(3))}. These constraints guarantee that the measurement operators are valid and form a valid measurement scheme. The output of the optimization will provide the optimal measurement operators that achieve the highest discrimination performance.
	
	We note that in the discrimination process, it is sometimes useful to consider scenarios where the weights of the different states are specified. By assigning specific weights to each state, the discrimination problem can be formulated in a more structured manner. This approach allows for a more targeted optimization of the average success metric, leading to enhanced discrimination capabilities. By leveraging knowledge about the weights of the states, researchers can design measurement schemes and strategies that are tailored to maximize the overall success rate in discriminating the non-orthogonal quantum states.
	
\end{document}